\documentclass[12pt]{article}

\usepackage{amsmath,amssymb,amsfonts,amsthm}
\usepackage{graphicx}
\usepackage{cite}
\usepackage[all]{xy}
\usepackage[toc,page]{appendix}
\usepackage{hyperref}
\hypersetup{colorlinks=true,  citecolor=red, linkcolor=blue}
\usepackage[mathcal]{eucal}
\usepackage{xcolor}

\newcommand{\draftnew}[1]{#1}

\newcounter{algorithma}
\newcommand{\algorithma}[2]{\refstepcounter{algorithma}\vspace{1.5ex}\noindent
\fbox{\parbox{0.965\textwidth}{\textbf{Algorithm \thealgorithma\ (#1).}\ #2}}\vspace{1.5ex}}

\allowdisplaybreaks[3]

\newcounter{propositiona}
\newcommand{\propositiona}[1]{\refstepcounter{propositiona}
\noindent
\textbf{Proposition \thepropositiona.}\, {\it #1}}
\newcounter{definitiona}

\newcounter{remarka}
\newcommand{\remarka}[1]{\refstepcounter{remarka}
\noindent
\textbf{Remark \theremarka.}\, #1}
\newcounter{examplea}
\newcommand{\examplea}[1]{\refstepcounter{examplea}
\noindent
\textbf{Example \theexamplea.}\, #1}
\newcounter{lemmaa}

\newcounter{theorema}
\newcommand{\theorema}[1]{\refstepcounter{theorema}
\noindent
\textbf{Theorem\, \thetheorema.}\, {\it #1}}
\newcounter{corollarya}

\renewcommand{\thefootnote}{\alph{footnote}}

\title{Computational Algorithms for Invariant Reduction of Variational Forms}

\author{ \renewcommand{\thefootnote}{\alph{footnote}}
Kostya Druzhkov\footnotemark[1],~~Alexey Shevyakov\footnotemark[2]\vspace{0.5cm}\\
\small $^{\rm a,b}$\emph{Department of Mathematics and Statistics, University of Saskatchewan, Saskatoon, Canada}\vspace{0.2cm}\\
}

\begin{document}

\footnotetext[1]{Corresponding author. Electronic mail: konstantin.druzhkov@gmail.com}
\footnotetext[2]{Electronic mail: shevyakov@math.usask.ca. Alexei Cheviakov is an earlier spelling of this author's name used in many of his publications.}

\maketitle \numberwithin{equation}{section}
\renewcommand{\thefootnote}{\arabic{footnote}}

\begin{abstract}
	Symmetry reductions of partial differential equations (PDEs) inherit more geometric structures than other types of reductions: invariant conservation laws, variational structures, and, under suitable conditions, Hamiltonian-type structures of the original model descend to the reduced model through the mechanism of invariant reduction. This paper develops computational algorithms that carry out such reductions explicitly. We use the interpretation of variational $p$-forms as conservation laws of an enlarged system --- the (degree-shifted) tangent system --- consisting of the original equations together with their linearizations, in which the perturbation variables are treated as anticommuting. This allows us to formulate the algorithms using well-known concepts from the theory of conservation laws, naturally adapted to the graded-commutative setting. The reduction of conservation laws, variational $1$-forms, and presymplectic structures thereby becomes a single algorithmic procedure. We present (i)~a homotopy-based reduction algorithm for systems of evolution equations, implemented in \texttt{Maple}; (ii) a descent reduction algorithm applicable to general $\ell$-normal systems for $p>0$; and (iii)~a simple reduction algorithm available for point symmetries under suitable conditions, including the description of reductions in terms of systems involving fewer independent variables. Examples include nonlinear evolution equations in one and two spatial dimensions, the Laplace equation, the incompressible Euler equations, and the cotangent system of Pavlov's equation.
\end{abstract}


\section{Introduction}

Symmetry reduction is one of the most widely used exact techniques for nonlinear partial differential equations (PDEs). Given a PDE system and a Lie symmetry, solutions invariant under the symmetry satisfy a simplified system with fewer independent variables; traveling waves, similarity solutions, and rotationally or helically symmetric flows are all instances of this construction~\cite{Ovsiannikov1982, BlumanKumei1989, Olver1993, BCA2010, Cantwell2002}. In the case of a higher symmetry, or when global aspects of the reduction are important, one can describe invariant solutions using an overdetermined system with the same number of independent variables as the original one. This system is obtained by imposing the supplementary condition that a characteristic of the symmetry vanishes. In both scenarios, the reduction of the \emph{equations} is classical. The subject of this paper is the reduction of the \emph{structures} attached to the equations: conservation laws, variational $1$-forms\footnote{They can be regarded as intrinsic counterparts of cosymmetries (or adjoint-symmetries).}, and presymplectic structures. These structures carry essential physical and analytical information --- conserved densities and fluxes, variational principles, the correspondence between Noether symmetries and conservation laws from Noether's theorem\footnote{Each local action functional gives rise to a unique presymplectic structure of the corresponding Euler--Lagrange equations. This presymplectic structure encodes the correspondence from Noether's theorem.}, integrability properties --- and it is natural to ask what part of this information is inherited by the systems that describe invariant solutions.

For reduction under Lie groups of point symmetries satisfying certain additional assumptions, the Anderson--Fels approach~\cite{AndersonFels1997} provides one possible answer to this question utilizing symmetry-invariant representatives of the corresponding structures understood in terms of ambient jet spaces. Subject to the additional assumptions, this approach is global and covers multi-dimensional Lie groups. A local variant of this approach for conservation laws and one-dimensional Lie groups, presented in a form more familiar in symmetry applications, is the well-known ``double reduction'' method~\cite{Sjoberg2007, Bokhari2010}. Its generalization by Anco and Gandarias~\cite{AncoGandarias2020} is less sensitive to invariant representatives and allows one to consider reductions of conservation laws under some multi-dimensional Lie groups of point symmetries in terms of conservation law characteristics (multipliers). All these methods show that a conservation law compatible with a symmetry descends to a conservation law (or a first integral) of the corresponding system with fewer independent variables. The following simple example is instructive. The Korteweg--de Vries equation $u_t = uu_x + u_{xxx}$ possesses the energy conservation law $D_t(u^2/2) = D_x(u^3/3 + uu_{xx} - u_x^2/2)$; for traveling-wave solutions $u = U(x - ct)$, this conservation law gives rise to the observation that
\begin{align*}
	\tfrac{1}{3}U^3 + UU'' - \tfrac{1}{2}(U')^2 + \tfrac{c}{2}\hspace{0.2ex} U^2
\end{align*}
is a first integral of the traveling-wave ODE --- one of the two integrals that reduce the third-order ODE to quadrature.

This work continues the series of papers~\cite{InvRedI, InvRedII, InvRedIII}, in which similar observations were developed into a general \emph{invariant reduction} mechanism: whenever a cohomological structure is invariant under a symmetry $X$, its Lie derivative along $X$ is trivial, which gives rise to a potential at the level of representatives. Such a potential, restricted to the system describing $X$-invariant solutions, represents the reduced structure. The mechanism applies uniformly to point, contact, and higher symmetries and provides reductions of various structures, including variational $p$-forms (with conservation laws as variational $0$-forms and presymplectic structures as variational $2$-forms\footnote{More precisely, presymplectic structures are variational $2$-forms that are $d_1$-closed (see Section~\ref{SectionDiffEq}).}), internal Lagrangians~\cite{InvRedII}, Poisson brackets~\cite{InvRedIII}, and other elements of suitable cohomology groups.

What has been missing is a \emph{computational} counterpart of the invariant reduction mechanism. For conservation laws of evolution systems, the required manipulations (integration by parts, homotopy formulas) were made algorithmic in~\cite{InvRedI, InvRedII}. For variational $1$-forms and presymplectic structures, however, the published computations rely more heavily on the geometric apparatus of Vinogradov's $\mathcal{C}$-spectral sequence~\cite{Vinogradov1984, Vinogradov1984II, KV1999}; they can be laborious and less accessible to practitioners of symmetry methods without a background in the geometry of PDEs. The purpose of this paper is to close this gap: we reformulate the reduction of variational $p$-forms (including conservation laws and presymplectic structures) as the reduction of \emph{conservation laws} of an enlarged system, and provide explicit algorithms, together with a \texttt{Maple} implementation, that carry it out.

The reformulation can be stated in the following terms (see, e.g.,~\cite{krasil2011geometry, KVV2017}). Given a system $F = 0$, append to it its linearization $l_F(q) = 0$, where the perturbation $q$ --- the familiar object of tangent linear models, sensitivity analysis, and adjoint methods --- is treated as an \emph{anticommuting} variable: $q^i q^j = -q^j q^i$. The result is the (degree-shifted) tangent system $\mathfrak{T}[1]\mathcal{E}$. Homogeneous polynomials of degree $p$ in the variables $q^i$ and their total derivatives encode certain differential $p$-forms, so that variational $p$-forms of the original system $\mathcal{E}$ become conservation laws of $\mathfrak{T}[1]\mathcal{E}$ of internal degree $p$: conservation laws of $\mathcal{E}$ correspond to $p = 0$, variational $1$-forms to $p = 1$, and presymplectic structures appear among the conservation laws of $\mathfrak{T}[1]\mathcal{E}$ with $p = 2$. {This representation places conservation laws, variational $1$-forms, and presymplectic structures within a common computational framework based on characteristics (multipliers), integration by parts, and homotopy formulas for reconstructing \draftnew{fluxes~\cite{anco2002direct, BCA2010, KVV2017, Olver1993, Anderson1989Var}}. The reduction algorithms developed below build on this framework. Section~\ref{SectionVarForms} reviews the underlying correspondence in the explicit, characteristic-based form used by the algorithms and illustrates the full chain of constructions with a detailed running example. Against this background, the main contributions of this paper are as follows.
\begin{itemize}
\item[1.] For systems of evolution equations, we present a three-step reduction algorithm (invariance check, integration by parts, homotopy) that computes the reduction of any symmetry-invariant structure of internal degree $p \geqslant 0$, together with its complete \texttt{Maple} implementation (Section~\ref{SectionEvolAlg}, Appendix~\ref{App:B}).
\item[2.] For general $\ell$-normal systems, not necessarily of evolution type, we prove a reduction theorem for $p\geqslant 1$ and derive from it a descent algorithm based on a generalization of integration by parts to total differential operators with values in horizontal forms of not necessarily top degree (Section~\ref{SectionDescent}, Theorem~\ref{Theoremgenerfun}).
\item[3.] For symmetries generating flows on jet spaces --- in particular, the point symmetries most common in applications --- we describe substantial practical simplifications of the reduction machinery under certain additional assumptions that are typically met in applications, including the computation of reductions directly on the quotients of reduced systems by the corresponding symmetry actions (Section~\ref{SectionFlows}).
\end{itemize}
}

The paper is organized as follows. Section~\ref{SectionNotat} fixes notation and, in Table~\ref{TableDictionary}, provides a dictionary between the geometric terminology used in this series and its application-oriented counterpart; readers primarily interested in applications may wish to read Sections~\ref{SectionNotat}, \ref{SectionEvolAlg}, \ref{SectionFlows}, and~\ref{SectionExamples} together with Appendix~\ref{App:B} first, referring to Section~\ref{SectionVarForms} as needed. Section~\ref{SectionVarForms} recalls the correspondence between variational forms and conservation laws of tangent systems. Section~\ref{SectionInvRed} recalls the invariant reduction mechanism. Sections~\ref{SectionEvolAlg} and~\ref{SectionDescent} contain the two main algorithms, and Section~\ref{SectionFlows} the simplifications available in some cases for symmetries that generate flows. Section~\ref{SectionExamples} contains brief descriptions of the examples considered in this paper, together with links to \texttt{Maple} code files that further demonstrate some of the steps involved in the computations. Proofs and details of a technical nature are collected in Appendices~\ref{App:C} and~\ref{App:D}; Appendix~\ref{App:A} summarizes the graded-commutative conventions, and Appendix~\ref{App:B} the \texttt{Maple} implementation.

\section{Notation, basic definitions, and a dictionary \label{SectionNotat}}

\subsection{Jets}

We now briefly recall the notion of jet bundles and related structures in local coordinates. For details, see, e.g.,~\cite{KV1999}.

\vspace{1ex}

Let us consider the infinite jet space $J^{\infty}(n, m)$ with coordinates\footnote{Note that $u^i_{\alpha}$ are not functions of $x$.} $(x, u_{\alpha})$. Here $x = (x^1, \ldots, x^n)\in \mathbb{R}^n$ and $u = (u^1, \ldots, u^m)\in \mathbb{R}^m$ are the vectors of independent and dependent variables, respectively; $\alpha = \alpha_1 x^1 + \ldots + \alpha_n x^n = \alpha_i x^i$ is a multi-index treated as a formal linear combination whose coefficients $\alpha_i$ are non-negative integers, $|\alpha| = \alpha_1 + \ldots + \alpha_n$.

\vspace{0.5ex}
\noindent
\textbf{Functions.} We denote by $\mathcal{F}(n, m)$ the algebra of smooth functions of finitely many coordinates on $J^{\infty}(n, m)$.

\vspace{0.5ex}
\noindent
\textbf{Cartan distribution.} The main structure on the jet space $J^{\infty}(n, m)$ is the Cartan distribution. Its planes are spanned by the total derivatives
$$
D_{x^i} = \partial_{x^i} + u^j_{\alpha + x^i}\partial_{u^j_{\alpha}}\qquad\quad i = 1, \ldots, n.
$$

\vspace{0.5ex}
\noindent
\textbf{Cartan forms and modules $E^{\hspace{0.1ex} p, \hspace{0.1ex} k}_0(n, m)$.} The Cartan distribution $\mathcal{C}$ determines the ideal $\mathcal{C}\Lambda^*(n, m)$
of the algebra $\Lambda^*(n, m)$
of differential forms on $J^{\infty}(n, m)$.
The ideal $\mathcal{C}\Lambda^*(n, m)$ is generated by Cartan forms, i.e., differential forms that vanish on all planes of the Cartan distribution.
A Cartan $(k+1)$-form $\omega\in\mathcal{C}\Lambda^{k+1}(n, m)$ can be written as a finite sum
$$
\omega = \theta^j_{\alpha}\wedge w_{j}^{\alpha},\qquad \theta^j_{\alpha} = du^j_{\alpha} - u^j_{\alpha + x^i}dx^i,\qquad w_{j}^{\alpha}\in \Lambda^{k}(n, m)\,.
$$
Denote by $\mathcal{C}^p\Lambda^*(n, m)$ the $p$-th exterior power of $\mathcal{C}\Lambda^*(n, m)$, $p\geqslant 0$. If $p=0$, then $\mathcal{C}^p\Lambda^*(n, m) = \Lambda^*(n, m)$; for $p\geqslant 1$, a form $\omega\in\mathcal{C}^p\Lambda^{p+k}(n, m)$ can be written as a finite sum
$$
\omega = \theta^{i_1}_{\alpha^1}\wedge \ldots \wedge \theta^{i_p}_{\alpha^p}\wedge w^{\alpha^1\ldots \alpha^p}_{i_1\ldots i_p},\qquad w^{\alpha^1\ldots \alpha^p}_{i_1\ldots i_p}\in \Lambda^{k}(n, m)\,.
$$

Consider the modules $E^{\hspace{0.1ex} p, \hspace{0.1ex} k}_0(n, m) = \mathcal{C}^p\Lambda^{p+k}(n, m)/\mathcal{C}^{p+1}\Lambda^{p+k}(n, m)$, $p\geqslant 0$. We identify\footnote{Note, however, that point (and contact) transformations act on $\mathcal{C}^p\Lambda^{p+k}(n, m)/\mathcal{C}^{p+1}\Lambda^{p+k}(n, m)$ but not on the module of such representatives in the general case.} elements of $E^{\hspace{0.1ex} p, \hspace{0.1ex} k}_0(n, m)$ with their (unambiguously defined) representatives of the form
\begin{align*}
	&p = 0\colon \qquad \omega_{i_1\ldots i_k} dx^{i_1}\wedge \ldots \wedge dx^{i_k};\\
	&p\geqslant 1\colon\qquad \omega^{\alpha^1\ldots \alpha^p}_{j_1\ldots j_p i_1\ldots i_k}\theta^{j_1}_{\alpha^1}\wedge \ldots \wedge \theta^{j_p}_{\alpha^p}\wedge dx^{i_1}\wedge \ldots \wedge dx^{i_k}.
\end{align*}
In terms of such representatives, the differential $d_0^{\hspace{0.1ex} p, \hspace{0.1ex} k}\colon E^{\hspace{0.1ex} p,\hspace{0.1ex} k}_0(n, m)\to E^{\hspace{0.1ex} p,\hspace{0.1ex} k+1}_0(n, m)$, induced by the de Rham differential $d$, takes the form
\begin{align*}
	d_0^{\hspace{0.1ex} p, \hspace{0.1ex} k}\omega = dx^i\wedge \mathcal{L}_{D_{x^i}}\omega\,,
\end{align*}
where $\mathcal{L}_{D_{x^i}}$ is the corresponding Lie derivative.
We use the notation $d_0$ where it does not lead to confusion.
The consideration of $E^{\hspace{0.1ex} p, \hspace{0.1ex} k}_0(n, m)$ in terms of representatives allows us to introduce the vertical (Cartan) differential $d_v = d - d_0$. Since $d\circ d = 0$, one has $d_0\circ d_0 = 0$, $d_v\circ d_v = 0$, $d_0\circ d_v = - d_v\circ d_0$. 

We refer to elements of $E^{\hspace{0.1ex} 0, \hspace{0.1ex} k}_0(n, m)$ as horizontal $k$-forms on $J^{\infty}(n, m)$ and also denote the module $E^{\hspace{0.1ex} 0, \hspace{0.1ex} k}_0(n, m)$ by $\Lambda^k_h(n, m)$. 

\vspace{0.5ex}
\noindent
\textbf{Infinitesimal symmetries.}
Elements of the module
\begin{align*}
	\varkappa(n, m) = \underbrace{\mathcal{F}(n, m)\times \ldots \times \mathcal{F}(n, m)}_{m}
\end{align*}
are characteristics of symmetries of $J^{\infty}(n, m)$. Any $\varphi = (\varphi^1, \ldots, \varphi^m)\in \varkappa(n, m)$ defines the corresponding evolutionary vector field $E_{\varphi} = D_{\alpha}(\varphi^i)\partial_{u^i_{\alpha}}$.
An infinitesimal symmetry of $J^{\infty}(n, m)$ is a sum of an evolutionary vector field and a trivial symmetry, i.e., a combination of the total derivatives $D_{x^1}$, $\ldots$, $D_{x^n}$.

\subsection{Differential equations \label{SectionDiffEq}}

Let us denote
\begin{align*}
	P(n, m) = \underbrace{\mathcal{F}(n, m)\times \ldots \times \mathcal{F}(n, m)}_{m_1}.
\end{align*}
for some $m_1 \geqslant 1$.
Then $F\in P(n, m)$ determines a differential operator\footnote{It acts on a vector function $(\sigma^1(x), \ldots, \sigma^m(x))$ by replacing $u^i_{\alpha}\mapsto \partial^{|\alpha|} \sigma^i/(\partial x^1)^{\alpha_1}\ldots (\partial x^n)^{\alpha_n}$ in $F$.}.
The \emph{infinite prolongation} of the system of differential equations $F = 0$ is the subset $\mathcal{E}\subset J^{\infty}(n, m)$ defined by
\begin{align*}
	\mathcal{E}\colon\qquad D_{\alpha}(F^i) = 0\,,\qquad |\alpha| \geqslant 0\,,\ i = 1, \ldots, m_1\,.
\end{align*}
It can be regarded as the set of \emph{formal} solutions of $F = 0$. We assume that $F = 0$ has no differential consequences involving only independent variables.

\vspace{0.5ex}
\noindent
\textbf{Functions.}
By $\mathcal{F}(\mathcal{E})$ we denote the algebra of smooth functions on $\mathcal{E}$,
$$
\mathcal{F}(\mathcal{E}) = \mathcal{F}(n, m)|_{\mathcal{E}} = \mathcal{F}(n, m)/I_{\mathcal{E}}.
$$
Here $I_{\mathcal{E}}$ denotes the ideal of the system $\mathcal{E}\subset J^{\infty}(n, m)$, $I_{\mathcal{E}} = \{f\in \mathcal{F}(n, m)\, \colon\, f|_{\mathcal{E}} = 0\}$. Tangent vectors/vector fields on $\mathcal{E}$ are defined in terms of derivations of the algebra $\mathcal{F}(\mathcal{E})$.

\vspace{0.5ex}
\noindent
\textbf{Cartan forms.} The ideal $\mathcal{C}\Lambda^*(n, m)\subset \Lambda^*(n, m)$ gives rise to the corresponding ideal $\mathcal{C}\Lambda^*(\mathcal{E})$ of the algebra $\Lambda^*(\mathcal{E}) = \Lambda^*(n, m)/(I_{\mathcal{E}}\cdot \Lambda^*(n, m) + \Lambda^*(n, m)\wedge dI_{\mathcal{E}})$ of differential forms on $\mathcal{E}$.

\vspace{0.5ex}
\noindent
\textbf{Regularity conditions.} We restrict ourselves to systems satisfying the following two conditions. 
\begin{enumerate}
	\item A function $f\in \mathcal{F}(n, m)$ vanishes on $\mathcal{E}$ if and only if there exists a differential operator $\Delta\colon P(n, m)\to \mathcal{F}(n, m)$ of the form $\Delta_i^{\alpha} D_{\alpha}$ ($\mathcal{C}$-differential operator, total differential operator) such that $f = \Delta(F)$.
	\item $\mathcal{E}$ is representable as the inverse limit\footnote{By this we mean that there exist dg-algebra morphisms $\Lambda^*(\mathcal{E}_k)\to \Lambda^*(\mathcal{E})$ exhibiting $\Lambda^*(\mathcal{E})$ as the corresponding direct limit in the category of dg-algebras. The local treatment of jet spaces does not remove the necessity of imposing conditions of this type, due to the way differential forms restrict to $\mathcal{E}$.} of a tower of finite-dimensional smooth manifolds
	\begin{align*}
		\xymatrix{
			\ldots \ar[r] & \mathcal{E}_{k+1} \ar[r] & \mathcal{E}_{k} \ar[r] & \ldots \ar[r] & \mathcal{E}_0,
		}
	\end{align*}
	where all maps are surjective submersions.
\end{enumerate}
The latter condition is satisfied, in particular, by all formally integrable systems
and any system that can be written in the form
\begin{align}
	u^j_{b_j t} - f^j = 0\,,\qquad j = 1, \ldots, m_1\,,
	\label{extKovunderdet}
\end{align}
where $t$ denotes $x^1$, $m_1\leqslant m$, all $b_j\geqslant 1$, and all $f^j$ are independent of the variables $u^i_{b_i t + \alpha}$ for $i = 1, \ldots, m_1$, $|\alpha| \geqslant 0$.

We also assume that the de Rham cohomology groups $H^i_{dR}$ of all considered systems are trivial for~$i > 0$.

\vspace{0.5ex}
\noindent
\textbf{Infinitesimal symmetries.} A \emph{symmetry} (more precisely, an infinitesimal symmetry) of an infinitely prolonged system of equations $\mathcal{E}$ is a vector field $X$ on $\mathcal{E}$ (a derivation of $\mathcal{F}(\mathcal{E})$) that preserves the Cartan distribution: $[X, \mathcal{C}D(\mathcal{E})]\subset \mathcal{C}D(\mathcal{E})$, where $\mathcal{C}D(\mathcal{E})$ denotes the module of \emph{Cartan derivations}, i.e., combinations of the total derivatives $\,\overline{\!D}_{x^i} = D_{x^i}|_{\mathcal{E}}$, $i = 1, \ldots, n$ (trivial symmetries).

If $\varphi\in \varkappa(n, m)$ is a characteristic such that $E_{\varphi}$ is tangent to $\mathcal{E}$ (i.e., $E_{\varphi}(F)|_{\mathcal{E}} = 0$), then the restriction $E_{\varphi}|_{\mathcal{E}}\colon \mathcal{F}(\mathcal{E})\to \mathcal{F}(\mathcal{E})$ is a symmetry of $\mathcal{E}$ (less formally, $E_{\varphi}$ can also be called a symmetry of $\mathcal{E}$).
Any symmetry $X$ of $\mathcal{E}\subset J^{\infty}(n, m)$ is equivalent to an evolutionary symmetry $E_{\varphi}|_{\mathcal{E}}$. Evolutionary symmetries of $\mathcal{E}\subset J^{\infty}(n, m)$ are in one-to-one correspondence with elements of the kernel of the (universal) linearization operator $l_{\mathcal{E}} = l_F|_{\mathcal{E}}\colon \varkappa(\mathcal{E})\to P(\mathcal{E})$, where $l_F\colon \varkappa(n, m)\to P(n, m)$, $\varphi \mapsto E_{\varphi}(F)$, $l_F(\varphi)^i = E_{\varphi}(F^i)$, and
$$
\varkappa(\mathcal{E}) = \varkappa(n, m)|_{\mathcal{E}}\,,\qquad P(\mathcal{E}) = P(n, m)|_{\mathcal{E}}\,.
$$

We say that $\mathcal{E}$ is \emph{$\ell$-normal} if every total differential operator $\Delta\colon P(\mathcal{E})\to \mathcal{F}(\mathcal{E})$ satisfying $\Delta \circ l_{\mathcal{E}} = 0$ necessarily vanishes. For example, all systems of the form~\eqref{extKovunderdet} are $\ell$-normal for $1\leqslant m_1\leqslant m$.

\noindent
\textbf{$\mathcal{C}$-spectral sequence.} The groups $E^{\hspace{0.1ex} p,\hspace{0.1ex} k}_0(n, m)$ and the differential $d_0$ restrict to $\mathcal{E}$. The restrictions give rise to the following cohomology groups and the differential $d_1$
$$
E^{\hspace{0.1ex} p, \hspace{0.1ex} k}_1(\mathcal{E}) = \ker d_0^{\hspace{0.1ex} p,\hspace{0.1ex} k}/ \hspace{0.15ex} \mathrm{im}\, d_0^{\hspace{0.1ex} p,\hspace{0.1ex} k-1}, \qquad
d_1^{\hspace{0.1ex} p,\hspace{0.1ex} k}\colon E^{\hspace{0.1ex} p, \hspace{0.1ex} k}_1(\mathcal{E}) \to E^{\hspace{0.1ex} p+1, \hspace{0.1ex} k}_1(\mathcal{E})\,.
$$
The differential $d_v$ also restricts to $\mathcal{E}$. If $\omega\in E^{\hspace{0.1ex} p, \hspace{0.1ex} k}_0(\mathcal{E})$ represents an element $\Omega\in E^{\hspace{0.1ex} p, \hspace{0.1ex} k}_1(\mathcal{E})$, then $d_v\hspace{0.1ex} \omega = d\omega$ represents $d_1\Omega$.

A \emph{variational $p$-form} of $\mathcal{E}$ is an element of the group $E^{\hspace{0.1ex} p,\hspace{0.1ex} n-1}_1(\mathcal{E})$. A \emph{conservation law} of $\mathcal{E}$ is a variational $0$-form, i.e., an element of the group $E^{\hspace{0.1ex} 0,\hspace{0.1ex} n-1}_1(\mathcal{E})$.
A \emph{presymplectic structure} of $\mathcal{E}$ is a $d_1$-closed variational $2$-form, i.e., an element of the kernel of the differential
$d_1^{\hspace{0.1ex} 2,\hspace{0.1ex} n-1}\colon E^{\hspace{0.1ex} 2, \hspace{0.1ex} n-1}_1(\mathcal{E})\to E^{\hspace{0.1ex} 3, \hspace{0.1ex} n-1}_1(\mathcal{E})$.

\vspace{0.5ex}
\noindent
\textbf{Cosymmetries and characteristics of conservation laws.} Let us introduce the modules
\begin{align*}
	\widehat{P}(n, m) = \mathrm{Hom}_{\mathcal{F}(n, m)}(P(n, m), E^{\hspace{0.1ex} 0, \hspace{0.1ex} n}_0(n, m))\,,\qquad \widehat{\varkappa}(n, m) = \mathrm{Hom}_{\mathcal{F}(n, m)}(\varkappa(n, m), E^{\hspace{0.1ex} 0, \hspace{0.1ex} n}_0(n, m))\,.
\end{align*}
For example, any homomorphism $\psi\in \widehat{P}(n, m)$ can be written in terms of its components
\begin{align*}
	\langle \psi, G \rangle = \psi_j\hspace{0.1ex} G^j dx^1\wedge \ldots \wedge dx^n,\qquad G\in P(n, m)\,.
\end{align*}
Here $\langle \cdot, \cdot \rangle$ denotes the natural pairing between a module and its adjoint. For simplicity, we identify $\psi\in \widehat{P}(n, m)$ with the tuple of its components $(\psi_1, \ldots, \psi_{m_1})$.

A \emph{characteristic} of a conservation law is a homomorphism $\psi\in \widehat{P}(n, m)$ such that
\begin{align*}
	\langle \psi, F \rangle = d_0\hspace{0.15ex} \mu\,,\qquad \mu\in E^{\hspace{0.1ex} 0, \hspace{0.1ex} n-1}_0(n, m)\,,
\end{align*}
where $\mu|_{\mathcal{E}}\in E_0^{\hspace{0.1ex} 0,\hspace{0.1ex} n-1}(\mathcal{E})$ represents the conservation law.

Elements of the kernel of the (formal) adjoint linearization operator $l_{\mathcal{E}}^{\, *}\colon \widehat{P}(\mathcal{E})\to \widehat{\varkappa}(\mathcal{E})$ are \emph{cosymmetries} of $\mathcal{E}$. Here $l_{\mathcal{E}}^{\, *} = l_{F}^{\, *}|_{\mathcal{E}}$, $\widehat{P}(\mathcal{E}) = \widehat{P}(n, m)|_{\mathcal{E}}$, $\widehat{\varkappa}(\mathcal{E}) = \widehat{\varkappa}(n, m)|_{\mathcal{E}}$. The restriction $\psi|_{\mathcal{E}}$ of any characteristic $\psi$ of a conservation law is a cosymmetry.

\subsection{A dictionary}

Table~\ref{TableDictionary} summarizes the correspondence between the geometric notions used in this paper and their counterparts arising in more applied areas utilizing symmetry-based methods. The translations are informal, mainly because our main object $\mathcal{E}$ cannot be regarded as the space of solutions of the corresponding PDE system\footnote{In this case, for example, it would be unclear how to interpret Cartan forms and notation of the form $\partial_{u^j_{\alpha}}$.}; precise statements appear in the sections indicated.
\begin{table}[htbp]
	\centering
	{%
		\renewcommand{\arraystretch}{1.35}
		\begin{tabular}{p{0.36\textwidth} p{0.57\textwidth}}
			\hline
			\textbf{Geometric notion} & \textbf{Symmetry-analysis translation} \\
			\hline
			characteristic $\chi\in\varkappa(n, m)$; evolutionary field $E_{\chi}$ & infinitesimal perturbation of $u$; symmetry in characteristic (evolutionary) form \\
			(universal) linearization $l_F$ & Fr\'echet derivative, operator determining ``tangent linear model'' of $F=0$ \\
			(formal) adjoint linearization $l_F^{\hspace{0.15ex}*}$ & formal (integration-by-parts) adjoint, operator determining adjoint model \\
			cosymmetry $\psi|_{\mathcal{E}}\in\ker l^{\hspace{0.15ex}*}_{\mathcal{E}}$; cosymmetry corresponding to a conservation law & adjoint-symmetry; conservation-law multiplier on solutions when the additional Euler-operator condition holds~\cite{anco2002direct, BCA2010} for some of its extensions off the solution space\\
			element of $E^{\hspace{0.1ex}0,\hspace{0.1ex}n-1}_1(\mathcal{E})$ & conservation law: divergence expression vanishing on solutions, modulo trivial ones \\
			Cartan forms $\theta^i_{\alpha}$ 
			&
			contact forms $du^i_{\alpha} - u^i_{\alpha+x^j}dx^j$, where $u^i_{\alpha}$ are understood as coordinates on the corresponding space (not as functions of $x$).\\
			variational $1$-form; variational $2$-form
			& 
			linear function that maps symmetries to conservation laws; skew-symmetric bilinear function that maps pairs of symmetries to conservation laws\\
			variational $1$-form represented by $\omega_{\psi}|_{\mathcal{E}}$, corresponding to a cosymmetry $\psi|_{\mathcal{E}}$
			& 
			total divergence produced by integrating $\psi_i\, l_F(\chi)^i$ by parts, where $\psi$ is an extension of an adjoint symmetry off the solution space --- adjoint symmetry regarded as linear function from symmetries to conservation laws
			(Section~\ref{SectionVarFormAndCosym}) \\
			$\,l^{\hspace{0.15ex}*}_{\mathcal{E}}\circ\nabla = \nabla^{*}\circ\hspace{0.1ex} l_{\mathcal{E}}$; presymplectic operator $\nabla$ & operator identity of multiplier type; ``symplectic operator'' of the integrable-systems literature (Sections~\ref{SectionVarFormsAndOperat};~\ref{SectionVar2FormsAsCL}) \\
			tangent system $\mathfrak{T}[1]\mathcal{E}$ and odd variables $q^i_{\alpha}$ & original system augmented by its linearized system $l_F(q)=0$, with the perturbation $q$ treated as anticommuting (Section~\ref{SectionVarFormsAsCL}) \\
			internal degree $p$ 
			& 
			polynomial degree in $q^i$ and their (total) derivatives --- encodes differential forms with precisely $p$ multipliers of the form $\theta^i_{\alpha}$ \\
			$d_0$ 
			& 
			generalization of total divergence to structures of other types (both off shell and on the system)\\
			$E_{\varphi}(\psi) + l_{\varphi}^{\hspace{0.15ex}*}(\psi) = 0$ & symmetry-invariance condition for a structure encoded by adjoint-symmetry provided that the system is in evolution form and all $t$-derivatives are eliminated (invariance of the conservation law if $\psi$ is a multiplier) \\
			$\ell$-normal system & ``not overdetermined'' system --- satisfied by evolution, extended Kovalevskaya-form and most other non-gauge systems encountered in applications\\
			\hline
	\end{tabular}}
	\caption{{Informal dictionary between geometric and symmetry-analysis terminology.}\label{TableDictionary}}
\end{table}

\subsection{Reading guide}

The computational core of the paper does not require familiarity with spectral sequences. All algorithms operate on three\footnote{Technically, the implementation (Appendix~\ref{App:B}) requires seven types of input data determined by these three.} kinds of input data: the left-hand sides $F^i$ of the equations $F^i = 0$, symmetry characteristic components $\varphi^i$, and components $\psi'^{\hspace{0.1ex} p}_i$, $\psi'^{\hspace{0.1ex} p-1}_i$ of a characteristic $\psi_l$ of a conservation law for the tangent system $F = 0$, $l_F(q) = 0$ (regarded as differential forms $\psi^p_i$, $\psi^{p-1}_i$ in terms of the original independent variables, dependent variables and their total derivatives).
	
The odd variables $q^i_{\alpha}$ may be treated as formal anticommuting symbols whose calculus is summarized in Appendix~\ref{App:A}. The cohomological language of Sections~\ref{SectionInvRed} and~\ref{SectionDescent} serves, among other purposes, to
state precisely in what sense the outputs of the algorithms are well-defined (independent of the choices made along the way).

\section{Variational forms as conservation laws of degree-shifted tangent systems \label{SectionVarForms}}

In simple terms, the message of this section is the following. All structures considered below originate in integration by parts. Multiplying the linearization of a system by an extension of a cosymmetry (an adjoint-symmetry) off the solution space and integrating by parts produces a boundary term; evaluated on solutions, that boundary term defines the variational $1$-form associated with the cosymmetry (Section~\ref{SectionVarFormAndCosym}). Self-adjointness-type operator identities produce variational $2$-forms in the same way (Section~\ref{SectionVarFormsAndOperat}), and presymplectic structures are a distinguished subclass (Section~\ref{SectionVar2FormsAsCL}). The central observation --- formulated in Sections~\ref{SectionVarFormsAsCL} and~\ref{SectionVar2FormsAsCL} --- is that all these objects can be handled \emph{uniformly} as conservation laws of a single enlarged system: the tangent system, obtained by appending to $F = 0$ its linearized equations $l_F(q) = 0$ with the perturbation $q$ treated as anticommuting. This reformulation allows us to extend the algorithm of conservation law reduction~\cite{InvRedII} to arbitrary variational $p$-forms of evolution systems (Section~\ref{SectionEvolAlg}).

\vspace{1ex}

The most common examples of variational $p$-forms arising in applications are conservation laws (variational $0$-forms), variational $1$-forms, and presymplectic structures ($d_1$-closed variational $2$-forms). Variational $1$-forms of $\ell$-normal systems
can be partitioned into two disjoint classes: those that yield nontrivial presymplectic structures (by means of the differential $d_1$) and those that are differentials of conservation laws. Nevertheless, the canonical variational $1$-forms of cotangent systems play a crucial role in the invariant reduction of Poisson brackets~\cite{InvRedIII}, one that extends beyond the corresponding presymplectic structures.

Presymplectic structures of systems of PDEs are related to their variational content. More precisely, each Lagrangian whose Euler--Lagrange equations are differential consequences of a given system gives rise to a unique presymplectic structure, and each presymplectic structure of an $\ell$-normal system can be obtained in this way~\cite{DRUZHKOV2023104848}. As with any variational $2$-form, presymplectic structures define homomorphisms from symmetries to variational $1$-forms. For a Lagrangian system $\mathcal{E}$, such a homomorphism defined by the corresponding presymplectic structure is always surjective. Moreover, in this case, one has:
\begin{align*}
\text{the kernel of the homomorphism is trivial}\ \Leftrightarrow\ \mathcal{E} \text{ is } \ell\text{-normal}\ \Leftrightarrow\ \mathcal{E} \text{ is not a gauge system}
\end{align*}
(if the kernel is non-trivial, then it consists of gauge symmetries). Thus, for $\ell$-normal Lagrangian systems, such homomorphisms are isomorphisms. Furthermore, in this case, each nontrivial conservation law produces a nontrivial variational $1$-form\footnote{More generally, each nontrivial conservation law of an arbitrary $\ell$-normal system produces a nontrivial variational $1$-form by means of the differential $d_1$.}, which corresponds to a unique symmetry. This correspondence reproduces the one from Noether's theorem, while symmetries that do not correspond to conservation laws produce presymplectic structures.

One important feature of variational $p$-forms is that they can be pulled back along maps of differential equations inducing maps of their Cartan distributions\footnote{Such maps relate solutions of the systems. In particular, this applies, but is not limited, to reductions and differential coverings (including Lax pairs, potential systems arising from conservation laws, Miura-type transformations, tangent and cotangent systems, and so on). Note that invariant reduction (Section~\ref{SectionInvRed}) is a cohomological descent rather than a mere pullback.}.
For example, the presymplectic structure produced by a local action functional (and encoding the correspondence from Noether's theorem) defines similar correspondences for all systems that are related to the original one by such maps to it.

\subsection{Variational $1$-forms and cosymmetries \label{SectionVarFormAndCosym}}

Variational $1$-forms of an $\ell$-normal system $\mathcal{E}$ are in one-to-one correspondence with its cosymmetries, i.e., elements of $\ker l_{\mathcal{E}}^{\hspace{0.15ex} *}$. More precisely, if $\mathcal{E}$ is the infinite prolongation of a system $F = 0$, then for each $\psi\in \widehat{P}(n, m)$, there exists a differential form $\omega_{\psi}\in E^{1,\hspace{0.1ex} n-1}_0(n, m)$ such that the relation
\begin{align}
\langle l_F(\chi), \psi \rangle = \langle \chi, l_F^{\hspace{0.15ex} *} (\psi) \rangle + d_0 (E_{\chi}\lrcorner\, \omega_{\psi})
\label{Cosymvarforms}
\end{align}
holds for any characteristic $\chi\in \varkappa(n, m)$, while the correspondence $\psi\mapsto \omega_{\psi}$ is a total differential operator. If $\psi|_{\mathcal{E}}$ is a cosymmetry, then $l_F^{\hspace{0.15ex} *} (\psi)$ vanishes on $\mathcal{E}$ and $\omega_{\psi}|_{\mathcal{E}}$ represents the corresponding variational $1$-form. This correspondence gives rise to the well-defined homomorphism $\ker l_{\mathcal{E}}^{\hspace{0.15ex} *}\to E^{1,\hspace{0.1ex} n-1}_1(\mathcal{E})$, which is always surjective. It is an isomorphism if and only if $\mathcal{E}$ is $\ell$-normal.
For a cosymmetry $\psi|_{\mathcal{E}}$, the restriction of~\eqref{Cosymvarforms} to $\mathcal{E}$ can be written in the form
\begin{align}
\psi|_{\mathcal{E}} \circ l_{\mathcal{E}} = d_0 (E_{\chi}\lrcorner\, \omega_{\psi})|_{\mathcal{E}}\,.
\label{Cosymmvarform}
\end{align}

\examplea{Let us consider the following $(1+2)$-dimensional equation \label{ExampleRunning}
\begin{align}
u_t = u_x^2 + u_{xxx} + u_{xxy}\,,
\label{2dpKdV}
\end{align}
where $t = x^1$, $x = x^2$, $y = x^3$, $u = u^1$, and $F = F^1 = u_t - (u_x^2 + u_{xxx} + u_{xxy})$.
We denote $D_t(\chi^1) = \chi^1_t$, $D_x(\chi^1) = \chi^1_x$, $\ldots$ for convenience. Here
\begin{align*}
\langle l_F(\chi), \psi \rangle = \big(\chi^1_t - 2u_x \chi^1_x - \chi^1_{xxx} - \chi^1_{xxy}\big)\psi_1 dt\wedge dx\wedge dy\,.
\end{align*}
Integrating by parts (with respect to $\chi = \chi^1$), we find that the potential $E_{\chi}\lrcorner\, \omega_{\psi}$ in~\eqref{Cosymvarforms} can be taken, for example, in the form
\begin{align*}
\chi^1 \psi_1 dx\wedge dy - \chi^1 \psi_{1\, xx} dt\wedge dx + (2u_x \chi^1 \psi_1 + \chi^1_{xx} \psi_1 - \chi^1_x\psi_{1\, x} + \chi^1\psi_{1\, xx} + \chi^1_{xy} \psi_1 - \chi^1_{y} \psi_{1\, x})dt\wedge dy
\end{align*}
Here we also denote $D_x(\psi_1) = \psi_{1\, x}$, $\ldots$ One finds
\begin{align}
\begin{aligned}
\omega_{\psi} =\ &\psi_1 \theta \wedge dx\wedge dy - \psi_{1\, xx} \theta \wedge dt\wedge dx\\
&+ (2u_x \psi_1 \theta + \psi_1 \theta_{xx} - \psi_{1\, x} \theta_x + \psi_{1\, xx} \theta + \psi_1 \theta_{xy} - \psi_{1\, x} \theta_y)\wedge dt\wedge dy\,,
\label{Omegapsi}
\end{aligned}
\end{align}
where $\theta = \theta^1 = du - u_t dt - u_x dx - u_y dy$, $\theta_x = \theta^1_x = du_x - u_{tx} dt - u_{xx} dx - u_{xy} dy$, $\ldots$

Equation~\eqref{2dpKdV} admits the cosymmetry
\begin{align}
\psi|_{\mathcal{E}} = \psi_1|_{\mathcal{E}} = (3tu_{tx} + xu_{xx} + y u_{xy} + 2u_x)|_{\mathcal{E}}\,.
\label{presstrpoten}
\end{align}
Let us take $\psi = 3tu_{tx} + xu_{xx} + y u_{xy} + 2u_x$. The restriction of the corresponding form $\omega_{\psi}$ to the infinite prolongation $\mathcal{E}$ of~\eqref{2dpKdV} represents the variational $1$-form.
}

\subsection{Variational $1$-forms as conservation laws \label{SectionVarFormsAsCL}}

In more applied terms, the construction introduced here is the following: augment the original system by its linearization --- the ``tangent linear model'' familiar from sensitivity analysis and adjoint methods --- with one modification: the perturbation variables $q^i$ anticommute, $q^i q^j = -q^j q^i$. The anticommutativity is needed for homogeneous polynomials of degree $p$ in the variables $q^i$ and their total derivatives to encode certain differential $p$-forms, so that variational $p$-forms become conservation laws. The sign conventions are collected in Appendix~\ref{App:A}.

\vspace{1ex}

A Cartan $1$-form on $\mathcal{E}$ can be interpreted as a degree-$1$ function on the degree-shifted tangent system $\mathfrak{T}[1]\mathcal{E}$. More precisely, one can introduce new dependent variables $q = (q^1, \ldots, q^m)$ using $q^i$ as another notation for $\theta^i = du^i - u^i_{x^j}dx^j = d_v u^i$. These variables are odd of degree $1$. In particular, $q^i q^j = -q^j q^i$. Similarly, $q^i_{\alpha}$ are odd variables of degree $1$ corresponding to $\theta^i_{\alpha} = d_v u^i_{\alpha}$. However, we choose the sign convention implying that $q^i_{\alpha}$ commute with $dx^j$ and $\theta^j_{\beta}$ (see Appendix~\ref{App:A}). Denote by $J^{\infty}(n, m;m)$ the corresponding infinite jet space with coordinates $x^j, u^i_{\alpha}, q^i_{\alpha}$.

The correspondence between $q^i_{\alpha}$ and $\theta^i_{\alpha}$ can be described through the contraction: $q^i_{\alpha} = E_{q}\lrcorner\, \theta^i_{\alpha}$, where $E_q = q^i_{\alpha}\partial_{u^i_{\alpha}}$ is the \emph{canonical vector field} on $J^{\infty}(n, m;m)$. Similarly, $\theta^i_{\alpha}\wedge dx^{i_1}\wedge \ldots \wedge dx^{i_k}$ corresponds to
\begin{align}
q^i_{\alpha} dx^{i_1}\wedge \ldots \wedge dx^{i_k} =  E_q\lrcorner\, (\theta^i_{\alpha}\wedge dx^{i_1}\wedge \ldots \wedge dx^{i_k})\,.
\label{thecorrespdeg1}
\end{align}

Replacing $\chi$ by $q$ in~\eqref{Cosymvarforms}, we see that the horizontal form $\omega_{\psi}' = E_{q}\lrcorner\, \omega_{\psi}$ of internal degree $1$, corresponding to an extended cosymmetry, determines a conservation law of the infinite prolongation $\mathfrak{T}[1]\mathcal{E}$ of the system
\begin{align*}
F = 0\,,\qquad l_F(q) = 0\,.
\end{align*}
Here the total derivatives are extended to act on $q^j_{\alpha}$ by $D_{x^i}(q^j_{\alpha}) = q^j_{\alpha + x^i}$ (abusing notation, we also denote the extended total derivatives by $D_{x^1}$, $\ldots$, $D_{x^n}$).
The system $\mathfrak{T}[1]\mathcal{E}$ is called the (degree-shifted) \emph{tangent system} of $\mathcal{E}$. Let us denote 
$$
F_0 = F\,,\qquad F_1 = l_F(q)\,,\qquad \widetilde{F} = (F_0, F_1)\,.
$$
Then a characteristic of this conservation law is given by the following formula, arising from integration by parts
\begin{align*}
-\langle \Psi^*(q), F_0 \rangle + \langle \psi, F_1 \rangle  = d_0(\hspace{0.15ex} \omega_{\psi}' + \ldots)\,.
\end{align*}
Here $\Psi\colon P(n, m)\to \widehat{\varkappa}(n, m)$ is a total differential operator such that $l_F^{\hspace{0.15ex} *} (\psi) = \Psi(F)$; a term vanishing on $\mathfrak{T}[1]\mathcal{E}$ is omitted under $d_0$. This relation can also be written in the form
\begin{align*}
\langle \psi_l, \widetilde{F} \rangle = d_0(\hspace{0.15ex} \omega_{\psi}' + \ldots)
\end{align*}
for the left characteristic $\psi_l = (-\Psi^*(q), \psi)$. Thus, the cosymmetry corresponding to a variational $1$-form of an $\ell$-normal system $\mathcal{E}$ can be obtained through the corresponding conservation law of $\mathfrak{T}[1]\mathcal{E}$ as the restriction to $\mathcal{E}$ of the component of its characteristic having internal degree $0$.

\vspace{1ex}

\remarka{A conservation law of $\mathfrak{T}[1]\mathcal{E}$ having internal degree $p \geqslant 1$ is trivial if and only if the component of its characteristic having internal degree $p-1$ vanishes on $\mathfrak{T}[1]\mathcal{E}$, provided $\mathcal{E}$ is $\ell$-normal.
}

\vspace{1ex}

\remarka{The canonical vector field $E_q$ acts on elements of $E^{0,\hspace{0.1ex} k}_0(n, m; m)$
by means of the corresponding Lie derivative. In terms of $J^{\infty}(n, m)$, this reproduces the action of $d_v$. The restriction $E_q|_{\mathfrak{T}[1]\mathcal{E}}$ is a degree-$1$ symmetry of the tangent system $\mathfrak{T}[1]\mathcal{E}$. Its action on $E^{0,\hspace{0.1ex} k}_1(\mathfrak{T}[1]\mathcal{E})$ reproduces the action of $d_1$. For an $\ell$-normal $\mathcal{E}$, a variational $1$-form is of the form $d_1\xi$ for some (unambiguously defined) $\xi\in E^{0,\hspace{0.1ex} n-1}_1(\mathcal{E})$ if and only if $E_q|_{\mathfrak{T}[1]\mathcal{E}}$ maps the corresponding conservation law of internal degree $1$ to the trivial conservation law (of internal degree $2$).
}

\vspace{1ex}

\remarka{In more algebraic terms, the tangent system of $\mathcal{E}$ corresponds to the exterior algebra $\oplus_{p\geqslant 0}\, \mathcal{C}^p\Lambda^p(\mathcal{E})$. It can be endowed with a dg-algebra structure by means of $d_v$. However, the differential $d_v$ depends on a projection of the form $\mathcal{E}\to M$, in contrast to $\oplus_{p\geqslant 0}\, \mathcal{C}^p\Lambda^p(\mathcal{E})$. Thus, point transformations of ambient jet spaces induce transformations of the tangent systems, but not as dg-algebras in the general case.
}

\vspace{1ex}

\examplea{Let us continue with Example~\ref{ExampleRunning}. \label{Example2}
The corresponding tangent system is the infinite prolongation of
\begin{align*}
u_t = u_x^2 + u_{xxx} + u_{xxy}\,,\qquad q_t = 2u_xq_x + q_{xxx} + q_{xxy}\,.
\end{align*}
Then for $\psi\in \widehat{P}(n, m)$, we obtain
\begin{align}
	\begin{aligned}
		\omega_{\psi}' =\ &\psi_1 q\, dx\wedge dy - \psi_{1\, xx} q\, dt\wedge dx\\
		&+ (2u_x \psi_1 q + \psi_1 q_{xx} - \psi_{1\, x} q_x + \psi_{1\, xx} q + \psi_1 q_{xy} - \psi_{1\, x} q_y) dt\wedge dy\,.
	\end{aligned}
    \label{omega_psi}
\end{align}
The restriction of the corresponding form evaluated at $\psi = \psi_1 = 3tu_{tx} + xu_{xx} + y u_{xy} + 2u_x$ represents the conservation law of $\mathfrak{T}[1]\mathcal{E}$ of internal degree $1$.
}

\subsection{Variational $2$-forms and operators \label{SectionVarFormsAndOperat}}

Similarly to the homomorphism $\ker l_{\mathcal{E}}^{\hspace{0.15ex} *}\to E^{1,\hspace{0.1ex} n-1}_1(\mathcal{E})$, we recall the correspondence between presymplectic operators and presymplectic structures~\cite{KV1999}. It is instructive to begin with the correspondence for arbitrary variational $2$-forms.

Consider a system of differential equations $F = 0$ and its infinite prolongation $\mathcal{E}$. 
Suppose that $\nabla\colon \varkappa(\mathcal{E})\to \widehat{P}(\mathcal{E})$ is a total differential operator satisfying 
\begin{align}
l_{\mathcal{E}}^{\hspace{0.15ex} *}\circ \nabla - \nabla^* \circ l_{\mathcal{E}} = 0\,.
\label{var2formop}
\end{align}
Then $\nabla$ yields a unique variational $2$-form as follows.
The operator $\nabla$ can be extended to $J^{\infty}(n, m)$: there exists a total differential operator $\nabla_e\colon \varkappa(n, m)\to \widehat{P}(n, m)$ such that $\nabla = \nabla_e|_{\mathcal{E}}$. Then one can put $\psi = \nabla_e(\chi_0)$ in~\eqref{Cosymvarforms}, $\chi_0\in \varkappa(n, m)$,
\begin{align}
\langle l_F(\chi), \nabla_e(\chi_0) \rangle - \langle \chi, l_F^{\,*} \nabla_e(\chi_0) \rangle = d_0 (E_{\chi}\lrcorner\, \omega_{\nabla_e(\chi_0)})\,.
\label{PseudoPresymvarforms}
\end{align}
The skew-symmetric operator $(\chi_0, \chi)\mapsto \frac{1}{2}(E_{\chi}\lrcorner\, \omega_{\nabla_e(\chi_0)} - E_{\chi_0}\lrcorner\, \omega_{\nabla_e(\chi)})$ coincides with $E_{\chi}\lrcorner\, (E_{\chi_0}\lrcorner\, \omega_{\nabla_e})$ for a unique $\omega_{\nabla_e}\in E^{2,\hspace{0.1ex} n-1}_0(n, m)$. The variational $2$-form corresponding to the operator $\nabla$ is represented by $\omega_{\nabla_e}|_{\mathcal{E}}$. 

\vspace{1ex}

\remarka{This homomorphism between operators satisfying~\eqref{var2formop} and variational $2$-forms is well-defined and surjective.
If $\mathcal{E}$ is $\ell$-normal, then its kernel consists of operators of the form $\square \circ l_{\mathcal{E}}$ for total differential operators $\square\colon P(\mathcal{E})\to \widehat{P}(\mathcal{E})$ such that $\square^* = \square$~\cite{KV1999}.

}

\vspace{1ex}

Let us note that formula~\eqref{PseudoPresymvarforms} implies
\begin{align}
\dfrac{1}{2}\big(\langle l_F(\chi), \nabla_e(\chi_0) \rangle\! -\! \langle \chi, l_F^{\,*} \nabla_e(\chi_0) \rangle\! -\! \langle \nabla_e(\chi), l_F(\chi_0) \rangle\! +\! \langle l_F^{\,*} \nabla_e(\chi), \chi_0 \rangle\big) = d_0 (E_{\chi}\lrcorner (E_{\chi_0}\lrcorner\hspace{0.25ex} \omega_{\nabla_e}))\,.
\label{Presymvarforms}
\end{align}

\vspace{1ex}

\examplea{Continuing with Example~\ref{ExampleRunning}, one can see that the operator\footnote{Recall that we identify elements of $\widehat{P}(\mathcal{E})$ with tuples of their components.} \label{ExampleRunning2}
\begin{align*}
\nabla = D_x|_{\mathcal{E}}
\end{align*}
satisfies~\eqref{var2formop}. Let us put $\nabla_e = D_x$. Note that we write all factors $\psi_1$, $\psi_{1\, x}$, $\ldots$ in~\eqref{Omegapsi} on the left. Then, taking into account $\nabla_e = D_x$, the corresponding form $\omega_{\nabla_e}$ can be obtained by replacing $\psi_1\mapsto D_x(\chi^1_0)\mapsto \frac{1}{2}\theta_x\hspace{0.15ex} \wedge $ in~\eqref{Omegapsi}:
\begin{align}
\begin{aligned}
\omega_{\nabla_e} = {} &\dfrac{1}{2}\theta_x\wedge \theta \wedge dx\wedge dy - \dfrac{1}{2} \theta_{xxx}\wedge \theta \wedge dt\wedge dx\\
&+ \dfrac{1}{2} (2u_x \theta_x\wedge \theta + \theta_x\wedge \theta_{xx} - \theta_{xx}\wedge \theta_x + \theta_{xxx}\wedge \theta + \theta_x\wedge \theta_{xy} - \theta_{xx}\wedge \theta_{y})\wedge dt\wedge dy\,.
\end{aligned}
\label{Omeganabla}
\end{align}
The corresponding variational $2$-form is represented by $\omega_{\nabla_e}|_{\mathcal{E}}$. As we show below, this variational $2$-form is $d_1$-closed, i.e., it is a presymplectic structure. In fact, under certain non-degeneracy conditions, all variational $2$-forms of evolution systems are presymplectic structures~\cite[Theorem~3]{GESSLER1997303}.
}

\subsection{Variational $2$-forms as conservation laws \label{SectionVar2FormsAsCL}}

Similarly to~\eqref{thecorrespdeg1}, a form $\theta^i_{\alpha}\wedge \theta^j_{\beta}\wedge dx^{i_1}\wedge \ldots \wedge dx^{i_k}$ corresponds to the horizontal form of internal degree $2$
\begin{align}
q^i_{\alpha} q^j_{\beta}dx^{i_1}\wedge \ldots \wedge dx^{i_k} = -\dfrac{1}{2} E_q\lrcorner\, (E_q\lrcorner\, (\theta^i_{\alpha}\wedge \theta^j_{\beta}\wedge dx^{i_1}\wedge \ldots \wedge dx^{i_k}))
\label{thecorrespdeg2}
\end{align}
on $J^{\infty}(n, m; m)$. 
Suppose that $\nabla_e$ is an extension of an operator from~\eqref{var2formop}.
Replacing both $\chi$ and $\chi_0$ by $q$ in~\eqref{Presymvarforms} while keeping track of signs, we obtain
\begin{align}
\langle l_F(q), \nabla_e(q) \rangle - \langle q, l_F^{\,*} \nabla_e(q) \rangle = d_0 (E_{q}\lrcorner\, (E_{q}\lrcorner\, \omega_{\nabla_e}))\,.
\label{almostprescl}
\end{align}
For some total differential operator $\Delta\colon P(n, m)\times \varkappa(n, m)\to \widehat{\varkappa}(n, m)$, one has
\begin{align*}
l_F^{\,*}\circ \nabla_e - \nabla_e^*\circ l_F = \Delta(F, \cdot)\,.
\end{align*}
It is convenient to denote $\Delta(\cdot, \chi) = \Delta_{\chi}(\cdot)$. Then~\eqref{almostprescl} takes the form
\begin{align}
\langle l_F(q), \nabla_e(q) \rangle - \langle q, \nabla_e^*\, l_F(q) + \Delta_q(F) \rangle = d_0 (E_{q}\lrcorner\, (E_{q}\lrcorner\,\omega_{\nabla_e}))\,.
\label{minustwo}
\end{align}
Note that $\langle l_F(q), \nabla_e(q) \rangle = - \langle \nabla_e(q), l_F(q) \rangle$ and $\langle q, \Delta_q(F) \rangle = - \langle \Delta_q(F), q \rangle$. Integrating by parts, we find that $(\Delta_q^*(q), -2\nabla_e (q))$ is a characteristic of a conservation law of the tangent system $\mathfrak{T}[1]\mathcal{E}$.
This conservation law is represented by the horizontal $(n-1)$-form $E_{q}\lrcorner\, (E_{q}\lrcorner\,\omega_{\nabla_e})$, which corresponds to $\omega_{\nabla_e}$ up to the coefficient (as in~\eqref{thecorrespdeg2}). Then the variational $2$-form of $\mathcal{E}$ defined by $\omega_{\nabla_e}|_{\mathcal{E}}$ corresponds to the conservation law of $\mathfrak{T}[1]\mathcal{E}$ having the characteristic $\psi_l = (-\frac{1}{2} \Delta_q^*(q), \nabla_e (q))$. Indeed,
\begin{align}
- \dfrac{1}{2}\langle \Delta_q^*(q), F_0 \rangle + \langle \nabla_e(q), F_1 \rangle = d_0 \Big(\!\! - \! \dfrac{1}{2}E_{q}\lrcorner\, (E_{q}\lrcorner\,\omega_{\nabla_e}) + \ldots\Big)\,,
\label{prescl0}
\end{align}
where a term vanishing on $\mathfrak{T}[1]\mathcal{E}$ is omitted under $d_0$. 

Applying $\mathcal{L}_{E_q}$ to $-\frac{1}{2}E_{q}\lrcorner\, (E_{q}\lrcorner\,\omega_{\nabla_e})$, we obtain a conservation law of $\mathfrak{T}[1]\mathcal{E}$ having internal degree~$3$. The variational $2$-form is a presymplectic structure if and only if this conservation law is trivial. In this case, $\nabla$ is a presymplectic operator.

\vspace{1ex}

\remarka{The differential $d_0$ of $\omega_{\nabla_e}$ vanishes on $\mathcal{E}$. This implies that $d_0\hspace{0.15ex} \omega_{\nabla_e}$ is an element of $\mathcal{C}\Lambda^{n+1}(n, m)\wedge d_v I_{\mathcal{E}} + I_{\mathcal{E}}\cdot \mathcal{C}^2\Lambda^{n+2}(n, m)$. Multiplied by a constant factor, formula~\eqref{minustwo} provides such a decomposition.
}

\vspace{1ex}

\remarka{If $F = 0$ has extended Kovalevskaya form\footnote{The form \eqref{extKovunderdet} with $m_1 = m$.} and $\nabla_e(q)$ does not involve $q^i_{b_i t}$, $u^i_{b_i t}$ and their total derivatives ($\nabla$ is a presymplectic operator), then $\nabla_e^*(F) = \operatorname{E}(L)$ for some Lagrangian $L\in E^{0, \hspace{0.1ex} n}_0(n, m)$ ($\operatorname{E}$ is the Euler operator). Moreover, if some differential consequence of $F = 0$ is an Euler--Lagrange equation, then the corresponding\footnote{If $\mathrm{E}(L) = 0$ is a differential consequence of $F = 0$, then $\mathcal{E}$ is a subsystem (reduction) of its infinite prolongation $\mathcal{N}$. The presymplectic structure of $\mathcal{N}$ produced by $L$ restricts to $\mathcal{E}\subseteq \mathcal{N}$ by means of the corresponding pullback. In terms of operators, this restriction is given by $\mathcal{A}^*|_{\mathcal{E}}$ where $\operatorname{E}(L) = \mathcal{A}(F)$.} presymplectic structure of $\mathcal{E}$ leads to a unique $\nabla_e(q)$ that does not involve $q^i_{b_i t}$, $u^i_{b_i t}$ and their total derivatives. Hence, it gives rise to some $L_0$ such that $\nabla_e^*(F) = \mathrm{E}(L_0)$. Finally, if 
$b_1 = \ldots = b_m = 1$ and $\mathcal{E}$ coincides with the infinite prolongation of some system of Euler--Lagrange equations $\operatorname{E}(L) = 0$, then it also coincides with the infinite prolongation of $\mathrm{E}(L_0) = 0$ for such a Lagrangian $L_0$, giving rise to the same presymplectic structure as $L$ (see~\cite{DRUZHKOV2021104013}). \label{RemPresToLagr}
}

\vspace{1ex}

\examplea{Continuing with Example~\ref{ExampleRunning2}, we see from~\eqref{Omeganabla} that the conservation law of the corresponding tangent system
is represented by \label{ExampleWithLagr}
\begin{align}
&\dfrac{1}{2}\Big(q_x q\, dx\wedge dy + (2u_x q_x q + q_xq_{xx} - q_{xx}q_x + q_{xxx} q + q_x q_{xy} - q_{xx}q_y) dt\wedge dy - q_{xxx} q\hspace{0.15ex} dt\wedge dx\Big).
\label{Examcl}
\end{align}
Its characteristic has the form $(-\frac{1}{2} \Delta_q^*(q), \nabla_e(q)) = (0, q_x)$. For any evolution system and any extension $\nabla_e$ that does not involve $D_t$ or $u^i, u^i_{x^j}, u^i_{x^jx^k}, \ldots$, the corresponding operator $\Delta$ can be taken as zero. Since the Lie derivative $\mathcal{L}_{E_q}$ of~\eqref{Examcl} is zero, the corresponding conservation law of internal degree $3$ is trivial, and hence, $\nabla = D_x|_{\mathcal{E}}$ is a presymplectic operator. Note that the Lie derivative $\mathcal{L}_{E_q}$ of $\omega_{\psi}'$ corresponding to~\eqref{presstrpoten} is a conservation law of internal degree $2$. Moreover, this conservation law has the same cosymmetry as~\eqref{Examcl}. Then the presymplectic structure under consideration is $d_1$ of the variational $1$-form represented by the corresponding $\omega_{\psi}|_{\mathcal{E}}$. Thus, this presymplectic structure originates from the cosymmetry~\eqref{presstrpoten}, corresponding to the scaling symmetry $-3t\partial_t - x\partial_x - y\partial_y + u\partial_u + \ldots$ A Lagrangian from Remark~\ref{RemPresToLagr} can be taken in the form
\begin{align}
L = \lambda\, dt\wedge dx\wedge dy\,,\qquad \lambda = \dfrac{u_t u_x + u_{xx}(u_{xx} + u_{xy})}{2} - \dfrac{u_x^3}{3}\,.
\label{LagrangianExample}
\end{align}
}

\vspace{1ex}

\remarka{\draftnew{Suppose that a system $\mathcal{E}$ admits the one-parameter scaling symmetry
$x^a\mapsto x^a$, $u^i\mapsto \lambda^{w_i}u^i$, where $\lambda>0$ and all $w_i>0$. Then
$E_2^{2,\hspace{0.1ex} n-1}(\mathcal{E})=0$~\cite[p.~111]{Vinogradov1984II}. Since
$E_2^{2,\hspace{0.1ex} n-1}=\ker d_1^{2,\hspace{0.1ex} n-1}/\operatorname{im}d_1^{1,\hspace{0.1ex} n-1}$,
this vanishing implies that every presymplectic structure is $d_1$-exact. Because every variational
$1$-form arises from a cosymmetry, all presymplectic structures of $\mathcal{E}$ are therefore
produced by cosymmetries. In particular, this applies to homogeneous linear systems (see
also~\cite{olver1986noether}).}
}

\section{Invariant reduction \label{SectionInvRed}}

The reduction mechanism of~\cite{InvRedII} is a one-line observation. If a structure represented by $\omega$ is invariant under a symmetry $X$, the Lie derivative $\mathcal{L}_X\omega$ represents the trivial structure: $\mathcal{L}_X\omega = d_0\vartheta$ for some potential $\vartheta$. Since $X$ vanishes on the system $\mathcal{E}_X$ describing $X$-invariant solutions, the restriction of $d_0\vartheta$ to $\mathcal{E}_X$ is zero, so $\vartheta|_{\mathcal{E}_X}$ is a cocycle; its class is a reduction of $\omega$. For a conservation law of a $(1+1)$-dimensional system, this is precisely the statement that $\vartheta$ is a constant of invariant motion~\cite{InvRedI} --- the traveling-wave first integrals of the Introduction arise this way. The rest of this section states the mechanism in more detail and reformulates it in terms of the tangent system.

\vspace{1ex}

Let $\mathcal{E}\subset J^{\infty}(n, m)$ be the infinite prolongation of a system of differential equations $F = 0$. If $X = E_{\varphi}|_{\mathcal{E}}$ is a symmetry of $\mathcal{E}$, $\varphi\in \varkappa(n, m)$, then $X$-invariant solutions of $\mathcal{E}$ are described by the infinite prolongation $\mathcal{E}_X$ of the system
\begin{align*}
F = 0\,,\qquad \varphi = 0\,.
\end{align*}
Suppose $\omega\in E^{\hspace{0.1ex} p, \hspace{0.1ex} k}_0(\mathcal{E})$ represents an $X$-invariant element of a group $E^{\hspace{0.1ex} p, \hspace{0.1ex} k}_1(\mathcal{E})$ of Vinogradov's $\mathcal{C}$-spectral sequence. Then there exists $\vartheta\in E^{\hspace{0.1ex} p, \hspace{0.1ex} k-1}_0(\mathcal{E})$ such that
\begin{align}
\mathcal{L}_X \omega = d_0 \vartheta.
\label{redformula}
\end{align}
The system $\mathcal{E}_X\subset \mathcal{E}$ is characterized by the condition that $X$ vanishes at its points. Then~\eqref{redformula} implies that
\begin{align*}
d_0 \vartheta|_{\mathcal{E}_X} = 0\,,
\end{align*}
and $\vartheta|_{\mathcal{E}_X}\in E^{\hspace{0.1ex} p, \hspace{0.1ex} k-1}_0(\mathcal{E}_X)$ represents an element of $E^{\hspace{0.1ex} p, \hspace{0.1ex} k-1}_1(\mathcal{E}_X)$. This mapping from $X$-invariant elements of $E^{\hspace{0.1ex} p, \hspace{0.1ex} k}_1(\mathcal{E})$ to $E^{\hspace{0.1ex} p, \hspace{0.1ex} k-1}_1(\mathcal{E}_X)$ is well-defined if and only if $E^{\hspace{0.1ex} p, \hspace{0.1ex} k-1}_1(\mathcal{E})|_{\mathcal{E}_X} = 0$. We also say that the reduction is well-defined in the case $p = 0$, $k = 1$, $E^{\hspace{0.1ex} 0, \hspace{0.1ex} 0}_1(\mathcal{E})|_{\mathcal{E}_X} \subset H^0_{dR}(\mathcal{E}_X)$, as we consider reductions of elements of the group $E^{\hspace{0.1ex} 0, \hspace{0.1ex} 1}_1(\mathcal{E})$ modulo additive locally constant functions on $\mathcal{E}_X$. In other words, the invariant reduction of $X$-invariant elements of $E^{\hspace{0.1ex} 0, \hspace{0.1ex} 1}_1(\mathcal{E})$ yields elements of $E^{\hspace{0.1ex} 0, \hspace{0.1ex} 0}_1(\mathcal{E}_X)/H^0_{dR}(\mathcal{E}_X)$. For an $\ell$-normal $\mathcal{E}$, the invariant reduction of $X$-invariant elements of $E^{\hspace{0.1ex} p, \hspace{0.1ex} n-1}_1(\mathcal{E})$ is well-defined for $p\geqslant 0$ and any symmetry $X$ such that $\mathcal{E}_X$ is regular.\footnote{In fact, all regularity assumptions can be removed, but then the reduction mechanism becomes more delicate. In particular, one can allow $\mathcal{E}$ and $\mathcal{E}_X$ to have differential consequences involving only independent variables. If $\mathcal{E}_X$ is empty, then a reduction outcome is also empty.}

The invariant reduction of elements of $E^{p, \hspace{0.1ex} k}_1(\mathcal{E})$ can be based on their identification with the corresponding elements of $E^{ 0, \hspace{0.1ex} k}_1(\mathfrak{T}[1]\mathcal{E})$ having internal degree $p$.
More specifically, there exists a symmetry $\mathcal{X}$
of $\mathfrak{T}[1]\mathcal{E}$ reproducing the action of $\mathcal{L}_X$ on Cartan $1$-forms. It has the characteristic $\widetilde{\varphi}|_{\mathfrak{T}[1]\mathcal{E}}\in \varkappa(\mathfrak{T}[1]\mathcal{E})$, where
\begin{align*}
\widetilde{\varphi} = (\varphi, E_q(\varphi))\in \varkappa(n, m; m) = \underbrace{\mathcal{F}(n, m; m)\times \ldots \times \mathcal{F}(n, m; m)}_{2m}\,.
\end{align*}
Let us denote $\varphi_0 = \varphi$ and $\varphi_1 = E_q(\varphi)$. Then $\mathcal{X} = E_{\widetilde{\varphi}}|_{\mathfrak{T}[1]\mathcal{E}}$,
\begin{align*}
E_{\widetilde{\varphi}} = D_{\alpha}(\varphi^i_0)\partial_{u^i_{\alpha}} + D_{\alpha}(\varphi^i_1)\partial_{q^i_{\alpha}}.
\end{align*}
We call $\mathcal{X}$ the \emph{lift} of $X$ to $\mathfrak{T}[1]\mathcal{E}$. Thus, the Lie derivative $\mathcal{L}_X \omega$ of any $\omega\in E_0^{p,\hspace{0.1ex} k}(\mathcal{E})$ corresponds to $\mathcal{L}_{\mathcal{X}}\hspace{0.15ex} \omega'$, where $\omega'\in E_0^{0,\hspace{0.1ex} k}(\mathfrak{T}[1]\mathcal{E})$ corresponds to $\omega$. The correspondence identifies $\theta^i_{\alpha}|_{\mathcal{E}}$ with $q^i_{\alpha}|_{\mathfrak{T}[1]\mathcal{E}}$ by means of the contractions with $E_q|_{\mathfrak{T}[1]\mathcal{E}}$ (as in Appendix~\ref{App:A}). Then in terms of the tangent system,~\eqref{redformula} takes the form
\begin{align*}
\mathcal{L}_{\mathcal{X}}\hspace{0.15ex} \omega' = d_0 ((-1)^p \vartheta')\,,
\end{align*}
where $\vartheta'\in E_0^{0,\hspace{0.1ex} k-1}(\mathfrak{T}[1]\mathcal{E})$ corresponds to $\vartheta$.

\section{A reduction algorithm for evolution systems \label{SectionEvolAlg}}

This section presents the first of the two main algorithms of the paper. It applies to systems of evolution equations and computes the reduction of any $X$-invariant structure represented as a conservation law of the tangent system $\mathfrak{T}[1]\mathcal{E}$ of internal degree $p$: conservation laws of $\mathcal{E}$ itself for $p = 0$, variational $1$-forms (cosymmetries) for $p = 1$, and presymplectic structures and other variational $2$-forms for $p = 2$. For $p = 0$, the algorithm reproduces the conservation-law reduction of~\cite{InvRedII}; the higher-degree cases are handled by the \emph{same} three steps, executed on the tangent system. A summary is given in Algorithm~\ref{algorithma} below; a complete \texttt{Maple} implementation is provided in Appendix~\ref{App:B}.

\vspace{1ex}

Let us consider a system of evolution equations $F = 0$, where $t$ denotes $x^1$,
\begin{align*}
F^i = u^i_t - f^i,\qquad i = 1, \ldots, m\,,
\end{align*}
and all functions $f^1$, \ldots, $f^m$ are independent of $u^j_{t + \alpha}$ for $|\alpha|\geqslant 0$, $1\leqslant j \leqslant m$. The tangent system $\mathfrak{T}[1]\mathcal{E}$ is the infinite prolongation of
\begin{align*}
u^i_t - f^i = 0 \,,\qquad q^i_t - E_q(f^i) = 0\,,\qquad i = 1,\ldots, m\,.
\end{align*}
We treat the variables $u^i_{t + \alpha}$, $q^i_{t+\alpha}$ as external coordinates associated with $\mathfrak{T}[1]\mathcal{E}$. Other adapted coordinates can be treated as local coordinates on the system $\mathfrak{T}[1]\mathcal{E}$. Then functions on $\mathfrak{T}[1]\mathcal{E}$ can be treated as functions on the ambient jet space $J^{\infty}(n, m;m)$,
\begin{align}
\mathcal{F}(\mathfrak{T}[1]\mathcal{E}) \subset \mathcal{F}(n, m;m)\,.
\label{Inclusions}
\end{align}
The same applies to differential forms $\Lambda^*(\mathfrak{T}[1]\mathcal{E}) \subset \Lambda^*(n, m;m)$, elements of $\varkappa(\mathfrak{T}[1]\mathcal{E})\subset \varkappa(n, m;m)$, and so on.
Since the system $\mathfrak{T}[1]\mathcal{E}$ also has an evolution form, the relation
\begin{align}
E_{\widetilde{\varphi}}(\widetilde{F}) = l_{\widetilde{\varphi}}(\widetilde{F})
\label{symmetrytotop}
\end{align}
holds for the characteristic $\varphi\in \varkappa(\mathcal{E})\subset \varkappa(n, m)$ of a symmetry of $\mathcal{E}$.

\subsection{The algorithm}

\algorithma{invariant reduction for evolution systems\label{algorithma}}{
\emph{Input:} an evolution system $F = 0$; the characteristic $\varphi$ of a symmetry $X$, written in internal coordinates; the cosymmetry $(\psi'^{\hspace{0.1ex}p}, \psi'^{\hspace{0.1ex}p-1})$ associated with a conservation law of $\mathfrak{T}[1]\mathcal{E}$, having internal degree $p\geqslant 0$.\\[0.6ex]
\emph{Output:} a potential $\widehat{\vartheta}'$ encoding an element of $E^{\hspace{0.1ex}p,\hspace{0.1ex}n-2}_0(n, m)$ whose restriction to $\mathcal{E}_X$ represents the reduction of the corresponding $X$-invariant element of $E^{\hspace{0.1ex}p,\hspace{0.1ex}n-1}_1(\mathcal{E})$.\\[0.6ex]
\emph{Step 1 (invariance check):} verify that $E_{\widetilde{\varphi}}(\psi_r) + l^{\hspace{0.15ex}*}_{\widetilde{\varphi}}(\psi_r) = 0$; if not, the structure is not $X$-invariant and no reduction is defined.\\[0.3ex]
\emph{Step 2 (integration by parts):} find the operator $A$ of~\eqref{IntByParts}.\\[0.3ex]
\emph{Step 3 (homotopy):} compute $\widehat{\vartheta}'$ from~\eqref{TheLastTotalHomotopy} via the total homotopy formula.\\[0.6ex]
At $p = 0$ no odd variables are needed: $\psi'^{\hspace{0.1ex}p}$ is a conservation-law multiplier of $F = 0$.
}

Suppose $\psi_l$ is the cosymmetry of a conservation law of $\mathfrak{T}[1]\mathcal{E}$ having internal degree $p\geqslant 0$. Then $\psi_l$ has components of degrees $p$ and $p-1$, $\psi_l = (\psi'^{\hspace{0.1ex} p}, \psi'^{\hspace{0.1ex} p-1})$. The conservation law is represented by the restriction $\omega'$ of a horizontal form $\widehat{\omega}'\in E^{0, \hspace{0.1ex} n-1}_0(n, m; m)$ of internal degree $p$ such that
\begin{align}
&\langle \psi_l, \widetilde{F}\rangle = d_0\hspace{0.2ex} \widehat{\omega}'\,.
\label{Firstpoten}
\end{align}
It is convenient to introduce the right cosymmetry $\psi_{r} = (\psi'^{\hspace{0.1ex} p}, (-1)^{p-1} \psi'^{\hspace{0.1ex} p-1})$. Then
\begin{align*}
d_0 \hspace{0.2ex} \widehat{\omega}' = \langle \psi_l, \widetilde{F}\rangle = \langle \psi'^{\hspace{0.1ex} p}, F_0 \rangle + \langle \psi'^{\hspace{0.1ex} p-1}, F_1 \rangle = \langle F_0, \psi'^{\hspace{0.1ex} p} \rangle + \langle F_1, (-1)^{p-1} \psi'^{\hspace{0.1ex} p-1} \rangle = \langle \widetilde{F}, \psi_{r} \rangle\,.
\end{align*}
Therefore, for the symmetry $\mathcal{X} = E_{\widetilde{\varphi}}$, one has
\begin{align}
d_0\hspace{0.2ex} \mathcal{L}_{E_{\widetilde{\varphi}}}\hspace{0.2ex} \widehat{\omega}' &= \mathcal{L}_{E_{\widetilde{\varphi}}}\hspace{0.2ex} d_0\hspace{0.2ex} \widehat{\omega}' = \mathcal{L}_{E_{\widetilde{\varphi}}} \langle \widetilde{F}, \psi_{r} \rangle = \langle l_{\widetilde{\varphi}}(\widetilde{F}), \psi_{r} \rangle + \langle \widetilde{F}, E_{\widetilde{\varphi}} (\psi_{r}) \rangle
\label{LieDerRightChar}
\end{align}
due to~\eqref{symmetrytotop}.

\noindent
\textbf{Step 1.} Check that the conservation law represented by $\omega'$ is $\mathcal{X}$-invariant. Since $\mathfrak{T}[1]\mathcal{E}$ is $\ell$-normal, this conservation law is $\mathcal{X}$-invariant if and only if the right characteristic $E_{\widetilde{\varphi}} (\psi_{r}) + l_{\widetilde{\varphi}}^{\hspace{0.15ex} *}(\psi_{r})$ corresponding to $\mathcal{L}_{\mathcal{X}} \hspace{0.2ex} \omega'$ vanishes on $\mathfrak{T}[1]\mathcal{E}$. This is equivalent to its vanishing on $J^{\infty}(n, m; m)$ because $\mathfrak{T}[1]\mathcal{E}$ is an evolution system and we use the inclusions determined by~\eqref{Inclusions}.

\noindent
\textbf{Step 2.} Using integration by parts, find a total differential operator $A\colon \varkappa(n, m; m)\to E^{\hspace{0.1ex} 0, \hspace{0.1ex} n-1}_0(n, m; m)$ satisfying
\begin{align}
\langle l_{\widetilde{\varphi}}(\widetilde{F}), \psi_{r} \rangle - \langle \widetilde{F}, l_{\widetilde{\varphi}}^{\hspace{0.15ex} *}(\psi_{r}) \rangle = d_0 A \widetilde{F}.
\label{IntByParts}
\end{align}

\noindent
\textbf{Step 3.}
Find an explicit potential $(-1)^p\widehat{\vartheta}'$ by applying the total homotopy formula~\cite{Olver1993} \draftnew{(for $p>0$, use its graded-commutative version; see, e.g., the horizontal homotopy operator in~\cite[Sec.~4A, p.~117, Eqs.~(4.13)--(4.14)]{Anderson1989Var})},
\begin{align}
	\mathcal{L}_{E_{\widetilde{\varphi}}} \hspace{0.2ex} \widehat{\omega}' - A \widetilde{F} = d_0 ((-1)^p \widehat{\vartheta}')\,.
	\label{TheLastTotalHomotopy}
\end{align}
Indeed, if the conservation law is $\mathcal{X}$-invariant, one obtains from~\eqref{LieDerRightChar} and~\eqref{IntByParts}
\begin{align}
d_0\hspace{0.2ex} \mathcal{L}_{E_{\widetilde{\varphi}}}\hspace{0.2ex} \widehat{\omega}' &= d_0 A \widetilde{F}.
\label{IntByPartsRes}
\end{align}
Restricting~\eqref{TheLastTotalHomotopy} to $\mathfrak{T}[1]\mathcal{E}$, we see that $\widehat{\vartheta}'|_{\mathfrak{T}[1]\mathcal{E}}$ corresponds to $\vartheta$ in~\eqref{redformula}. An implementation of this algorithm in \texttt{Maple} is provided in Appendix~\ref{App:B}. It is given in terms of the corresponding objects on $J^{\infty}(n, m)$.

\vspace{1ex}

\remarka{In coordinates, the operators $l_{\widetilde{\varphi}}$ and $l_{\widetilde{\varphi}}^{\hspace{0.15ex} *}$ are defined by
\begin{align*}
&l_{\widetilde{\varphi}}(\widetilde{F})_0^j = E_{\widetilde{F}}(\varphi_0^j) = D_{\alpha}(F_0^i)\dfrac{\partial \varphi_0^j}{\partial u^i_{\alpha}}\,,\qquad
l_{\widetilde{\varphi}}(\widetilde{F})_1^j = E_{\widetilde{F}}(\varphi_1^j) = D_{\alpha}(F_0^i)\dfrac{\partial \varphi_1^j}{\partial u^i_{\alpha}} + D_{\alpha}(F_1^i)\dfrac{\partial \varphi_1^j}{\partial q^i_{\alpha}}\,, \\
&\langle \widetilde{F}, l_{\widetilde{\varphi}}^{\hspace{0.15ex} *}(\psi_{r}) \rangle = \Big(F_0^j\, l_{\widetilde{\varphi}}^{\hspace{0.15ex} *}(\psi_{r})^p_j + F_1^j\, (-1)^{p-1} l_{\widetilde{\varphi}}^{\hspace{0.15ex} *}(\psi_{r})^{p-1}_j\Big) dt\wedge dx^2\wedge \ldots\wedge dx^n,\\
&l_{\widetilde{\varphi}}^{\hspace{0.15ex} *}(\psi_{r})^p_j = \sum_{\alpha}(-1)^{|\alpha|} D_{\alpha} \Big(\dfrac{\partial \varphi_0^i}{\partial u^j_{\alpha}} \psi'^{\hspace{0.1ex} p}_i + \dfrac{\partial \varphi_1^i}{\partial u^j_{\alpha}} (-1)^{p-1} \psi'^{\hspace{0.1ex} p-1}_i\Big),\quad
l_{\widetilde{\varphi}}^{\hspace{0.15ex} *}(\psi_{r})^{p-1}_j = \sum_{\alpha}(-1)^{|\alpha|} D_{\alpha} \Big(\dfrac{\partial \varphi_1^i}{\partial q^j_{\alpha}} \psi'^{\hspace{0.1ex} p-1}_i\Big),
\end{align*}
where $\psi_{r} = (\psi'^{\hspace{0.1ex} p}, (-1)^{p-1} \psi'^{\hspace{0.1ex} p-1})$, $\varphi_0 = \varphi$, and $\varphi_1 = E_q(\varphi)$; the invariance condition $E_{\widetilde{\varphi}} (\psi_{r}) + l_{\widetilde{\varphi}}^{\hspace{0.15ex} *}(\psi_{r}) = 0$ can be written in the form
\begin{align}
\label{InvCond1}
&D_{\alpha}(\varphi_0^i)\dfrac{\partial \psi'^{\hspace{0.1ex} p}_j}{\partial u^i_{\alpha}} + D_{\alpha}(\varphi_1^i)\dfrac{\partial \psi'^{\hspace{0.1ex} p}_j}{\partial q^i_{\alpha}} + \sum_{\alpha}(-1)^{|\alpha|} D_{\alpha} \Big(\dfrac{\partial \varphi_0^i}{\partial u^j_{\alpha}} \psi'^{\hspace{0.1ex} p}_i + \dfrac{\partial \varphi_1^i}{\partial u^j_{\alpha}} (-1)^{p-1} \psi'^{\hspace{0.1ex} p-1}_i\Big) = 0\,,\\
\label{InvCond2}
&D_{\alpha}(\varphi_0^i)\dfrac{\partial \psi'^{\hspace{0.1ex} p-1}_j}{\partial u^i_{\alpha}} + D_{\alpha}(\varphi_1^i)\dfrac{\partial \psi'^{\hspace{0.1ex} p-1}_j}{\partial q^i_{\alpha}} + \sum_{\alpha}(-1)^{|\alpha|} D_{\alpha} \Big(\dfrac{\partial \varphi_1^i}{\partial q^j_{\alpha}} \psi'^{\hspace{0.1ex} p-1}_i\Big) = 0\,,
\end{align}
where $j = 1, \ldots, m$. If $p = 0$, then $\psi'^{\hspace{0.1ex} p-1} = 0$ and it suffices to check only~\eqref{InvCond1}; for $p\geqslant 1$, it suffices to check only~\eqref{InvCond2}.
}

\vspace{1ex}

\examplea{\label{Example3}}Let us illustrate the algorithm using Example~\ref{ExampleRunning2}. The potentials presented here are obtained via the \texttt{Maple} implementation provided in Appendix~\ref{App:B}.

Equation~\eqref{2dpKdV} admits the following point symmetry
\begin{align}
-4t^2\partial_t - 4t(x-y)\partial_x + (x^2 - 2xy)\partial_u + \ldots
\label{PointSymmExamp}
\end{align}
The characteristic of its lift to the tangent system is given by
\begin{align*}
&\varphi_0 = x^2 - 2xy + 4t^2(u_x^2 + u_{xxx} + u_{xxy}) + 4t(x-y)u_x\,,\\
&\varphi_1 = 4t^2(2u_x q_x + q_{xxx} + q_{xxy}) + 4t(x-y)q_x\,.
\end{align*}
The presymplectic structure related to the operator $\nabla$ corresponds to
$p = 2$, $\psi'^{\hspace{0.1ex} p} = 0$, $\psi'^{\hspace{0.1ex} p-1} = \psi'^{\hspace{0.1ex} p-1}_1 = q_x$.

\noindent
\textbf{1.} A direct computation shows that $E_{\widetilde{\varphi}} (\psi_{r}) + l_{\widetilde{\varphi}}^{\hspace{0.15ex} *}(\psi_{r}) = 0$. 
One has
\begin{align*}
\langle l_{\widetilde{\varphi}}(\widetilde{F}), \psi_{r} \rangle = -E_{\widetilde{F}}(\varphi_1)q_x dt\wedge dx\wedge dy\,,\qquad
\langle \widetilde{F}, l_{\widetilde{\varphi}}^{\hspace{0.15ex} *}(\psi_{r}) \rangle = F_1 D_x(\varphi_1) dt\wedge dx\wedge dy\,.
\end{align*}
\noindent
\textbf{2.} As $A\widetilde{F}$ in~\eqref{IntByParts}, we take
\begin{align*}
\begin{aligned}
A\widetilde{F} = &\ \dfrac{4t}{3}\Big(3tF_1\, q_{xxx} + 2tF_1\, q_{xxy} + 3t F_{1\, xx}\, q_x + 2t F_{1\, xy}\, q_x 
- tF_{1\, y}\, q_{xx} - 3tF_{1\, x}\, q_{xx} - tF_{1\, x}\, q_{xy}\\
&+ 3(2tu_x + x - y)F_1\, q_x
\Big)dt\wedge dy + \dfrac{4t^2}{3}\Big(F_{1\, x}\, q_{xx} - F_{1\, xx}\, q_{x} - F_1\, q_{xxx}\Big) dt\wedge dx\,.
\end{aligned}
\end{align*}
Here we denote $D_x(F_{1}) = F_{1\, x}$, $D_x^2(F_{1}) = F_{1\, xx}$, $\ldots$

Let us choose the following potential for~\eqref{Firstpoten}
\begin{align*}
\widehat{\omega}' = \dfrac{1}{6}\Big((qq_{xxx} - 2q_x q_{xx})dt\wedge dx + (- 3qq_t + qq_{xxy}\! + 6q_x q_{xx}\! + 3q_x q_{xy}\! + q_y q_{xx})dt\wedge dy - 3qq_x dx\wedge dy\Big).
\end{align*}

\noindent
\textbf{3.}
Finally, one can take
\begin{align*}
\widehat{\vartheta}' &= \frac{t}{6} \Big(5t q q_{txx} - 4t q_x q_{tx} - 3t q_t q_{xx} - 8t u_{xxx} q q_x + 2(1 - 8t u_{xx}) q q_{xx} - 4(2t u_x + x - y) q q_{xxx}\\
&\quad   + 8(2t u_x + x - y) q_x q_{xx} - 4t q (q_{5x} + q_{4x+y})
+ 4t q_x (q_{xxxx} + q_{xxxy}) - 4t q_{xx} (q_{xxx} + q_{xxy}) \Big) dt\\
&\quad + \frac{t^2}{6} \Big( 5 q q_{xxx} - 7 q_x q_{xx} \Big) dx\\
&\quad + \frac{t}{6} \Big(24t q_x q_{xx} + 12t q_x q_{xy} + 5t q_y q_{xx} - 12(2t u_x + x - y) q q_x - 12t q q_{xxx} - 7t q q_{xxy} \Big) dy.
\end{align*}
as a potential $(-1)^p \widehat{\vartheta}' = \widehat{\vartheta}'$ in~\eqref{TheLastTotalHomotopy}.
Its restriction to the system $\mathfrak{T}[1]\mathcal{E}$ has the form
\begin{align*}
\vartheta' &= \frac{t}{6} \Big( 2t u_{xxx} q q_x + 2(2t u_{xx} + 1) q q_{xx} + 2(t u_x - 2x + 2y) q q_{xxx}
+ t q (q_{5x} + q_{4x+y})\\
&\quad - t q_{xx} (q_{xxx} + q_{xxy}) + 2(t u_x + 4x - 4y) q_x q_{xx} \Big) dt
+ \frac{t^2}{6} \Big( 5 q q_{xxx} - 7 q_x q_{xx} \Big) dx\\
&\quad + \frac{t}{6} \Big(24t q_x q_{xx} + 12t q_x q_{xy} + 5t q_y q_{xx} - 12(2t u_x + x - y) q q_x - 12t q q_{xxx} - 7t q q_{xxy} \Big) dy.
\end{align*}
In terms of $\mathcal{E}$, it defines $\vartheta\in E^{\hspace{0.1ex} p, \hspace{0.1ex} 1}_0(\mathcal{E})$ in~\eqref{redformula} by replacing $q\mapsto \,\overline{\!\theta} \, \wedge $\,, $q_x\mapsto \,\overline{\!\theta}_{x} \, \wedge $\,, $\ldots$,
\begin{align*}
\vartheta &= \frac{t}{6} \Big( 2t u_{xxx} \,\overline{\!\theta} \wedge \,\overline{\!\theta}_{x} + 2(2t u_{xx} + 1) \,\overline{\!\theta} \wedge \,\overline{\!\theta}_{xx} + 2(t u_x - 2x + 2y) \,\overline{\!\theta} \wedge \,\overline{\!\theta}_{xxx}
+ t \,\overline{\!\theta} \wedge (\,\overline{\!\theta}_{5x} + \,\overline{\!\theta}_{4x+y})\\
&\quad - t \,\overline{\!\theta}_{xx} \wedge (\,\overline{\!\theta}_{xxx} + \,\overline{\!\theta}_{xxy}) + 2(t u_x + 4x - 4y) \,\overline{\!\theta}_{x} \wedge \,\overline{\!\theta}_{xx} \Big)\wedge dt
+ \frac{t^2}{6} \Big( 5 \,\overline{\!\theta} \wedge \,\overline{\!\theta}_{xxx} - 7 \,\overline{\!\theta}_{x} \wedge \,\overline{\!\theta}_{xx} \Big)\wedge dx\\
&\quad +\! \frac{t}{6} \Big(24t \,\overline{\!\theta}_{x}\! \wedge \,\overline{\!\theta}_{xx}\! + \! 12t \,\overline{\!\theta}_{x}\! \wedge \,\overline{\!\theta}_{xy}\! + \! 5t \,\overline{\!\theta}_{y} \wedge \,\overline{\!\theta}_{xx}\! - \! 12(2t u_x\! +\! x \!-\! y) \,\overline{\!\theta}\! \wedge \,\overline{\!\theta}_{x}\! - \! 12t \,\overline{\!\theta} \wedge \,\overline{\!\theta}_{xxx}\! - \! 7t \,\overline{\!\theta}\! \wedge \,\overline{\!\theta}_{xxy} \Big)\! \wedge dy,
\end{align*}
where $\,\overline{\!\theta} = \theta|_{\mathcal{E}} = du - (u_x^2 + u_{xxx} + u_{xxy})dt - u_x dx - u_y dy$, $\,\overline{\!\theta}_x = \theta_x|_{\mathcal{E}}$, $\ldots$ The restriction of $\vartheta$ to $\mathcal{E}_X$ represents the reduction of the presymplectic structure corresponding to $\nabla = D_x|_{\mathcal{E}}$.

\vspace{1ex}

\examplea{Let us illustrate the algorithm in the case of a higher symmetry. \label{ExampleHigher} The potentials presented here are also obtained via the \texttt{Maple} implementation provided in Appendix~\ref{App:B}.
	
Consider the cotangent covering
\begin{align*}
	w_{tt} - w_{xy} - v_t w_{xx} + v_x w_{tx} - 2(v_{tx} w_x - v_{xx} w_t) = 0\,,\qquad v_{tt} - v_{xy} - v_t v_{xx} + v_x v_{tx} = 0 
\end{align*}
of Pavlov's equation\footnote{Also called Pavlov--Mikhalev or r-th dispersionless KP equation.}~\cite{mikhalev1992hamiltonian, pavlov2003integrable} and denote by $\mathcal{E}$ its infinite prolongation. The characteristic $(v_x, w_x)$ gives rise to its conservation law, which is invariant under the following higher symmetry~\cite{Baran_2014}
\begin{align}
	X = 0\, \partial_v + (v_{xxy} + v_{xxx}(v_t + v_x^2) + v_{xx}(v_{tx} + v_x v_{xx}) + v_x v_{txx})\partial_w + \ldots
	\label{SymmetryHigher}
\end{align}
To apply the algorithm, we rewrite the system in the evolution form $F = 0$ for $u^1 = v$, $u^3 = w$ and the auxiliary variables $u^2 = u^1_t$, $u^4 = u^3_t$. Then one has
\begin{align}
	\begin{aligned}
		&F^1 = u^1_t - u^2,\qquad
		F^2 = u^2_{t} - u^1_{xy} - u^2 u^1_{xx} + u^1_x u^2_{x}\,,\\
		&F^3 = u^3_t - u^4,\qquad 
		F^4 = u^4_{t} - u^3_{xy} - u^2 u^3_{xx} + u^1_x u^4_{x} - 2(u^2_{x} u^3_x - u^1_{xx} u^4).
	\end{aligned}
	\label{EquationsPavlovCotan}
\end{align}
The conservation law characteristic takes the form $\psi'^{\hspace{0.1ex} p} = \psi^p = (-u^4_x , u^3_x, -u^2_x, u^1_x)$. Here $\psi^{p-1} = 0$ because $p = 0$. Then the symmetry characteristic reads
\begin{align*}
	\varphi^1 = 0,\quad \varphi^2 = 0,\quad \varphi^3 = u^1_{xxy} + u^1_{xxx}(u^2 + (u^1_x)^2) + u^1_{xx}(u^2_{x} + u^1_x u^1_{xx}) + u^1_x u^2_{xx},\quad \varphi^4 = \,\overline{\!D}_t(\varphi^3)
\end{align*}
for $\,\overline{\!D}_t = D_t|_{\mathcal{E}}$.
Let us choose the following potential for~\eqref{Firstpoten}
\begin{align*}
	\widehat{\omega} = -u^1_x u^3_x\, dt\wedge dx + (u^1_t u^4 - u^2_t u^3 - u^2 u^4 + u^1_x u^2 u^3_x - (u^1_x)^2 u^4) dt\wedge dy + (u^1_x u^4 - u^2_x u^3) dx\wedge dy.
\end{align*}
Here $\widehat{\omega}' = \widehat{\omega}$ since $p = 0$. Applying the algorithm, we find
\begin{align*}
	\vartheta= {}&\Bigl(
	v_x^3 v_{xxx}
	+\frac12 v_x^2 v_{xx}^2
	+\frac1{18}v v_x v_{txxx}
	+\frac7{18}v v_{tx} v_{xxx}
	-\frac1{18}v v_{xx} v_{txx}
	+\frac12 v_x v_{xxy}
	+\frac16 v v_{xxxy} \\
	&\qquad
	-\frac19 v_t v_{xx}^2
	+\frac{35}{36} v_x^2 v_{txx}
	+\frac19 v_x v_{tx} v_{xx}
	+\frac16 v v_t v_{xxxx}
	+\frac{13}{18}v_t v_x v_{xxx}
	-\frac16 v_{xx} v_{xy}
	\Bigr)\,dt \\
	&+\Bigl(
	-\frac{11}{36}v_x^2 v_{xxx}
	+\frac29 v v_x v_{xxxx}
	+\frac13 v v_{xx} v_{xxx}
	+\frac16 v v_{txxx}
	-\frac13 v_x v_{txx}
	+\frac13 v_{tx} v_{xx}
	-\frac16 v_t v_{xxx}
	\Bigr)\,dx \\
	&+\Bigl(
	v_t v_x^2 v_{xxx}
	+\frac29 v v_x v_{xxxy}
	+\frac19 v v_{xx} v_{xxy}
	+\frac29 v v_{xy} v_{xxx}
	+\frac{17}{36} v_x^2 v_{xxy}
	+\frac29 v_x v_y v_{xxx} \\
	&\qquad
	-\frac1{18}v_x v_{xx} v_{xy}
	+v_t v_x v_{txx}
	+\frac1{18} v_y v_{xx}^2
	-v_t v_{tx} v_{xx}
	+\frac16 v v_{txxy}
	+\frac12 v_x v_{txy}
	+\frac16 v_y v_{txx} \\
	&\qquad
	-\frac16 v_{xx} v_{ty}
	-\frac12 v_{tx} v_{xy}
	-\frac16 v_t v_{xxy}
	\Bigr)\,dy.
\end{align*}
The restriction of $\vartheta$ to $\mathcal{E}_X$ represents the corresponding element of the group $E_1^{\hspace{0.1ex} 0, 1}(\mathcal{E}_X)$.
Moreover, the restriction of $\vartheta$ to the infinite prolongation $\mathcal{S}$ of the reduced Pavlov equation
\begin{align*}
	v_{tt} - v_{xy} - v_t v_{xx} + v_x v_{tx} = 0,\qquad v_{xxy} + v_{xxx}(v_t + v_x^2) + v_x v_{txx} + v_{xx}(v_{tx} + v_x v_{xx}) = 0
\end{align*}
represents the corresponding element of its group $E_1^{\hspace{0.1ex} 0, 1}$.
}

\vspace{1ex}

\remarka{One can introduce a one-dimensional differential covering over $\mathcal{S}$ by adjoining a nonlocal variable $h$ subject to the relation
$\vartheta|_{\mathcal{S}} = d_0 h$. \label{RemarkDiffCovering}}

\vspace{1ex}

\remarka{Since $X$ is a symmetry of the evolution system $\mathcal{E} \simeq (\mathbb{R}\times J^{\infty}(2, 4), \,\overline{\!D}_t)$, the corresponding subsystem $\mathcal{E}_X\subset \mathcal{E}$, specified by the condition $X = 0$, is the infinite prolongation of the family of systems $\varphi^1 = \ldots = \varphi^4 = 0$ on the space $J^{\infty}(2, 4)$, where the total derivative $\,\overline{\!D}_t$ is not present and $t\in \mathbb{R}$ plays the role of a parameter. A simple analysis shows that this infinite prolongation can be solved in $u^1_{xxy}$, $u^2_{xxy}$ and their total derivatives (on $J^{\infty}(2, 4)$).
Thus, as intrinsic coordinates on $\mathcal{S}\subset J^{\infty}(3, 1)$, one can take $t, x, y, v$ and all total derivatives of $v$ except $v_{tt}, v_{xxy}$ and their total derivatives. In particular, 
\begin{align*}
	v_{txxy}|_{\mathcal{S}} = {}&2 v_x v_{xx}^3 - (2 v_{tx} - 3 v_x v_{xx}) v_{txx} - (v_t v_{xx} - 5 v_x^2 v_{xx} + v_{xy}) v_{xxx} - (v_t - v_x^2) v_{txxx} + v_x^3 v_{xxxx}\,.
\end{align*}
}

\section{A descent algorithm for general $\ell$-normal systems \label{SectionDescent}}

The algorithm of Section~\ref{SectionEvolAlg} requires an evolution representation of the system. The approach of this section removes that restriction for $p\geqslant 1$: it applies to any $\ell$-normal system and symmetry for which the additional assumption in Theorem~\ref{Theoremgenerfun} holds. As noted below, this assumption is satisfied in broad classes of cases. Computationally, the main ingredient is a \emph{descent} procedure that generalizes integration by parts from total divergences to total differential operators with values in arbitrary horizontal forms; each stage of the descent lowers the order of the operator by one, in direct analogy with removing one derivative at a time when integrating by parts.

\vspace{1ex}

Let $X = E_{\varphi}|_{\mathcal{E}}$ be a symmetry of an $\ell$-normal system $\mathcal{E}$. Suppose that $\psi_l$ is a left characteristic of an $\mathcal{X}$-invariant conservation law of $\mathfrak{T}[1]\mathcal{E}$ having internal degree $p\geqslant 1$. The conservation law is represented by the restriction $\omega'$ of a horizontal form $\widehat{\omega}'\in E^{\hspace{0.1ex} 0, \hspace{0.1ex} n-1}_0(n, m; m)$ of internal degree $p$ such that
\begin{align*}
&\langle \psi_l, \widetilde{F}\rangle = d_0\hspace{0.2ex} \widehat{\omega}'.
\end{align*}
The corresponding form $\widehat{\omega}$ defines the operator $\gamma_{\hspace{0.15ex} \widehat{\omega}}\colon \varkappa(n, m)\to E^{p-1,\hspace{0.1ex} n-1}_0(n, m)$, $\chi\mapsto E_{\chi} \lrcorner\, \widehat{\omega}$.

The characteristic $\psi_l$ defines the homomorphism $\psi^{p-1}\colon P(n, m)\to E_0^{p-1,\hspace{0.1ex} n}(n, m)$ corresponding to its component $\psi'^{\hspace{0.1ex} p-1}$. In what follows, we omit the superscript $p-1$ and denote $\psi^{p-1}$ by $\psi$. 
Moreover, the symmetry $X$ gives rise to a total differential operator $\Phi\colon P(n, m)\to P(n, m)$ such that $E_{\varphi}(F) = \Phi(F)$. 
Integrating by parts in $\psi(\Phi(G))$, one obtains a total differential operator $\nu\colon P(n, m)\to E_0^{p-1,\hspace{0.1ex} n-1}(n, m)$
satisfying 
\begin{align}
\psi(\Phi(F)) = (\Phi^*_{(p-1)}\psi)F + d_0\hspace{0.15ex} \nu(F)\,.
\label{Relation5}
\end{align}
Here $\Phi^*_{(p-1)}$ is the total differential operator arising from integration by parts. 

Let us recall that a total differential operator $\gamma\colon \varkappa(n, m)\to E_0^{p-1,\hspace{0.1ex} k}(n, m)$ identifies\footnote{One can write all $D_{\alpha}(\chi^i)$ on the left and then replace them by $\theta^i_{\alpha}\otimes $.} with an element of $\mathcal{C}\Lambda^1(n, m)\otimes E_0^{p-1,\hspace{0.1ex} k}(n, m) = \mathcal{C}\Lambda^1(n, m)\otimes \mathcal{C}^{p-1}\Lambda^{p-1}(n, m) \otimes E_0^{0,\hspace{0.1ex} k}(n, m)$, whose direct summand in $E_0^{p,\hspace{0.1ex} k}(n, m) = \mathcal{C}^{p}\Lambda^{p}(n, m) \otimes E_0^{0,\hspace{0.1ex} k}(n, m)$ arises through the projection (alternatization) $\operatorname{Alt}\colon \mathcal{C}\Lambda^1(n, m)\otimes \mathcal{C}^{p-1}\Lambda^{p-1}(n, m)\to \mathcal{C}^{p}\Lambda^{p}(n, m)$. We denote the corresponding direct summand in $E_0^{p,\hspace{0.1ex} k}(n, m)$ by $\widehat{\omega}_\gamma$. Then the reduction algorithm is based on the following result.

\vspace{1ex}

\theorema{There exists a total differential operator $\gamma\colon \varkappa(n, m)\to E_0^{p-1,\hspace{0.1ex} n-2}(n, m)$
such that
\begin{align}
(\nu\circ l_{F} - \gamma_{\hspace{0.15ex} \widehat{\omega}} \circ l_{\varphi})|_{\mathcal{E}_X} = d_0 \circ \gamma|_{\mathcal{E}_X}.
\label{Theoremrelation}
\end{align}
The restriction $\widehat{\omega}_\gamma|_{\mathcal{E}_X}$ represents an element of $E_1^{p,\hspace{0.1ex} n-2}(\mathcal{E}_X)$. If the equation $\square\circ l_F|_{\mathcal{E}_X} = 0$ for a total differential operator $\square\colon P(\mathcal{E}_X) \to \mathcal{F}(\mathcal{E}_X)$ has no nonzero solutions, then $\widehat{\omega}_\gamma|_{\mathcal{E}_X}\in E_0^{p,\hspace{0.1ex} n-2}(\mathcal{E}_X)$ represents the reduction of the given element of $E_1^{p,\hspace{0.1ex} n-1}(\mathcal{E})$.\label{Theoremgenerfun}
}

\vspace{1ex}

\noindent
The proof is routine but somewhat technical (see Appendix~\ref{App:C}).

\vspace{1ex}

\remarka{The assumption that the equation $\square\circ l_F|_{\mathcal{E}_X} = 0$ has no nonzero solutions is not restrictive. In particular, it is satisfied by any\footnote{We assume, however, that the system $\mathcal{E}_X$ satisfies the regularity conditions.} symmetry if $F = 0$ is of the form~\eqref{extKovunderdet} or if $F = 0$ is a scalar equation with $l_F|_{\mathcal{E}_X} \neq 0$.
}

\vspace{1ex}

To find $\widehat{\omega}_{\gamma}$, it suffices to apply a simple computational procedure to the operator
\begin{align*}
\nu\circ l_{F} - \gamma_{\hspace{0.15ex} \widehat{\omega}} \circ l_{\varphi}\,.
\end{align*}
The procedure is a descent on the symbol (the highest-order\footnote{A total differential operator $\Delta = \Delta_i^{\alpha}\,\overline{\!D}_{\alpha}$, $\Delta_i^{\alpha}\in \Lambda^*(\mathcal{E})$, has order $h$ if $\Delta_i^{\alpha}$ does not vanish for some $\alpha$ with $|\alpha| = h$ but vanishes whenever $|\alpha| > h$.} terms) of the restriction of this operator to $\mathcal{E}_X$ using the homotopy~\eqref{HomotopyOperator}. It generalizes integration by parts to the case of total differential operators with values in horizontal forms of degree $k \leqslant n$; here $D_{x^i}$ play the role of $e_i$, while $dx^i$ play the role of $e^i$. The presence of Cartan forms only affects signs through contraction with total derivatives.
As with integration by parts, it suffices to consider operators on $J^{\infty}(n, m)$. 

Thus, the algorithm can be formulated as follows.

\noindent
\textbf{Step 1.} Find an operator $\nu$ satisfying~\eqref{Relation5} via integration by parts in $\psi(\Phi(G))$.

\noindent
\textbf{Step 2.} Determine the order of $(\nu\circ l_{F} - \gamma_{\hspace{0.15ex} \widehat{\omega}} \circ l_{\varphi})|_{\mathcal{E}_X}$. We denote it by $r + n - 1$.

\noindent
\textbf{Step 3.} Apply the homotopy~\eqref{HomotopyOperator} to terms of order\footnote{\draftnew{The homotopy is applied separately to the derivatives of each component $\chi^j$.}} $r + n - 1$ in $\nu\circ l_{F} - \gamma_{\hspace{0.15ex} \widehat{\omega}} \circ l_{\varphi}$ to find an operator $\gamma_{n+r-2}$ such that the order of the restriction of
$\nu\circ l_{F} - \gamma_{\hspace{0.15ex} \widehat{\omega}} \circ l_{\varphi} - d_0 \circ (\frac{1}{n+r}\gamma_{n+r-2})$
to $\mathcal{E}_X$ is less than $r + n - 1$.

\noindent
\textbf{Step 4.} Apply the homotopy~\eqref{HomotopyOperator} to terms of order $r + n - 2$ in $\nu\circ l_{F} - \gamma_{\hspace{0.15ex} \widehat{\omega}} \circ l_{\varphi} - d_0 \circ (\frac{1}{n+r}\gamma_{n+r-2})$ to find an operator $\gamma_{n+r-3}$ such that the order of the restriction of
$$
\nu\circ l_{F} - \gamma_{\hspace{0.15ex} \widehat{\omega}} \circ l_{\varphi} - d_0 \circ \Big(\frac{1}{n+r}\,\gamma_{n+r-2} + \frac{1}{n+r-1}\,\gamma_{n+r-3}\Big)
$$
to $\mathcal{E}_X$ is less than $r + n - 2$.

\noindent
\textbf{...}
Continuing, one obtains $\gamma = \frac{1}{n+r}\,\gamma_{n+r-2} + \frac{1}{n+r-1}\,\gamma_{n+r-3} + \ldots + \frac{1}{2}\hspace{0.1ex} \gamma_{0}$. It remains to reconstruct $\widehat{\omega}_{\gamma}$ by writing all $D_{\alpha}(\chi^i)$ in $\gamma$ on the left and replacing them by $\frac{1}{p} \theta^i_{\alpha}\hspace{0.1ex} \wedge $\,.

Let us illustrate the computational aspects of this approach with a simple example.

\vspace{1ex}

\examplea{Consider the infinite prolongation $\mathcal{E}$ of the Laplace equation \label{ExampleLaplace}}
\begin{align*}
    u_{xx} + u_{yy} = 0
\end{align*}
and its presymplectic structure corresponding to the following conservation law of the tangent system $\mathfrak{T}[1]\mathcal{E}$ in a characteristic form.
\begin{align*}
	q(q_{xx} + q_{yy})dx\wedge dy = d_0 (qq_x dy - qq_y dx).
\end{align*}
Here $\widehat{\omega} = \theta \wedge \theta_x \wedge dy - \theta \wedge \theta_y \wedge dx$ and
\begin{align*}
	&\psi\colon G\mapsto \theta\wedge G\, dx\wedge dy\,,\qquad
	\gamma_{\widehat{\omega}}\colon \chi \mapsto \chi \theta_x \wedge dy - \chi_x \theta \wedge dy - \chi \theta_y \wedge dx + \chi_y \theta \wedge dx\,.
\end{align*}

The presymplectic structure is invariant under the point symmetry
\begin{align*}
	x\partial_y - y\partial_x + 0\, \partial_u + \ldots\,,
\end{align*}
generating rotations and having the characteristic $\varphi = yu_x - xu_y$. Then
\begin{align*}
	l_{\varphi} = yD_x - xD_y = \Phi\,.
\end{align*}
\textbf{1.} From
\begin{align*}
	\psi(\Phi(G)) = \theta\wedge (y\, G_x - x\, G_y) dx\wedge dy = d_0 (-G\, \theta\wedge (x dx + y dy)) + G (x\theta_y - y\theta_x) \wedge dx\wedge dy
\end{align*}
it follows that we can set
\begin{align*}
    \nu(G) = - G\, \theta \wedge (xdx + ydy)\,.
\end{align*}
\textbf{2.} The order of $(\nu\circ l_{F} - \gamma_{\hspace{0.15ex} \widehat{\omega}} \circ l_{\varphi})|_{\mathcal{E}_X}$ is $2 = r+n-1$,
\begin{align*}
	\nu\circ l_{F} - \gamma_{\hspace{0.15ex} \widehat{\omega}} \circ l_{\varphi}\colon \chi \mapsto & -(y\chi_{xy} + x\chi_{xx} + \chi_x) \theta \wedge dx - (y\chi_{yy} + x \chi_{xy} + \chi_y) \theta \wedge dy\\
	& + (y\chi_x - x\chi_y) (\theta_y \wedge dx - \theta_x \wedge dy)\,.
\end{align*}
\textbf{3.} Applying the homotopy~\eqref{HomotopyOperator} to $-(y\chi_{xy} + x\chi_{xx}) \theta \wedge dx - (y\chi_{yy} + x \chi_{xy}) \theta \wedge dy$, we find
\begin{align*}
	\gamma_1(\chi) = 3(y\chi_{y} + x \chi_{x}) \theta\,.
\end{align*}
\textbf{4.} The terms of order $r + n - 2 = 1$ in $\nu\circ l_{F} - \gamma_{\hspace{0.15ex} \widehat{\omega}} \circ l_{\varphi} - d_0\circ(\frac{1}{3}\gamma_1)$ are given by
\begin{align*}
	(y\chi_x - x\chi_y) (\theta_y \wedge dx - \theta_x \wedge dy) + (y\chi_{y} + x \chi_{x}) (\theta_x\wedge dx + \theta_y\wedge dy)\,.
\end{align*}
The homotopy~\eqref{HomotopyOperator} yields
\begin{align*}
	\gamma_0(\chi) = -2\chi(x\theta_x + y\theta_y)\,.
\end{align*}
Finally, one obtains
\begin{align*}
	\gamma(\chi) = (y\chi_{y} + x \chi_{x}) \theta - \chi(x\theta_x + y\theta_y)\,,\qquad \widehat{\omega}_{\gamma} = (y\theta_{y} + x \theta_x)\wedge \theta\,.
\end{align*}
The restriction $\widehat{\omega}_{\gamma}|_{\mathcal{E}_X}$ represents the corresponding reduction.

\section{Symmetries that generate flows}\label{SectionFlows}

The symmetries most common in applications --- translations, scalings, rotations, Galilean boosts --- are point symmetries generating flows on the jet space. For such symmetries the reduction machinery can be simplified considerably: under the mild condition~\eqref{InvarianceonJets} below, a reduction is obtained directly from a Lie derivative and the total homotopy formula.
In the simplest and frequent situation $\mathcal{L}_{\mathcal{Y}}\hspace{0.15ex}\widehat{\omega}' = 0$ (see Remark~\ref{RemarkAndersonFelsAncoGandarias}), a reduction is represented by the contraction $-(-1)^p\hspace{0.2ex}\mathcal{Y}\lrcorner\, \widehat{\omega}'$ --- for a conservation law ($p = 0$), precisely the flux form contracted with the symmetry generator taken with minus (see Example~\ref{ExampleEuler} for the energy conservation inheritance in rotationally invariant Euler flows). In essence, the approach based on the relation $\mathcal{L}_{\mathcal{Y}}\hspace{0.15ex}\widehat{\omega}' = 0$ is the Anderson--Fels approach~\cite{AndersonFels1997} restricted to the case of one-dimensional Lie groups (up to sign). From a computational point of view, it is highly practical (when applicable): it reduces to taking linear combinations involving only known functions.

\vspace{1ex}

Let $Y$ be a symmetry of $J^{\infty}(n, m)$ whose restriction to an infinitely prolonged system $\mathcal{E}\subset J^{\infty}(n, m)$ is also a symmetry. Denote by $\mathcal{Y}$ the lift of $Y$ to $J^{\infty}(n, m; m)$. Here the action of $\mathcal{Y}$ on $\mathcal{F}(n, m; m)$ reproduces the action of $\mathcal{L}_Y$ on $\oplus_{p\geqslant 0}\, \mathcal{C}^p\Lambda^p(n, m)$. Consider a horizontal form $\widehat{\omega}'\in E^{0, \hspace{0.1ex} n-1}_0(n, m; m)$ of internal degree $p$ whose restriction to $\mathfrak{T}[1]\mathcal{E}$ represents a $\mathcal{Y}$-invariant conservation law. Suppose additionally that
\begin{align}
\mathcal{L}_{\mathcal{Y}}\hspace{0.15ex} d_0 \widehat{\omega}' = 0\,.
\label{InvarianceonJets}
\end{align}
Then one has $d_0 \mathcal{L}_{\mathcal{Y}}\hspace{0.15ex} \widehat{\omega}' = 0$. It follows that there exists $\widehat{\vartheta}'_{\mathcal{Y}}\in E^{0, \hspace{0.1ex} n-2}_0(n, m; m)$ of internal degree $p$ satisfying
\begin{align*}
\mathcal{L}_{\mathcal{Y}}\hspace{0.15ex} \widehat{\omega}' = d_0 ((-1)^p \widehat{\vartheta}'_{\mathcal{Y}})\,.
\end{align*}
It can be constructed via the total homotopy formula. Suppose that $X$ is the evolutionary symmetry equivalent to $Y|_{\mathcal{E}}$. One can show that the restriction of $\widehat{\vartheta}'_{\mathcal{Y}} - (-1)^p\, \mathcal{Y} \lrcorner\, \widehat{\omega}'$ to $\mathfrak{T}[1]\mathcal{E}_{\mathcal{X}}$ identifies with an element of $E_0^{p, \hspace{0.1ex} n-2}(\mathcal{E}_X)$ representing a\footnote{Recall that if $\mathcal{E}$ is $\ell$-normal, this reduction is unambiguously defined.} reduction of the $X$-invariant element of $E_1^{p, \hspace{0.1ex} n-1}(\mathcal{E})$ determined by $\omega'$.

\vspace{1ex}

\remarka{Relation~\eqref{InvarianceonJets} seems rather typical in applications involving point symmetries $Y$. If one additionally has $\mathcal{L}_{\mathcal{Y}}\hspace{0.15ex} \widehat{\omega}' = 0$, then, for $\widehat{\vartheta}'_{\mathcal{Y}} = 0$ and under suitable assumptions on $Y$, this approach aligns, up to sign, with the restriction of the Anderson--Fels method~\cite{AndersonFels1997} to the case of one-dimensional Lie algebras.
In this situation, the reduction can be \emph{globally} described in terms of the quotient of $\mathcal{E}_X$ by the flow of $Y|_{\mathcal{E}_X}$. For a scalar PDE $F = 0$ and a point symmetry $Y$, under suitable assumptions on $F$, the reduction based on~\eqref{InvarianceonJets} in the case $p = 0$ aligns with the method in~\cite{AncoGandarias2020} restricted to one-dimensional Lie algebras. \label{RemarkAndersonFelsAncoGandarias}}

\vspace{1ex}

\examplea{Let us consider the $(1+2)$-dimensional incompressible Euler equations
\begin{align*}
&u_t + uu_x + vu_y + P_x = 0\,,\qquad 
v_t + uv_x + vv_y + P_y = 0\,,\qquad
u_x + v_y = 0\,.
\end{align*}
Here $u^1 = u$, $u^2 = v$ are the velocity components, $u^3 = P$ is the pressure. The energy conservation law is determined by
\begin{align}
\widehat{\omega} = K\, dx\wedge dy - u(K + P) dt\wedge dy + v(K + P) dt\wedge dx\,,\qquad K = \dfrac{u^2 + v^2}{2}\,.
\label{EnergyCLRepres}
\end{align}
Since $p = 0$, condition~\eqref{InvarianceonJets} reduces to $\mathcal{L}_Y d_0 \widehat{\omega} = 0$. This condition holds for the rotation symmetry
\begin{align*}
Y = x\partial_y - y\partial_x + u\partial_v - v\partial_u + 0\,\partial_P + \ldots
\end{align*}
Moreover, one can see that $\mathcal{L}_Y \widehat{\omega} = 0$.
Then one can put $\widehat{\vartheta} = 0$ in $\mathcal{L}_Y \widehat{\omega} = d_0 \widehat{\vartheta}$, and the reduction is represented by the horizontal form
\begin{align*}
- Y\lrcorner\, \widehat{\omega} = K (x dx + y dy) - (xu + yv)(K + P) dt\,.
\end{align*}
The restriction of its differential to a rotationally invariant solution is zero.

We can introduce local coordinates $r, \phi, w, s$ such that
\begin{align*}
	x = r \cos \phi\,,\qquad y = r \sin \phi\,,\qquad w = xu + yv\,,\qquad
	s = \sqrt{u^2 + v^2}\,.
\end{align*}
Then the conserved quantity is determined by
\begin{align*}
	- Y\lrcorner\, \widehat{\omega} = K rdr - w(K + P) dt\,,\qquad K = \dfrac{s^2}{2}\,.
\end{align*}
It represents a conservation law of the (local) quotient of $\mathcal{E}_X$ by the action $\phi \mapsto \phi + \epsilon$.
\label{ExampleEuler}
}

\subsection{Reductions and the quotient systems \label{SectionRedandQuot}}

Let us consider point symmetries and discuss the invariant reduction locally in terms of characteristics of conservation laws of the tangent systems. Unlike in the previous sections, the locality of the present discussion is essential.

Let $X$ be equivalent to the restriction of the point symmetry $Y = \partial_{\tau}$ of $J^{\infty}(n, m)$, where $\tau$ denotes $x^1$. Suppose that the components of $F$ are independent\footnote{It is essential in the condition $Y(F^i) = 0$ that the components $F^i$ are scalar functions. In a more general situation when $F$ is a section of some nontrivial vector bundle, $Y(F)$ can be undefined if $Y$ is not an evolutionary symmetry.} of $\tau$ and denote by $H$ the vector function obtained from $F$ by substituting $u^i_{\tau + \alpha} = 0$ for $i = 1, \ldots, m$. Then the system $H = 0$ can be considered independently on the corresponding jet space $J^{\infty}(n-1, m)$ with the independent variables $(x^2, \ldots, x^n)$ and the same dependent variables $(u^1, \ldots, u^m)$. This system also arises as the reduction. More precisely, its infinite prolongation $\mathcal{S}\subset J^{\infty}(n-1, m)$ is the quotient of the action of the Lie group determined by $\partial_{\tau}$ on $\mathcal{E}_X$.
The canonical projection $\mathcal{E}_X\to \mathcal{S}$ induces the corresponding homomorphisms $E_1^{p, \hspace{0.1ex} n-2}(\mathcal{S})\to E_1^{p, \hspace{0.1ex} n-2}(\mathcal{E}_X)$.

From now on, we assume that both systems $\mathcal{E}$ and $\mathcal{S}$ are $\ell$-normal. Then the canonical homomorphisms $E_1^{p, \hspace{0.1ex} n-2}(\mathcal{S})\to E_1^{p, \hspace{0.1ex} n-2}(\mathcal{E}_X)$ are isomorphisms, $p\geqslant 0$ (see Appendix~\ref{App:C}). In this case, one can put $\tau = \tau_0\in \mathbb{R}$ and $d\tau = 0$ in a representative of an element of $E_1^{p, \hspace{0.1ex} n-2}(\mathcal{E}_X)$ and obtain the corresponding element of $E_1^{p, \hspace{0.1ex} n-2}(\mathcal{S})$, the same for all $\tau_0\in \mathbb{R}$.

Another practically useful approach for constructing reductions of invariant structures locally can be based on characteristics of conservation laws.
Let $\psi_l$ be a (left) characteristic of an $\mathcal{X}$-invariant conservation law of $\mathfrak{T}[1]\mathcal{E}$ corresponding to an element $\Omega\in E^{p,\hspace{0.1ex} n-1}_1(\mathcal{E})$. Choose $\tau_0\in \mathbb{R}$ and denote by $\rho$ the inclusion of the plane $\{\tau - \tau_0 = u^i_{\tau + \alpha} = q^i_{\tau + \alpha} = 0\} \subset J^{\infty}(n, m; m)$ into $J^{\infty}(n, m; m)$. In the coordinate form, this plane identifies with $J^{\infty}(n-1, m; m)$, and $d_0 \circ \rho^* = \rho^*\circ d_0$. 
Note that the lift $\mathcal{X}$ is equivalent to the restriction of the point symmetry $\mathcal{Y} = \partial_{\tau}$ of $J^{\infty}(n, m; m)$.
The composition $\mathcal{Y}\, \lrcorner\, \psi_l$ defines the corresponding homomorphism $\rho^*(\mathcal{Y}\, \lrcorner\, \psi_l)$ on $J^{\infty}(n-1, m; m)$, while the infinite prolongation of $\rho^*(\widetilde{F}) = 0$ is the system $\mathfrak{T}[1]\mathcal{S}$.
The following result extends the part of the method in~\cite{AncoGandarias2020} concerning one-dimensional Lie groups to systems of PDEs and arbitrary $p\geqslant 0$.

\vspace{1ex}

\propositiona{If $\mathcal{L}_\mathcal{Y}\hspace{0.15ex} \langle \psi_l, \widetilde{F} \rangle = 0$, then $(-1)^{p} \rho^*(\mathcal{Y}\, \lrcorner\, \psi_l)$ is a characteristic of a conservation law of the system $\mathfrak{T}[1]\mathcal{S}\subset J^{\infty}(n-1, m; m)$. The corresponding element of $E_1^{p, \hspace{0.1ex} n-2}(\mathcal{S})$ maps, via the canonical homomorphism $E_1^{p, \hspace{0.1ex} n-2}(\mathcal{S})\to E_1^{p, \hspace{0.1ex} n-2}(\mathcal{E}_X)$, to the reduction of $\Omega$.
}

\vspace{0.5ex}

\noindent
\textbf{Proof.} Let $\widehat{\omega}'\in E^{\hspace{0.1ex} 0, \hspace{0.1ex} n-1}_0(n, m; m)$ be a horizontal form of internal degree $p$ such that 
$\langle \psi_l, \widetilde{F} \rangle = d_0\hspace{0.15ex} \widehat{\omega}'$. Then one has
\begin{align*}
\mathcal{Y}\, \lrcorner\, \langle \psi_l, \widetilde{F} \rangle = D_{\tau}\, \lrcorner\, \langle \psi_l, \widetilde{F} \rangle = D_{\tau}\, \lrcorner\, d_0\hspace{0.15ex} \widehat{\omega}' = \mathcal{L}_{D_{\tau}} \widehat{\omega}' - d_0\hspace{0.15ex} (D_{\tau}\, \lrcorner\,  \widehat{\omega}')\,.
\end{align*}
Since $d_0 \mathcal{L}_\mathcal{Y}\hspace{0.15ex}\widehat{\omega}' = \mathcal{L}_\mathcal{Y}\hspace{0.15ex} d_0 \widehat{\omega}' = 0$, there exists a horizontal form $\widehat{\vartheta}'_\mathcal{Y}\in E^{\hspace{0.1ex} 0, \hspace{0.1ex} n-2}_0(n, m; m)$ of internal degree $p$ such that
\begin{align*}
\mathcal{L}_\mathcal{Y}\hspace{0.15ex} \widehat{\omega}' = d_0 ((-1)^{p} \widehat{\vartheta}_\mathcal{Y}')\,.
\end{align*}
The pullbacks $\rho^*$ of $\mathcal{L}_{D_{\tau}} \widehat{\omega}' - d_0\hspace{0.15ex} (D_{\tau}\, \lrcorner\,  \widehat{\omega}')$ and $\mathcal{L}_{\partial_{\tau}} \widehat{\omega}' - d_0\hspace{0.15ex} (\partial_{\tau}\, \lrcorner\,  \widehat{\omega}') = d_0 ((-1)^{p}\widehat{\vartheta}_\mathcal{Y}' - \mathcal{Y} \lrcorner\,  \widehat{\omega}')$ coincide. Hence, on $J^{\infty}(n-1, m; m)$, we have
\begin{align*}
\langle \rho^*(\mathcal{Y}\, \lrcorner\, \psi_l), \rho^*(\widetilde{F}) \rangle = d_0\hspace{0.15ex} \rho^*((-1)^{p} \widehat{\vartheta}_\mathcal{Y}' - \mathcal{Y} \lrcorner\,  \widehat{\omega}')\,.
\end{align*}
But the restriction of $\widehat{\vartheta}_\mathcal{Y}' - (-1)^{p}\, \mathcal{Y} \lrcorner\,  \widehat{\omega}'$ to the system $\mathfrak{T}[1]\mathcal{E}_{\mathcal{X}}$ corresponds to a representative of the reduction of $\Omega$. This observation completes the proof.

\vspace{1ex}

\examplea{Let us consider the point symmetry~\eqref{PointSymmExamp}. We introduce the following adapted coordinates in the region $t\neq 0$. \label{ExampleReductionViaChar}
\begin{align*}
\tau = \dfrac{1}{4t}\,,\qquad \mu = \dfrac{x-y}{t}\,,\qquad \nu = y\,,\qquad w = u + \dfrac{(x-y)^2}{4t} + \dfrac{y^2}{4t}\,,\qquad \ldots
\end{align*}
Here $\tau$, $\hat{x}^2 = \mu$, and $\hat{x}^3 = \nu$ are the new independent variables ($\tau\neq 0$); $\hat{u} = w$ is the new dependent variable. The symmetry~\eqref{PointSymmExamp} takes the form $\partial_{\tau}$.
The corresponding change of the odd variables can be written in the form
\begin{align*}
q = s\,,\qquad q_t = -4\tau(\tau s_{\tau} + \mu s_{\mu})\,,\qquad q_x = 4\tau s_{\mu}\,,\qquad q_y = -4\tau s_{\mu} + s_{\nu}\,,\qquad \ldots\,,
\end{align*}
where $s$ denotes the corresponding new odd variable.
The scalar function $F = u_t - u_x^2 - u_{xxx} - u_{xxy}$ writes $F = -16\tau^2(w_{\mu\mu\nu} - \nu^2/4 + w_{\mu}^2 +  w_{\tau}/4)$. Instead, we can take
\begin{align*}
\hat{F} = w_{\mu\mu\nu} - \dfrac{\nu^2}{4} + w_{\mu}^2 +  \dfrac{w_{\tau}}{4}\,.
\end{align*}
Then for the form $\widehat{\omega}'$ from Example~\ref{Example3}, one has
\begin{align*}
d_0\hspace{0.15ex} \widehat{\omega}' = q_x(q_t - 2u_xq_x - q_{xxx} - q_{xxy})dt\wedge dx\wedge dy = 4s_{\mu}\Big(s_{\mu\mu\nu} + 2w_{\mu}s_{\mu} +  \dfrac{s_{\tau}}{4}\Big)d\tau\wedge d\mu \wedge d\nu\,.
\end{align*}
Here $(0, 4s_{\mu})$ is the corresponding characteristic for the system $(\hat{F}, l_{\hat{F}}(s)) = (0, 0)$.
Then the infinite prolongation $\mathfrak{T}[1]\mathcal{S}$ of the system $H = 0$, $l_H(s) = 0$,
\begin{align*}
H = w_{\mu\mu\nu} - \dfrac{\nu^2}{4} + w_{\mu}^2\,,
\end{align*}
admits the cosymmetry $(0, 4s_{\mu}|_{\mathfrak{T}[1]\mathcal{S}})$, and the reduction of the presymplectic structure from Example~\ref{ExampleRunning2} to $\mathcal{E}_X$ is induced by the presymplectic operator $(-1)^p 4D_{\mu}|_{\mathcal{S}} = 4D_{\mu}|_{\mathcal{S}}$ of $\mathcal{S}$.
}

\vspace{1ex}

\remarka{The relation between symmetries of $\mathcal{E}$ commuting with $\partial_\tau$ and its variational $1$-forms, established by the operator $D_x|_{\mathcal{E}}$, is inherited by $-4D_{\mu}|_{\mathcal{S}}$ in terms of projections to $\mathcal{S}$ of the symmetry restrictions and the corresponding reductions induced by the isomorphism $E_1^{p, \hspace{0.1ex} n-2}(\mathcal{S})\to E_1^{p, \hspace{0.1ex} n-2}(\mathcal{E}_X)$ (see~\cite[Theorem 2]{InvRedII}).
}

\newcommand{\lc}{\lrcorner\,}
\newcommand{\TE}{\mathfrak{T}[1]\mathcal{E}}
\newcommand{\E}{\mathcal{E}}

\section{Worked examples and Maple worksheets}\label{SectionExamples}
	
This section collects complete, end-to-end reductions carried out with the
machinery of Sections~\ref{SectionVarForms}--\ref{SectionFlows}, and documents the accompanying electronic
supplementary material: six self-contained \texttt{Maple} worksheets, one per
subsection. The inline Examples~\ref{ExampleRunning}--\ref{ExampleEuler} of the preceding sections introduce the
objects as they appear in the theory; here the same computations are organized
by \emph{application}: for each model we state the input data, the algorithm
used, the output, and its interpretation. Table~\ref{TableExamples} gives
an overview.
	
Each worksheet requires seven kinds of input data. For Algorithm~\ref{algorithma} from Section~\ref{SectionEvolAlg}: the number $m$ of the dependent variables, the left-hand sides $F^i$ of the system $F = 0$ in evolution form, the components $\varphi^i$ of the symmetry characteristic, written in internal coordinates,\footnote{For an evolution system, internal coordinates on $\E$ are $x^j$ and the jet variables	$u^i_{\alpha}$ with $\alpha$ free of $t$-derivatives; ``written in internal	coordinates'' means that $u_t, u_{tx}, \ldots$ have been eliminated using $F = 0$.} the degree $p$, the cosymmetry components $\psi_i'^p$, $\psi_i'^{p-1}$ regarded as differential forms (for $p = 0$, a conservation-law multiplier), a jet order $K$, and the list of independent variables. For the descent algorithm (Section~\ref{SectionDescent}): the left-hand sides $F^i$ of the system $F = 0$, the number $m_1$ of the equations, the components $\varphi^i$ of the symmetry characteristic on $J^{\infty}(n, m)$, the degree $p$, the components $\psi_i'^p$, $\psi_i'^{p-1}$ of the conservation law characteristic, regarded as differential forms on $J^{\infty}(n, m)$, a jet order $K$, and the list of independent variables.

The
worksheets are self-verifying: every step ends in a numbered checkpoint whose
expected output is zero (38 checkpoints across the suite), so that a failed
step is localized immediately. The files are:
\texttt{ws1\_kdv\_conservation\_law\_p0.mw},
\texttt{ws2\_eq33\_examples1to4.mw},\\
\texttt{ws3\_example5\_evolution\_algorithm\_p2.mw},
\texttt{ws4\_example6\_pavlov\_p0.mw},\\
\texttt{ws5\_example7\_laplace\_descent.mw},
\texttt{ws6\_example8\_euler\_rotation.mw}.\\
They are available at \cite{DruzhkovShevyakov2026code}.

\begin{table}[htbp]
	\centering
	\renewcommand{\arraystretch}{1.3}
	\small
	\begin{tabular}{llllll}
		\hline
		\S & System & Structure (degree) & Symmetry & Method & Worksheet \\
		\hline
		\ref{ExKdV}    & KdV & conservation laws ($p=0$) & traveling wave & Alg.~\ref{algorithma}; \S\ref{SectionFlows} & \texttt{ws1} \\
		\ref{ExChain}  & Eq.~\eqref{2dpKdV} & cosymmetry, var.\ 1-form, & scaling & \S\ref{SectionVarForms} & \texttt{ws2} \\
		&           & presympl.\ operator ($p=1,2$) & & & \\
		\ref{ExP2}     & Eq.~\eqref{2dpKdV} & presympl.\ structure ($p=2$) & point symm.~\eqref{PointSymmExamp} & Alg.~\ref{algorithma} & \texttt{ws3} \\
		\ref{ExPavlov} & Pavlov (cotangent) & conservation law ($p=0$) & higher symm.~\eqref{SymmetryHigher} & Alg.~\ref{algorithma} & \texttt{ws4} \\
		\ref{ExLap}    & Laplace & presympl.\ structure ($p=2$) & rotation & descent (\S\ref{SectionDescent}) & \texttt{ws5} \\
		\ref{ExEuler}  & Euler (2D) & energy cons.\ law ($p=0$) & rotation & contraction (\S\ref{SectionFlows}) & \texttt{ws6} \\
		\hline
	\end{tabular}
	\caption{Overview of the worked examples and supplementary worksheets.}
	\label{TableExamples}
\end{table}
	
\subsection{KdV: traveling waves and first integrals}\label{ExKdV}

The simplest instance of invariant reduction is the one outlined in the
Introduction, and we present it in full. Consider the Korteweg--de Vries
equation and the traveling-wave symmetry,
\begin{align*}
	F \equiv u_t - uu_x - u_{xxx} = 0, \qquad
	\varphi = (u_t + cu_x)\big|_{\E} = uu_x + u_{xxx} + cu_x ,
\end{align*}
so that $X$-invariant solutions $u = U(x - ct)$ satisfy the traveling-wave ODE
$U''' + UU' + cU' = 0$ obtained by solving $u_t + cu_x = 0$ and substituting the solution to the KdV equation. The energy
conservation law is generated by the multiplier $\psi = u$,
\begin{align}
	u\hspace{0.15ex} F \;=\; D_t\!\left(\frac{u^2}{2}\right)
	- D_x\!\left(\frac{u^3}{3} + uu_{xx} - \frac{u_x^2}{2}\right),
	\label{KdVCL}
\end{align}
represented by the flux form
$\widehat{\omega}
= \bigl(u^3/3 + uu_{xx} - u_x^2/2\bigr)\,dt + \bigl(u^2/2\bigr)\,dx$
with $d_0\hspace{0.15ex} \widehat{\omega} = u F\, dt\wedge dx$.

\emph{Invariance (Step 1 of Algorithm~\ref{algorithma}).} Here
$l_{\varphi} = u_x + (u + c)D_x + D_x^3$, so
\begin{align*}
	E_{\varphi}(\psi) + l^{\,*}_{\varphi}(\psi)
	= \varphi + \Bigl( u_x u - D_x\bigl((u+c)u\bigr) - D_x^3(u) \Bigr)
	= \varphi - uu_x - cu_x - u_{xxx} = 0
\end{align*}
identically --- the invariance condition holds.
	
\emph{Reduction.} Since $dt\wedge dx$ is invariant under the flow of $Y$ and the coefficients of $\widehat{\omega}$ do not depend
explicitly on $t$ and $x$, the point symmetry $Y = -(\partial_t + c\,\partial_x)$
satisfies $\mathcal{L}_Y \widehat{\omega} = 0$, and the Anderson--Fels approach (Section~\ref{SectionFlows}) gives the
reduction with no further computation: it is represented by
\begin{align}
	-\,Y \lc \widehat{\omega}
	= \frac{u^3}{3} + uu_{xx} - \frac{u_x^2}{2}
	+ \frac{c\,u^2}{2} ,
	\label{KdVI2}
\end{align}
i.e., by the classical second first integral
$I_2 = \tfrac13 U^3 + UU'' - \tfrac12 (U')^2 + \tfrac{c}{2}U^2$
of the traveling-wave ODE, up to an additive
constant.\footnote{For $p = 0$, $n = 2$, the reduction is an element of
$E^{0,0}_1(\E_X)/H^0_{dR}(\E_X)$: a function on $\E_X$ constant on
solutions and defined modulo additive constants. Indeed,
$D_{\zeta} I_2 = U\,(U''' + UU' + cU')$, $\zeta = x - ct$, so $I_2$ is constant on solutions of the ODE.}
Worksheet \texttt{ws1} computes a representative of the same class by the
three steps of Algorithm~\ref{algorithma} (invariance check, integration by parts, horizontal
homotopy) and verifies at its final checkpoints that 
$D_x \widehat{\vartheta}\,\big|_{\E_X} = 0$ and $D_t \widehat{\vartheta}\,\big|_{\E_X} = 0$. Running it with the mass
multiplier $\psi = 1$ instead yields the first integral
$I_1 = \tfrac12 U^2 + U'' +  cU$; together, $I_1$ and $I_2$ reduce the
third-order traveling-wave ODE to a quadrature. This is the $p = 0$ face of
the machinery for point symmetries and symmetry-invariant representatives: nothing beyond classical objects appears, and the algorithm
reproduces textbook results.
	
\subsection{Equation~\eqref{2dpKdV}: from a cosymmetry to a presymplectic structure}
\label{ExChain}
	
The running example of Section~\ref{SectionVarForms}, the equation
$u_t = u_x^2 + u_{xxx} + u_{xxy}$ of~\eqref{2dpKdV}, illustrates the full chain of
correspondences: cosymmetry $\to$ variational $1$-form $\to$ conservation law
of $\TE$ $\to$ presymplectic operator and structure. Worksheet \texttt{ws2}
verifies each link. First, the restricted $\psi$ in~\eqref{presstrpoten},
$\psi = 3tu_{tx} + xu_{xx} + yu_{xy} + 2u_x$, satisfies the
adjoint-symmetry equation $l^{\,*}_F\big|_{\mathcal{E}}(\psi|_{\E}) = 0$ for
$l_F = D_t - 2u_xD_x - D_x^3 - D_x^2D_y$, and the associated horizontal form
$\omega'_{\psi}$~\eqref{omega_psi} satisfies the off-shell divergence identity
obtained from~\eqref{Cosymvarforms} with $\chi = q$. Second, the operator $\nabla = D_x|_{\E}$
satisfies the presymplectic-operator identity~\eqref{var2formop},
$l^{\,*}_{\E}\circ\nabla - \nabla^{*}\circ l_{\E} = 0$, verified with a
symbolic characteristic. Third, the Lagrangian~\eqref{LagrangianExample} satisfies the
identity of Remark~\ref{RemPresToLagr} in the explicit form
\begin{align*}
	\dfrac{\delta \lambda}{\delta u} = -D_x(F), \qquad
	\lambda = \frac{u_t u_x + u_{xx}(u_{xx} + u_{xy})}{2} - \frac{u_x^3}{3},
\end{align*}
where ${\delta }/{\delta u}$ denotes the variational derivative, i.e.,
$\nabla^{*}_{e}(F) = {\delta \lambda}/{\delta u}$ for $\nabla_{e} = D_x$.
	
Finally, the statement of Example~\ref{ExampleWithLagr} that the presymplectic structure
``originates from the scaling symmetry'' follows from an exact identity at the
level of characteristics in this case. The scaling symmetry
$-3t\,\partial_t - x\,\partial_x - y\,\partial_y + u\,\partial_u + \ldots$
has the characteristic
$\varphi_{s} = u + 3tu_t + xu_x + yu_y$, and a one-line computation gives, identically on the ambient jet space,
\begin{align}
	D_x(\varphi_{s})
	= 2u_x + 3tu_{tx} + xu_{xx} + yu_{xy}
	= \psi.
	\label{ScalingOrigin}
\end{align}
Thus the presymplectic operator $\nabla = D_x|_{\E}$ maps the scaling symmetry
precisely to the cosymmetry~\eqref{presstrpoten}, realizing on this example the homomorphism
from symmetries to variational $1$-forms of Section~\ref{SectionVarForms}, and the presymplectic
structure of Example~\ref{ExampleRunning2} is $d_1$ of the corresponding variational $1$-form.
	
\subsection{Equation~\eqref{2dpKdV}: reduction of the presymplectic structure}
\label{ExP2}
	
The first computation requiring the essential use of differential forms is the reduction of the
presymplectic structure of Example~\ref{ExampleRunning2} under the point
symmetry~\eqref{PointSymmExamp}. This is Example~\ref{Example3}, computed by Algorithm~\ref{algorithma} at $p = 2$: the
input consists of the lifted characteristic $(\varphi_0, \varphi_1)$
displayed there and the cosymmetry $(\psi'^{\,p}, \psi'^{\,p-1}) = (0, q_x)$
of the conservation law of $\TE$ representing the structure; the output is the
potential $\widehat{\vartheta}$ of Example~\ref{Example3}, whose restriction to $\E_X$
represents the reduced presymplectic structure. Worksheet \texttt{ws3} is the
implementation of Appendix~\ref{App:B} restructured into input blocks and eight
checkpoints; already for this scalar equation and one symmetry the
intermediate expressions (the operator $A\widetilde{F}$ and the homotopy
potential) are impractical to produce by hand, which is precisely the case the
implementation is designed for. The same reduction is obtained in closed form
by the quotient-coordinate method of Section~\ref{SectionRedandQuot} in Example~\ref{ExampleReductionViaChar}, providing an
independent check: the reduced structure is induced by the presymplectic
operator $4D_{\mu}|_{\mathcal{S}}$ of the quotient system.
	
\subsection{Pavlov's equation: a higher symmetry}\label{ExPavlov}

The final example demonstrating Algorithm~\ref{algorithma} is the one where the Anderson--Fels approach (and the double reduction) does
not reach regardless of representative choices: a \emph{higher} (third-order) symmetry of the cotangent covering
of Pavlov's equation, Example~\ref{ExampleHigher}. In evolution variables
$(u^1, u^2, u^3, u^4) = (v, v_t, w, w_t)$ the system consists of the four
equations $F^1 = 0, \ldots, F^4 = 0$ of \eqref{EquationsPavlovCotan}; the conservation law has the
four-component multiplier
$\psi^{\,p} = (-u^4_x,\; u^3_x,\; -u^2_x,\; u^1_x)$ ($p = 0$), and the
symmetry characteristic is $\varphi^1 = \varphi^2 = 0$ with $\varphi^3$ as
displayed there and $\varphi^4 = D_t(\varphi^3)\big|_{\E}$, which worksheet
\texttt{ws4} computes automatically by on-shell elimination of
$t$-derivatives. The output is the horizontal $1$-form $\vartheta$ of
Example~\ref{ExampleHigher}, whose restrictions to $\E_X$ and to the reduced Pavlov system
$\mathcal{S}$ represent the reduced conservation law;\footnote{Potentials
produced by the homotopy are defined modulo $d_0$-exact forms, so a worksheet
run may return a representative differing from the printed $\vartheta$ by an
exact term; the represented class is the same.} by Remark~\ref{RemarkDiffCovering}, it generates a
one-dimensional differential covering over $\mathcal{S}$ through
$\vartheta|_{\mathcal{S}} = d_0 h$.

\subsection{Laplace equation: the descent algorithm}\label{ExLap}
	
Example~\ref{ExampleLaplace} runs the descent algorithm of Section~\ref{SectionDescent} on the Laplace equation
$u_{xx} + u_{yy} = 0$ with the rotation symmetry of characteristic
$\varphi = yu_x - xu_y$, reducing the presymplectic structure represented by
$\widehat{\omega} = \theta\wedge\theta_x\wedge dy -
\theta\wedge\theta_y\wedge dx$. The computation requires no evolution form:
one integration by parts produces the operator $\nu$, and two sweeps of the
descent --- each a generalized integration by parts lowering the operator
order by one --- terminate with
\begin{align*}
	\gamma(\chi) = (y\chi_y + x\chi_x)\,\theta - \chi\,(x\theta_x + y\theta_y),
	\qquad
	\widehat{\omega}_{\gamma} = (y\theta_y + x\theta_x)\wedge\theta ,
\end{align*}
and $\widehat{\omega}_{\gamma}|_{\E_X}$ represents the reduced structure on the space of
rotationally invariant harmonic functions. Worksheet \texttt{ws5} verifies the
three displayed stages of Example~\ref{ExampleLaplace}: the defining identity of $\nu$, the
explicit form of the operator $\nu\circ l_F - \gamma_{\widehat{\omega}}\circ
l_{\varphi}$, and the final descent identity.
	
\subsection{Euler equations: rotation and the contraction shortcut}
\label{ExEuler}
	
For the $(1+2)$-dimensional incompressible Euler equations of Example~\ref{ExampleEuler}, the
energy conservation law admits the multiplier form
\begin{align}
	d_0\hspace{0.15ex} \widehat{\omega}
	= \Bigl( u\hspace{0.15ex} F^1 + v\hspace{0.15ex} F^2 + (K + P)\hspace{0.15ex} F^3 \Bigr)\, dt\wedge dx\wedge dy,
	\qquad K = \frac{u^2 + v^2}{2},
	\label{EulerMult}
\end{align}
with $\widehat{\omega}$ given by~\eqref{EnergyCLRepres} and $F^1, F^2, F^3$ defining the two momentum
equations and the incompressibility constraint: the energy multiplier is
$(u,\, v,\, K + P)$. Since $\mathcal{L}_Y\widehat{\omega} = 0$ for the
rotation generator $Y = x\partial_y - y\partial_x + u\partial_v -
v\partial_u$, the reduction requires no integration by parts at all
(Section~\ref{SectionFlows}): it is the contraction
$-Y\lc\widehat{\omega} = K(x\,dx + y\,dy) - (xu + yv)(K + P)\,dt$, which in
polar-type coordinates becomes $Kr\,dr - w(K + P)\,dt$ --- a conservation law
of the quotient of $\E_X$ by the rotation flow, expressing the balance of
energy for rotationally invariant flows. Worksheet \texttt{ws6} verifies the
multiplier identity \eqref{EulerMult}, the invariance
$\mathcal{L}_Y\widehat{\omega} = 0$, and the contraction formula; this is the
shortest worksheet of the suite (three checkpoints), reflecting how little
computation the point-symmetry shortcut leaves to do in the case of symmetry-invariant representatives understood in terms of ambient spaces.

\section{Discussion}

\textbf{Summary.} We have presented computational algorithms for the invariant reduction of conservation laws, variational $1$-forms, and presymplectic structures of systems of PDEs. The unifying device for evolution systems (Algorithm~\ref{algorithma}) is the tangent system $\mathfrak{T}[1]\mathcal{E}$: appending the linearized equations with anticommuting perturbation variables turns each of these structures into a conservation law of internal degree $p = 0, 1, 2$, respectively, so that a single reduction procedure covers all cases. The procedure consists of an invariance check, one integration by parts, and one application of the total homotopy formula; it is implemented in \texttt{Maple} (Appendix~\ref{App:B}). For general $\ell$-normal systems and $p\geqslant 1$, Theorem~\ref{Theoremgenerfun} yields a descent algorithm that requires no evolution representation. For point symmetries, under moderate assumptions, the reduction can be described in terms of characteristics of conservation laws of the tangent system (Section \ref{SectionRedandQuot}). Moreover, the reduction often reduces to a contraction of the flux form with the symmetry generator, and can be rewritten directly on the quotient of the reduced (invariant) system. 

\noindent
\textbf{Limitations.} Algorithm~\ref{algorithma} requires an evolution representation of the system. Although the vast majority of non-gauge systems arising in applications can be written in an evolution form, such representations can be cumbersome and lead to lengthy outcomes. For $p = 0$, applying the total homotopy formula in Maple may require substantial computational resources, and the corresponding computations may be time-consuming. The total homotopy potentials grow rapidly in size for higher symmetries and larger internal degrees $p$.
The descent algorithm of Section~\ref{SectionDescent} applies only to the case $p\geqslant 1$. It relies on $\ell$-normality together with the additional assumption of Theorem~\ref{Theoremgenerfun} (which is, however, not restrictive).

\noindent
\textbf{Future directions.} Natural extensions include the reduction with respect to multi-dimensional symmetry algebras (multi-reduction), the reduction of nonlocal structures, the algorithmic reduction of Hamiltonian (Poisson) structures via cotangent coverings along the lines of~\cite{InvRedIII}, integration of the present implementation with existing symbolic packages for symmetry and conservation-law computation~\cite{cheviakov2007gem, cheviakov2010computation,cheviakov2010symbolic,cheviakov2017symbolic, KVV2017}, and a systematic library of reductions for physically significant models.

\subsubsection*{Acknowledgements.}
The authors acknowledge support from NSERC of Canada through the Discovery grant RGPIN-2024-04308.

\subsubsection*{AI disclosure.}
ChatGPT Sol 5.6 was used to proofread the final draft and examine the consistency of formulas. AI has not been used to develop any mathematical results, algorithms, or examples.

\subsubsection*{Conflict of interest.}
Authors declare no conflict of interest.

\bibliographystyle{ieeetr}
{\small
	\bibliography{references24e}
}

\appendix

\section{Elements of the graded-commutative geometry}\label{App:A}

The algebra $\mathcal{F}(n, m; m)$ is non-negatively $\mathbb{Z}$-graded. It identifies with the exterior algebra of the $\mathcal{F}(n, m)$-module $\mathcal{C}\Lambda^1(n, m)$. This identification gives rise to the inclusion $\mathcal{F}(n, m)\subset \mathcal{F}(n, m; m)$, which, in
geometric terms, corresponds to the pullback along the natural projection $J^{\infty}(n, m; m)\to J^{\infty}(n, m)$.

All differential forms on $J^{\infty}(n, m; m)$ are polynomial in the variables $q^i_{\alpha}$. The algebra $\Lambda^{*}(n, m; m)$ is bigraded, with the bigrading assigned as follows:
\begin{align*}
	x^i(0, 0)\,,\qquad u^i_{\alpha}(0, 0)\,,\qquad q^i_{\alpha}(1, 0)\,,\qquad dx^i(0, 1)\,,\qquad du^i_{\alpha}(0, 1)\,,\qquad dq^i_{\alpha}(1, 1)\,.
\end{align*}
The first component is the internal degree (inherited from $\mathcal{F}(n, m; m)$), and the second is the differential form degree. Similarly, the projection $J^{\infty}(n, m; m)\to J^{\infty}(n, m)$ induces the inclusion $\Lambda^*(n, m)\subset \Lambda^*(n, m; m)$.

We denote the internal degree by $|\cdot|$; for instance, $|u^i_{\alpha}| = 0$, $|q^i_{\alpha}| = 1$, $|du^{i}_{\alpha}| = 0$, $|dq^i_{\alpha}| = 1$. The signs in algebraic expressions are governed by the inner product of the bigradings. In particular,
\begin{align*}
	q^i\hspace{0.15ex} du^j = du^j\hspace{0.15ex} q^i\,,\quad q^i\hspace{0.15ex} dq^j = - dq^j\hspace{0.2ex} q^i\,,\quad du^i\wedge dq^j = - dq^j\wedge du^i\,,\quad dq^{i}\wedge dq^j = dq^j\wedge dq^i\,.
\end{align*}

The contraction of vector fields with differential forms is defined by bringing partial derivatives and differentials together, with vector fields placed on the left. For example, if $\omega$ is a differential $1$-form of the form $du^i_{\alpha}\, \omega^{\alpha}_i + dq^j_{\beta}\, \omega_j^{\beta}$ and $X = X^i_{\alpha} \partial_{u^i_{\alpha}} + X^j_{\beta} \partial_{q^j_{\beta}}$, then
\begin{align*}
	X\lrcorner\, \omega = X^i_{\alpha}\,\omega^{\alpha}_i + X^j_{\beta}\,\omega_j^{\beta}\,.
\end{align*}
A form $\omega = \theta^{i_1}_{\alpha^1} \wedge \ldots \wedge \theta^{i_p}_{\alpha^p}\wedge \xi$, $\xi\in E^{0, \hspace{0.1ex} k}_0(n, m)$, $p > 0$, corresponds to
\begin{align*}
	\omega' = q^{i_1}_{\alpha^1} \ldots q^{i_p}_{\alpha^p}\, \xi' = {\dfrac{(-1)^{p(p-1)/2}}{p!}}\underbrace{E_q\, \lrcorner\, \ldots E_q\, \lrcorner\,}_{p} (\theta^{i_1}_{\alpha^1} \wedge \ldots \wedge \theta^{i_p}_{\alpha^p}\wedge \xi)\,,
\end{align*}
while $\xi' = \xi$. The operator $E_q \lrcorner\,$, having the bigrading $(1, -1)$, anticommutes with the operator of multiplication by $q^i_{\alpha}$.

\section{An implementation of the algorithm from Section~\ref{SectionEvolAlg}}\label{App:B}

Let us demonstrate an implementation\footnote{It is also available at\cite{DruzhkovShevyakov2026code}.} of the algorithm using Example~\ref{Example3}, where the number of dependent variables is $m=1$, the internal degree of the corresponding conservation law of $\mathfrak{T}[1]\mathcal{E}$ is $p=2$. We begin with
\begin{verbatim}
	restart;
	with(DifferentialGeometry): with(JetCalculus):
\end{verbatim}
and the following input data, which should be specified for each particular problem.
\begin{verbatim}
	m:=1;
	p:=2;
	K:=8;
	independent_variables:=[t,x,y];
\end{verbatim}
Here \verb|K| is a sufficiently large integer (exceeding orders of all derivatives that can appear); $\verb|t|$, $\verb|x|$, $\verb|y|$ are names of the independent variables. Note that the list of independent variables is ordered.
We assign the number of independent variables to \verb|n| and introduce the dependent variables: $m$ variables of the form \verb|u1|, \verb|u2|, $\ldots$ and $m$ additional variables \verb|q1|, \verb|q2|, $\ldots$
\begin{verbatim}
	n:=nops(independent_variables):
	dependent_variables:=seq(u||i, i=1..m):
	odd_variables:=seq(q||i, i=1..m):
\end{verbatim}
Note that \verb|q1|, \verb|q2|, $\ldots$ are even variables, although they play the role of the odd ones $q^1$, $q^2$, $\ldots$ To maintain this, we replace them by the corresponding Cartan forms where needed.

It is convenient to introduce auxiliary dependent variables\footnote{The role of the auxiliary variables is to simplify the integration by parts in~\eqref{IntByParts}.} and create a new variable for $m+1$.
\begin{verbatim}
	lhs_variables:=seq(F0||i, i=1..m), seq(F1||i, i=1..m):
	psi_variables:=seq(pmo_psi||i, i=1..m), seq(p_psi||i, i=1..m):
	mpo:=m+1:
\end{verbatim}
The corresponding jet space and the volume form $dx^1\wedge \ldots\wedge dx^n$ arise.
\begin{verbatim}
	DGsetup(independent_variables, [dependent_variables, odd_variables, 
	lhs_variables, psi_variables], J, K):
	Volume_form:=1:
	for var in independent_variables do
	   Volume_form:=evalDG(Volume_form &wedge D||var):
	end do:
\end{verbatim}

Now we need the left-hand sides $F^i = u^i_t - f^i$ of the equations $F^i = 0$. We assign them to the variables \verb|F0_form1|, \verb|F0_form2|, $\ldots$ This input should also be specified for each particular problem.
\begin{verbatim}
	F0_form1:= u1[1] - u1[2]^2 - u1[2,2,2] - u1[2,2,3];
\end{verbatim}
For example, \verb|u1[2,2,3]| corresponds to $u^1_{x^2x^2x^3} = u_{xxy}$. Check that the number of equations is $m$.
\begin{verbatim}
	`if`(assigned(F0_form||m)=true and assigned(F0_form||mpo)=false, 0, 
	print("WARNING: m does not match the number of equations"));
\end{verbatim}
If the number of equations is $m$ indeed, the output is $0$, whereas the warning output indicates that the number of equations is incorrect.

We assign the Cartan forms $d_v F^1$, $d_v F^2$, $\ldots$ to the variables \verb|F1_form1|, \verb|F1_form2|, $\ldots$, respectively. They correspond to the left-hand sides $E_q(F^1)$, $E_q(F^2)$, $\ldots$ of the equations $l_F(q) = 0$.
\begin{verbatim}
	for i from 1 to m do
	   F1_form||i:=VerticalExteriorDerivative(F0_form||i):
	end do:
\end{verbatim}
Now we need the cosymmetry of the conservation law of $\mathfrak{T}[1]\mathcal{E}$, written in terms of its components $\psi^p_i$, $\psi^{p-1}_i$ (regarded as Cartan forms). We assign $\psi^{p}_1$, $\psi^{p}_2$, $\ldots$ to the variables \verb|p_psi_form1|, \verb|p_psi_form2|, $\ldots$ and $\psi^{p-1}_1$, $\psi^{p-1}_2$, $\ldots$ to \verb|pmo_psi_form1|, \verb|pmo_psi_form2|, $\ldots$, respectively. This input should also be specified for each particular problem.
\begin{verbatim}
	p_psi_form1:= 0;
	pmo_psi_form1:= Cu1[2];
\end{verbatim}
Here \verb|Cu1[2]| denotes $\theta^1_{x^2} = \theta_x$.
Check that the number of the cosymmetry components is correct.
\begin{verbatim}
	`if`(assigned(pmo_psi_form||m)=true and assigned(p_psi_form||m)=true and 
	assigned(pmo_psi_form||mpo)=false and assigned(p_psi_form||mpo)=false, 0, 
	"WARNING: the number of the cosymmetry components is incorrect");
\end{verbatim}
Now we find $\widehat{\omega}$ satisfying $\langle \psi_l, \widetilde{F}\rangle = d_0\hspace{0.2ex} \widehat{\omega}'$. We use Cartan forms instead of the odd variables.
\begin{verbatim}
	psi_of_F:=
	evalDG(expand(
	   add(`if`(pmo_psi_form||i = 0, 0, 
	   pmo_psi_form||i &wedge F1_form||i &wedge Volume_form), i=1..m) + 
	   add(`if`(p_psi_form||i = 0, 0, 
	   p_psi_form||i &wedge F0_form||i &wedge Volume_form), i=1..m)
	)):
	omega_hat:=(-1)^p*HorizontalHomotopy(psi_of_F):
\end{verbatim}
Check that the result is correct. The following output is supposed to be $0$.
\begin{verbatim}
	evalDG(HorizontalExteriorDerivative((-1)^p*omega_hat) - psi_of_F);
\end{verbatim}

Assign the symmetry components $\varphi^1$, $\varphi^2$, $\ldots$ to the variables \verb|phi01|, \verb|phi02|, $\ldots$, respectively. This is the last type of input data that should be specified for each particular problem. Recall that $\varphi^i$ are independent of $u^1_t$, $u^2_t$, $\ldots$ and their total derivatives. Check that the number of components is correct.
\begin{verbatim}
	phi01:= x^2 - 2*x*y + 4*t^2*(u1[2]^2+u1[2,2,2]+u1[2,2,3]) + 4*t*(x-y)*u1[2];
	`if`(assigned(phi0||m)=true and assigned(phi0||mpo)=false, 0, 
	"WARNING: m does not match the number of symmetry components");
\end{verbatim}
Evaluate the Lie derivative $\mathcal{L}_{E_{\varphi}} \widehat{\omega}$.
\begin{verbatim}
	E_phi:=
	Prolong(
	   add(evalDG(phi0||i*D_u||i[]), i=1..m),
	K):
	Lie_omega_hat:=
	evalDG(
	   Hook(E_phi, VerticalExteriorDerivative(omega_hat)) + 
	   `if`(p=0, 0, VerticalExteriorDerivative(Hook(E_phi, omega_hat)))
	):
\end{verbatim}

Find the component of $\langle l_{\widetilde{\varphi}}(\widetilde{F}), \psi_r \rangle$.
\begin{verbatim}
	E_q:=
	Prolong(
	   add(evalDG(q||i[]*D_u||i[]), i=1..m), 
	K):
	for i from 1 to m do
	   phi1||i:=LieDerivative(E_q, phi0||i):
	end do:
	E_F:=
	Prolong(
	   add(evalDG(F0||i[]*D_u||i[] + F1||i[]*D_q||i[]), i=1..m), 
	K):
	lphi_F_psi_r:=add(LieDerivative(E_F, phi1||i)*(-1)^(p-1)*pmo_psi||i[], i=1..m) 
	+ add(LieDerivative(E_F, phi0||i)*p_psi||i[], i=1..m):
\end{verbatim}

To perform the integration by parts~\eqref{IntByParts}, we replace the auxiliary variables \verb|F01|, \verb|F02|, $\ldots$, \verb|F11|, \verb|F12|, $\ldots$ by their vertical differentials and apply the total homotopy formula.
\begin{verbatim}
	ForIntegrationByParts_coeff:=simplify(
	   lphi_F_psi_r 
	   - add(F0||i[]*EulerLagrange(lphi_F_psi_r)[2*m+i], i=1..m)
	   - add(F1||i[]*EulerLagrange(lphi_F_psi_r)[3*m+i], i=1..m)
	):
	Replace_F_by_CF:=proc(expr)
	   local res, j, i, var;
	   res:= 0;
	   for j from 0 to 1 do
	      for i from 1 to m do
	         for var in indets(expr, specindex(F||j||i)) do
	            res:= res + diff(expr, var)*CF||j||i[op(var)];
	         end do;
	      end do;
	   end do;
	   return evalDG(res);
	end proc:
	ForIntegrationByParts_coeff_form:= Replace_F_by_CF(ForIntegrationByParts_coeff):
	AF_CF:= HorizontalHomotopy(ForIntegrationByParts_coeff_form &wedge Volume_form):
\end{verbatim}
The following output is supposed to be $0$.
\begin{verbatim}
	evalDG(ForIntegrationByParts_coeff_form &wedge Volume_form 
	- HorizontalExteriorDerivative(AF_CF));
\end{verbatim}
Now we restore the auxiliary variables \verb|F01|, \verb|F02|, $\ldots$, \verb|F11|, \verb|F12|, $\ldots$ from their vertical differentials
\begin{verbatim}
	Replace_CF_by_F_field:=
	Prolong(
	   add(evalDG(F0||i[]*D_F0||i[] + F1||i[]*D_F1||i[]), i=1..m), 
	K):
	AF_incomplete:= evalDG(-Hook(Replace_CF_by_F_field, AF_CF)):
\end{verbatim}
and check that the result is correct. The following output is supposed to be $0$.
\begin{verbatim}
	evalDG(ForIntegrationByParts_coeff &wedge Volume_form 
	- HorizontalExteriorDerivative(AF_incomplete));
\end{verbatim}
Finally, we replace $q^i$ by the corresponding Cartan forms and all auxiliary variables by their particular expressions.
\begin{verbatim}
	Replace_q_by_Cu_F_by_F_form_psi_by_psi_form:=proc(expr)
	   local res, i_F, i_q, i_psi, var_q, var_F0, var_F1, var_pmo_psi, var_p_psi;
	   res:= 0;
	   for i_F from 1 to m do
	      for i_psi from 1 to m do
	         for var_F0 in indets(expr, specindex(F0||i_F)) do
	            for var_p_psi in indets(expr, specindex(p_psi||i_psi)) do
	               if diff(expr, var_F0, var_p_psi) <> 0 and 
	               p_psi_form||i_psi <> 0 then
	                  res:= res + diff(expr, var_F0, var_p_psi) 
	                  *TotalDiff(F0||_form||i_F, [op(var_F0)]) 
	                  &wedge TotalDiff(p_psi_form||i_psi, [op(var_p_psi)]);
	               end if;
	            end do;
	         end do;
	         for var_F1 in indets(expr, specindex(F1||i_F)) do
	            for var_pmo_psi in indets(expr, specindex(pmo_psi||i_psi)) do
	               if diff(expr, var_F1, var_pmo_psi) <> 0 and 
	               pmo_psi_form||i_psi <> 0 then
	                  res:= res + diff(expr, var_F1, var_pmo_psi) 
	                  *TotalDiff(F1||_form||i_F, [op(var_F1)]) 
	                  &wedge TotalDiff(pmo_psi_form||i_psi, [op(var_pmo_psi)]);
	               end if;
	            end do;
	         end do;
	         for i_q from 1 to m do
	            for var_F0 in indets(expr, specindex(F0||i_F)) do
	               for var_q in indets(expr, specindex(q||i_q)) do
	                  for var_pmo_psi in indets(expr, specindex(pmo_psi||i_psi)) do
	                     if diff(expr, var_F0, var_q, var_pmo_psi) <> 0 and 
	                     pmo_psi_form||i_psi <> 0 then
	                        res:= res + diff(expr, var_F0, var_q, var_pmo_psi) 
	                        *TotalDiff(F0||_form||i_F, [op(var_F0)])
	                        &wedge Cu||i_q[op(var_q)] 
	                        &wedge TotalDiff(pmo_psi_form||i_psi,[op(var_pmo_psi)]);
	                     end if;
	                  end do;
	               end do;
	            end do;
	         end do;
	      end do;
	   end do;
	   return evalDG(res);
	end proc:
\end{verbatim}
Restore the form corresponding to $A\widetilde{F}$.
\begin{verbatim}
	AF:= add(
	   Replace_q_by_Cu_F_by_F_form_psi_by_psi_form(op([1,2,i,2], AF_incomplete)) 
	   &wedge _DG([["biform",J,[n-1, 0]],[[op([1,2,i,1],AF_incomplete),1]]]), 
	   i=1..nops(op([1,2], AF_incomplete))
	):
\end{verbatim}
Check that the result is correct (the first output is supposed to be the zero form --- this is equivalent to the $X$-invariance of the structure represented by $\omega = \widehat{\omega}|_{\mathcal{E}}$) and find $\widehat{\vartheta}$.
\begin{verbatim}
	evalDG(HorizontalExteriorDerivative(Lie_omega_hat - AF));
	vartheta_hat:=HorizontalHomotopy(Lie_omega_hat - AF);
\end{verbatim}
The following form is also supposed to be zero.
\begin{verbatim}
	evalDG(HorizontalExteriorDerivative(vartheta_hat) - (Lie_omega_hat - AF));
\end{verbatim}
The reduction of $\widehat{\omega}|_{\mathcal{E}}$ is represented by $\widehat{\vartheta}|_{\mathcal{E}_X}$.

To summarize, seven types of input data depend on the particular situation: the numbers $m$, $p$, $K$, names of the independent variables, the left-hand sides $F^i$, the components $\psi^p_i$, $\psi^{p-1}_i$ of the cosymmetry, and the components $\varphi^i$ of the symmetry characteristic. Checks are supposed to result in eight zeros.

\section{Theorem~\ref{Theoremgenerfun} and the isomorphism $E_1^{p,\hspace{0.1ex} n-2}(\mathcal{S})\to E_1^{p,\hspace{0.1ex} n-2}(\mathcal{E}_X)$}\label{App:C}

For a symmetry $X = E_{\varphi}|_{\mathcal{E}}$ and an operator $\Phi\colon P(n, m)\to P(n, m)$ satisfying $l_F (\varphi) = \Phi(F)$, the restriction of the identity $E_{\chi}(l_F (\varphi)) = E_{\chi}(\Phi(F))$ to the system $\mathcal{E}_X$ yields the relation
\begin{align}
l_F|_{\mathcal{E}_X} \circ l_{\varphi}|_{\mathcal{E}_X} = \Phi|_{\mathcal{E}_X} \circ l_F|_{\mathcal{E}_X}.
\label{Relation4}
\end{align}

\subsection{Proof of Theorem~\ref{Theoremgenerfun}}

\vspace{0.5ex}

\noindent
\textbf{Proof.} First, all operators
\[
\gamma|_{\mathcal{E}_X}\colon
\varkappa(\mathcal{E}_X)\longrightarrow
E_0^{p-1,\hspace{0.1ex}n-2}(\mathcal{E}_X)
\]
satisfying~\eqref{Theoremrelation} give rise to forms
\(\widehat{\omega}_{\gamma}|_{\mathcal{E}_X}\) representing the same
element of \(E_1^{p,\hspace{0.1ex}n-2}(\mathcal{E}_X)\).
Indeed, if\footnote{In this notation, the lower indices are unrelated to the orders of the operators.} \(\gamma_1\) and \(\gamma_2\) both
satisfy~\eqref{Theoremrelation}, then
\[
d_0\circ(\gamma_1-\gamma_2)|_{\mathcal{E}_X}=0.
\]
The horizontal complex of \(\mathcal{C}\)-differential operators with
values in \(E_0^{p-1,\hspace{0.1ex}\bullet}(\mathcal{E}_X)\) is exact
in horizontal degrees strictly less than \(n\); this is\footnote{\draftnew{This result is based on the same descent on the symbol that yields the algorithm in Section~\ref{SectionDescent} (see~\cite{Vinogradov1984, Vinogradov1984II, KV1999}).}} the column
exactness used below in the bicomplex~\eqref{bicomplex2} (see~\cite{Verbovetsky1998}).
Hence, if \(n>2\), there exists a
\(\mathcal{C}\)-differential operator
\[
\beta\colon
\varkappa(n, m)\longrightarrow
E_0^{p-1,\hspace{0.1ex}n-3}(n, m)
\]
such that
\[
(\gamma_1-\gamma_2)|_{\mathcal{E}_X}=d_0\circ\beta|_{\mathcal{E}_X}.
\]
When \(n=2\), the same exactness implies
\[
(\gamma_1-\gamma_2)|_{\mathcal{E}_X}=0.
\]
Since alternatization commutes with the corresponding version of \(d_0\), it follows that
\[
\widehat{\omega}_{\gamma_1}|_{\mathcal{E}_X}
-
\widehat{\omega}_{\gamma_2}|_{\mathcal{E}_X}
=
-d_0\hspace{0.15ex} \widehat{\omega}_{\beta}|_{\mathcal{E}_X}
\]
for \(n>2\), while the difference vanishes for \(n=2\). Thus the
corresponding classes in
\(E_1^{p,\hspace{0.1ex}n-2}(\mathcal{E}_X)\) coincide.

Let $\widehat{\vartheta}\in E^{p,\hspace{0.1ex} n-2}_0(n, m)$ be a form such that $\widehat{\vartheta}|_{\mathcal{E}} = \vartheta$ for $\vartheta$ from~\eqref{redformula}. Then there exists a $\mathcal{C}$-differential operator $\Delta\colon P(\mathcal{E})\to E_0^{p-1,\hspace{0.1ex} n-1}(\mathcal{E})$ such that
\begin{align*}
(E_{\chi} \lrcorner\,\mathcal{L}_{E_{\varphi}} \widehat{\omega})|_{\mathcal{E}} = -d_0 (E_{\chi} \lrcorner\, \widehat{\vartheta})|_{\mathcal{E}} + \Delta( l_{\mathcal{E}}(\chi|_{\mathcal{E}}))\,,\qquad \chi\in \varkappa(n, m)\,.
\end{align*}
Denote the operator  $\chi\mapsto E_{\chi} \lrcorner\, \widehat{\vartheta}$ by $\gamma_{\hspace{0.15ex} \widehat{\vartheta}}$.
Since $E_{\chi} \lrcorner\,\mathcal{L}_{E_{\varphi}} \widehat{\omega} = \mathcal{L}_{E_{\varphi}}(E_{\chi} \lrcorner\, \widehat{\omega}) - [E_{\varphi}, E_{\chi}]\lrcorner\, \widehat{\omega}$, one has
$(E_{l_{\varphi}(\chi)} \lrcorner\, \widehat{\omega})|_{\mathcal{E}_X} = -d_0 (E_{\chi} \lrcorner\, \widehat{\vartheta})|_{\mathcal{E}_X} + \Delta( l_{\mathcal{E}}(\chi|_{\mathcal{E}}))|_{\mathcal{E}_X}$, and hence,
\begin{align}
\gamma_{\hspace{0.15ex} \widehat{\omega}}|_{\mathcal{E}_X} \circ l_{\varphi}|_{\mathcal{E}_X} = -d_0 \circ \gamma_{\hspace{0.15ex} \widehat{\vartheta}}|_{\mathcal{E}_X} + \Delta|_{\mathcal{E}_X} \circ l_{\mathcal{E}}|_{\mathcal{E}_X}.
\label{Relation1}
\end{align}

Since $\mathcal{E}$ is $\ell$-normal and the element of $E^{p,\hspace{0.1ex} n-1}_1(\mathcal{E})$ represented by $\omega$ is $X$-invariant, the operator $\Phi^*_{(p-1)}\psi \colon P(n, m) \to E_0^{p-1,\hspace{0.1ex} n}(n, m)$ vanishes on $\mathcal{E}_X$.
Applying $E_{\chi}$ to~\eqref{Relation5} and restricting to $\mathcal{E}_X$, one finds
\begin{align}
\psi|_{\mathcal{E}_X}\circ \Phi|_{\mathcal{E}_X}\circ l_{\mathcal{E}}|_{\mathcal{E}_X} = d_0\circ \nu|_{\mathcal{E}_X} \circ l_{\mathcal{E}}|_{\mathcal{E}_X}\,.
\label{Relation3}
\end{align}
Identity~\eqref{Relation1} implies that
$d_0\circ \gamma_{\hspace{0.15ex} \widehat{\omega}}|_{\mathcal{E}_X} \circ l_{\varphi}|_{\mathcal{E}_X} = d_0 \circ \Delta|_{\mathcal{E}_X} \circ l_{\mathcal{E}}|_{\mathcal{E}_X}$.

Note that $\psi|_{\mathcal{E}}$ and $\gamma_{\hspace{0.15ex} \widehat{\omega}}|_{\mathcal{E}}$ are related by the following analogue of~\eqref{Cosymmvarform}.
\begin{align}
\psi|_{\mathcal{E}} \circ l_{\mathcal{E}} = d_0\circ \gamma_{\hspace{0.15ex} \widehat{\omega}}|_{\mathcal{E}}\,.
\label{Relation2}
\end{align}
Combining~\eqref{Relation2} with the preceding identity and~\eqref{Relation4}, we have
\begin{align*}
\psi|_{\mathcal{E}_X} \circ l_{\mathcal{E}}|_{\mathcal{E}_X} \circ l_{\varphi}|_{\mathcal{E}_X} = d_0 \circ \Delta|_{\mathcal{E}_X} \circ l_{\mathcal{E}}|_{\mathcal{E}_X},\qquad 
\psi|_{\mathcal{E}_X} \circ \Phi|_{\mathcal{E}_X} \circ l_\mathcal{E}|_{\mathcal{E}_X} = d_0 \circ \Delta|_{\mathcal{E}_X} \circ l_{\mathcal{E}}|_{\mathcal{E}_X}.
\end{align*}
Using~\eqref{Relation3}, one obtains
\begin{align}
d_0\circ \nu|_{\mathcal{E}_X} \circ l_\mathcal{E}|_{\mathcal{E}_X} = d_0 \circ \Delta|_{\mathcal{E}_X} \circ l_{\mathcal{E}}|_{\mathcal{E}_X}.
\label{Relation6}
\end{align}
Then $\Delta|_{\mathcal{E}_X} \circ l_{\mathcal{E}}|_{\mathcal{E}_X} = \nu|_{\mathcal{E}_X} \circ l_\mathcal{E}|_{\mathcal{E}_X} + d_0 \circ \mu|_{\mathcal{E}_X}$ for some $\mathcal{C}$-differential operator $\mu\colon \varkappa(n, m) \to E_0^{p-1,\hspace{0.1ex} n-2}(n, m)$.
Therefore,~\eqref{Relation1} has the form
\begin{align*}
\gamma_{\hspace{0.15ex} \widehat{\omega}}|_{\mathcal{E}_X} \circ l_{\varphi}|_{\mathcal{E}_X} = -d_0 \circ (\gamma_{\hspace{0.15ex} \widehat{\vartheta}} - \mu)|_{\mathcal{E}_X} + \nu|_{\mathcal{E}_X} \circ l_{\mathcal{E}}|_{\mathcal{E}_X}.
\end{align*}
It follows that for $\gamma = \gamma_{\hspace{0.15ex} \widehat{\vartheta}} - \mu$, the form $\widehat{\omega}_{\gamma}|_{\mathcal{E}_X}$ represents an element of $E_1^{p,\hspace{0.1ex} n-2}(\mathcal{E}_X)$.

If the equation $\square\circ l_\mathcal{E}|_{\mathcal{E}_X} = 0$ has no nonzero solutions, then~\eqref{Relation6} yields $d_0\circ \nu|_{\mathcal{E}_X}  = d_0 \circ \Delta|_{\mathcal{E}_X}$ and thus $\Delta|_{\mathcal{E}_X} = \nu|_{\mathcal{E}_X} + d_0 \circ \mu_1|_{\mathcal{E}_X}$ for some $\mathcal{C}$-differential operator $\mu_1\colon P(n, m)\to E_0^{p-1,\hspace{0.1ex} n-2}(n, m)$. In this case,~\eqref{Relation1} takes the form
\begin{align*}
\gamma_{\hspace{0.15ex} \widehat{\omega}}|_{\mathcal{E}_X} \circ l_{\varphi}|_{\mathcal{E}_X} = -d_0 \circ (\gamma_{\hspace{0.15ex} \widehat{\vartheta}} - \mu_1\circ l_{F})|_{\mathcal{E}_X} + \nu|_{\mathcal{E}_X} \circ l_{\mathcal{E}}|_{\mathcal{E}_X}.
\end{align*}
Here for $\gamma = \gamma_{\hspace{0.15ex} \widehat{\vartheta}} - \mu_1 \circ l_{F}$, one has $\widehat{\omega}_{\gamma}|_{\mathcal{E}} = \vartheta$.

\subsection{The isomorphism $E_1^{p,\hspace{0.1ex} n-2}(\mathcal{S})\to E_1^{p,\hspace{0.1ex} n-2}(\mathcal{E}_X)$}

Relation~\eqref{Relation4} can be written in the form
\begin{align*}
\square_1 \circ l_{\mathcal{E}_X} = 0\,,\qquad
\square_1 = 
\begin{pmatrix}
-\Phi|_{\mathcal{E}_X} & l_\mathcal{E}|_{\mathcal{E}_X}
\end{pmatrix},
\qquad
l_{\mathcal{E}_X} = 
\begin{pmatrix}
l_\mathcal{E}|_{\mathcal{E}_X}\\
l_{\varphi}|_{\mathcal{E}_X}
\end{pmatrix}.
\end{align*}
Therefore, there exists the following complex of length $3$
\begin{align}
\xymatrix{
\varkappa(\mathcal{E}_X) \ar[r]^-{l_{\mathcal{E}_X}} & P(\mathcal{E}_X) \oplus \varkappa(\mathcal{E}_X) \ar[r]^-{\square_1} & P(\mathcal{E}_X) \ar[r] & 0\,.
}
\label{CompatComp}
\end{align}
Denote $P_0 = \varkappa(\mathcal{E}_X)$, $P_1 = P(\mathcal{E}_X) \oplus \varkappa(\mathcal{E}_X)$, $P_2 = P(\mathcal{E}_X)$, $\square_0 = l_{\mathcal{E}_X}$, and $\mathcal{F} = \mathcal{F}(\mathcal{E}_X)$. Then the complex takes the form
\begin{align*}
\xymatrix{
P_0 \ar[r]^-{\square_0} & P_1 \ar[r]^-{\square_1} & P_2 \ar[r] & 0\,.
}
\end{align*}
We say that it is a \emph{compatibility complex} for $l_{\mathcal{E}_X}$ if the following two conditions hold:
(i) if a $\mathcal{C}$-differential operator $\Delta_0\colon P_1 \to \mathcal{F}$ satisfies $\Delta_0 \circ \square_0 = 0$, then it is of the form $\Delta_1\circ \square_1$ for some $\mathcal{C}$-differential operator $\Delta_1$, (ii) a $\mathcal{C}$-differential operator $\square_2\colon P_2 \to \mathcal{F}$ such that $\square_2 \circ \square_1 = 0$ necessarily vanishes.

Under the assumptions of Section~\ref{SectionRedandQuot}, $X$ is equivalent to $\partial_{\tau}|_{\mathcal{E}}$, the components of $F$ are independent of $\tau$, and both $\mathcal{E}$ and $\mathcal{S}$ are $\ell$-normal. Then we can put
\begin{align*}
\varphi^i = -u^i_{\tau}\,,\qquad l_{\varphi}|_{\mathcal{E}_X} = -\operatorname{diag}(\partial_{\tau})\,,\qquad \Phi|_{\mathcal{E}_X} = -\operatorname{diag}(\partial_{\tau})\,,
\end{align*}
where both $\operatorname{diag}(\partial_{\tau})$ are square matrix operators\footnote{More precisely, $l_{\varphi}|_{\mathcal{E}_X}$ is the $m\times m$ matrix and $\Phi|_{\mathcal{E}_X}$ is the $m_1\times m_1$ matrix.} with $\partial_{\tau}$ on the main diagonal and zeros elsewhere. In addition, the coordinate systems on $J^{\infty}(n, m)$ and $J^{\infty}(n-1, m)$ yield the corresponding inclusions $\mathcal{F}(n-1, m)\subset \mathcal{F}(n, m)$, $\varkappa(n-1, m)\subset \varkappa(n, m)$, $P(n-1, m)\subset P(n, m)$, $\ldots$ and $\mathcal{F}(\mathcal{S})\subset \mathcal{F}(\mathcal{E}_X)$, $\varkappa(\mathcal{S})\subset \varkappa(\mathcal{E}_X)$, $\ldots$ Then for $H\in P(n-1, m)\subset P(n, m)$, one has
\begin{align*}
l_F|_{\mathcal{E}_X} = l_H|_{\mathcal{E}_X} + \Gamma\circ \operatorname{diag}(\partial_{\tau})\,,
\end{align*}
where $\Gamma$ is unambiguously defined. One can show that the corresponding complex is a compatibility complex for $l_{\mathcal{E}_X}$.

We adopt the notation $\mathcal{C}(Q_1, Q_2)$ for the $\mathcal{F}$-module of $\mathcal{C}$-differential operators $Q_1\to Q_2$, where the module structure is given by $f\cdot \nabla\colon r\mapsto f\nabla(r)$. The compatibility complex yields the following resolution of the $\mathcal{F}$-module $\mathcal{C}\Lambda^1$ of Cartan $1$-forms on $\mathcal{E}_X$, $\mathcal{C}\Lambda^1 = \mathcal{C}\Lambda^1(\mathcal{E}_X)$.
\begin{align*}
\xymatrix{
0 \ar[r] & \mathcal{C}(P_2, \mathcal{F}) \ar[r]^-{\circ\,\square_1} & \mathcal{C}(P_1, \mathcal{F}) \ar[r]^-{\circ\, \square_0} & \mathcal{C}(P_0, \mathcal{F}) \ar[r] & \mathcal{C}\Lambda^1\ar[r] & 0\,.
}
\end{align*}
The homomorphism $\mathcal{C}(P_0, \mathcal{F})\to \mathcal{C}\Lambda^1$ maps an operator $A^{\alpha}_i D_{\alpha}|_{\mathcal{E}_X}$ to the Cartan form $A^{\alpha}_i \theta^i_{\alpha}|_{\mathcal{E}_X}$. Following~\cite{Verbovetsky1998}, we describe the groups $E^{p,\hspace{0.1ex} n-2}_1(\mathcal{E}_X)$ using bicomplexes arising from this resolution.

\subsubsection{Bicomplexes}

Tensoring the deleted resolution with $\Lambda^k_h = \Lambda^k_h(\mathcal{E}_X)$, $1\leqslant k\leqslant n$, yields the bicomplex
\begin{align}
\xymatrix{
& 0 & 0 & 0 \\
0 \ar[r] & \mathcal{C}(P_2, \Lambda_h^n) \ar[u] \ar[r]^-{\circ\,\square_1} & \mathcal{C}(P_1, \Lambda_h^n) \ar[u] \ar[r]^-{\circ\, \square_0} & \mathcal{C}(P_0, \Lambda_h^n) \ar[u] \ar[r] & 0\\
0 \ar[r] & \mathcal{C}(P_2, \Lambda_h^{n-1}) \ar[u]^-{-d_0\circ} \ar[r]^-{\circ\,\square_1} & \mathcal{C}(P_1, \Lambda_h^{n-1}) \ar[u]^-{-d_0\circ} \ar[r]^-{\circ\, \square_0} & \mathcal{C}(P_0, \Lambda_h^{n-1}) \ar[u]^-{-d_0\circ} \ar[r] & 0\\
0 \ar[r] & \mathcal{C}(P_2, \Lambda_h^{n-2}) \ar[u]^-{-d_0\circ} \ar[r]^-{\circ\,\square_1} & \mathcal{C}(P_1, \Lambda_h^{n-2}) \ar[u]^-{-d_0\circ} \ar[r]^-{\circ\, \square_0} & \mathcal{C}(P_0, \Lambda_h^{n-2}) \ar[u]^-{-d_0\circ} \ar[r] & 0\\
& &\ldots\\
0 \ar[r] & \mathcal{C}(P_2, \mathcal{F}) \ar[r]^-{\circ\,\square_1} & \mathcal{C}(P_1, \mathcal{F}) \ar[r]^-{\circ\, \square_0} & \mathcal{C}(P_0, \mathcal{F}) \ar[r]  & 0\\
& 0 \ar[u] & 0 \ar[u] & 0 \ar[u] 
}
\label{bicomplex1}
\end{align}
Tensoring the rows with $\mathcal{C}^{p-1}\Lambda^{p-1} = \mathcal{C}^{p-1}\Lambda^{p-1}(\mathcal{E}_X)$, $p\geqslant 1$, we obtain the corresponding bicomplexes 
\begin{align}
\xymatrix{
& 0 & 0 & 0 \\
0 \ar[r] & \mathcal{C}^{p-1}\Lambda^{p-1}\otimes\mathcal{C}(P_2, \Lambda_h^n) \ar[u] \ar[r] & \mathcal{C}^{p-1}\Lambda^{p-1} \otimes \mathcal{C}(P_1, \Lambda_h^n) \ar[u] \ar[r] & \mathcal{C}^{p-1}\Lambda^{p-1} \otimes\mathcal{C}(P_0, \Lambda_h^n) \ar[u] \ar[r] & 0\\
0 \ar[r] & \mathcal{C}^{p-1}\Lambda^{p-1} \otimes \mathcal{C}(P_2, \Lambda_h^{n-1}) \ar[u] \ar[r] & \mathcal{C}^{p-1}\Lambda^{p-1} \otimes \mathcal{C}(P_1, \Lambda_h^{n-1}) \ar[u] \ar[r] & \mathcal{C}^{p-1}\Lambda^{p-1} \otimes \mathcal{C}(P_0, \Lambda_h^{n-1}) \ar[u] \ar[r] & 0\\
0 \ar[r] & \mathcal{C}^{p-1}\Lambda^{p-1} \otimes \mathcal{C}(P_2, \Lambda_h^{n-2}) \ar[u] \ar[r] & \mathcal{C}^{p-1}\Lambda^{p-1} \otimes \mathcal{C}(P_1, \Lambda_h^{n-2}) \ar[u] \ar[r] & \mathcal{C}^{p-1}\Lambda^{p-1} \otimes \mathcal{C}(P_0, \Lambda_h^{n-2}) \ar[u] \ar[r] & 0\\
& &\ldots\\
0 \ar[r] & \mathcal{C}^{p-1}\Lambda^{p-1} \otimes \mathcal{C}(P_2, \mathcal{F}) \ar[r] & \mathcal{C}^{p-1}\Lambda^{p-1} \otimes \mathcal{C}(P_1, \mathcal{F}) \ar[r] & \mathcal{C}^{p-1}\Lambda^{p-1} \otimes \mathcal{C}(P_0, \mathcal{F}) \ar[r]  & 0\\
& 0 \ar[u] & 0 \ar[u] & 0 \ar[u] 
}
\label{bicomplex2}
\end{align}
whose vertical arrows are defined by
\begin{align*}
\omega\otimes \nabla\ \, \mapsto\ \, (-1)^p \mathcal{L}_{\,\widetilde{D}_{x^i}}\omega\otimes dx^i\wedge \nabla + (-1)^p \omega\otimes d_0\circ \nabla,
\end{align*}
where $\,\widetilde{D}_{x^i} = D_{x^i}|_{\mathcal{E}_X}$, $\omega\in \mathcal{C}^{p-1}\Lambda^{p-1}$, while $dx^i\wedge \nabla$ denotes the composition of $\nabla$ and $dx^i\wedge$. The horizontal arrows are the tensor products (over $\mathcal{F}$) of the identity morphism $1_{\mathcal{C}^{p-1}\Lambda^{p-1}}$ and the corresponding morphisms from~\eqref{bicomplex1}.

\subsubsection{The first pages}

Let us describe the first pages of the two spectral sequences associated with the two filtrations of the total complex (see, e.g.,~\cite{rotman2009introduction}) arising from~\eqref{bicomplex2}.

The rows of~\eqref{bicomplex2} are exact everywhere except for the terms in the right nonzero column. The arising homology groups
give rise to the corresponding complex $\mathcal{C}^{p-1}\Lambda^{p-1}(\mathcal{E}_X)\otimes E^{1,\hspace{0.1ex} \bullet}_0(\mathcal{E}_X)$. Note that $E^{p,\hspace{0.1ex} \bullet}_0(\mathcal{E}_X)$ is a direct summand of $\mathcal{C}^{p-1}\Lambda^{p-1}(\mathcal{E}_X)\otimes E^{1,\hspace{0.1ex} \bullet}_0(\mathcal{E}_X)$. The cohomology groups of the total complex are isomorphic to the respective cohomology groups of $\mathcal{C}^{p-1}\Lambda^{p-1}(\mathcal{E}_X)\otimes E^{1,\hspace{0.1ex} \bullet}_0(\mathcal{E}_X)$.

The columns of~\eqref{bicomplex2} are exact everywhere except for the terms in the top nonzero row. The arising cohomology groups
give rise to the corresponding complex
\begin{align}
\xymatrix{
0 \ar[r] & \mathcal{C}^{p-1}\Lambda^{p-1}\otimes \widehat{P}_2 \ar[r]^-{\square^{\hspace{0.15ex} *}_{1\, (p-1)}} & \mathcal{C}^{p-1}\Lambda^{p-1}\otimes \widehat{P}_1 \ar[r]^-{\square^{\hspace{0.15ex} *}_{0\, (p-1)}} & \mathcal{C}^{p-1}\Lambda^{p-1}\otimes \widehat{P}_0 \ar[r] & 0\,,
}
\label{Theothercomp}
\end{align}
where the homomorphisms are obtained through integration by parts. For example, an element of $\mathcal{C}^{p-1}\Lambda^{p-1}\otimes \widehat{P}_2$ can be interpreted as an operator $\psi|_{\mathcal{E}_X}$, where $\psi\colon P(n, m)\to E^{p-1,\hspace{0.1ex} n}_0(n, m)$ is a homomorphism of $\mathcal{F}(n, m)$-modules. Integrating by parts with respect to $a$ and $b$ in 
\begin{align*}
\psi|_{\mathcal{E}_X} \circ \square_1(a, b) \equiv \psi|_{\mathcal{E}_X} (-\Phi|_{\mathcal{E}_X}(a) + l_F|_{\mathcal{E}_X}(b))\,,
\end{align*}
one finds $\square^{\hspace{0.15ex} *}_{1\, (p-1)}(\psi|_{\mathcal{E}_X})$.
A homomorphism of the form $\psi|_{\mathcal{E}_X}$ lies in the kernel of $\square^{\hspace{0.15ex} *}_{1\, (p-1)}$ if and only if $(\Phi^*_{(p-1)} \psi)|_{\mathcal{E}_X} = 0$ and $(l_{H\, (p-1)}^{\hspace{0.15ex} *} \psi)|_{\mathcal{E}_X} = 0$, because $l_F|_{\mathcal{E}_X} = l_H|_{\mathcal{E}_X} - \Phi|_{\mathcal{E}_X} \circ \Gamma$. Note that $\psi|_{\mathcal{E}_X}$ can be written as $\psi_i|_{\mathcal{E}_X}\wedge dx^1\wedge \ldots \wedge dx^n$ for suitable $\psi_i\in \mathcal{C}^{p-1}\Lambda^{p-1}(n, m)$. Since $\Phi|_{\mathcal{E}_X} = -\operatorname{diag}(\partial_{\tau})$, the first condition $(\Phi^*_{(p-1)} \psi)|_{\mathcal{E}_X} = 0$ implies that $\psi_i|_{\mathcal{E}_X}$ are invariant under the flow of $\partial_{\tau}$. Moreover, $\partial_{\tau} \lrcorner\, \psi_i|_{\mathcal{E}_X} = 0$ because $\partial_{\tau}\in \mathcal{C}D(\mathcal{E}_X)$. Then one can show that $\psi_i|_{\mathcal{E}_X}$ are the pullbacks of some elements of $\mathcal{C}^{p-1}\Lambda^{p-1}(\mathcal{S})$. It follows that the second condition $(l_{H\, (p-1)}^{\hspace{0.15ex} *} \psi)|_{\mathcal{E}_X} = 0$ yields the isomorphisms
$\ker l_{\mathcal{S}\, (p-1)}^{\hspace{0.15ex} *} \to \ker \square^{\hspace{0.15ex} *}_{1\, (p-1)}$\,,
\begin{align*}
\eta \mapsto d\tau \wedge \eta\,,\qquad p\geqslant 1\,,
\end{align*}
where $d\tau \wedge \eta$ is the corresponding composition.

The cohomology groups of the total complex are isomorphic to the respective homology groups of~\eqref{Theothercomp}. In particular, the group $\ker \square^{\hspace{0.15ex} *}_{1\, (p-1)}$ is isomorphic to $H^{n-2}(\mathcal{C}^{p-1}\Lambda^{p-1}(\mathcal{E}_X)\otimes E^{1,\hspace{0.1ex} \bullet}_0(\mathcal{E}_X))$, whose direct summand is $E^{p,\hspace{0.1ex} n-2}_1(\mathcal{E}_X)$. More specifically, for $\psi|_{\mathcal{E}_X}\in \ker \square^{\hspace{0.15ex} *}_{1\, (p-1)}$, there exists a $\mathcal{C}$-differential operator $\nu_1\colon P_1\to E^{p-1, \hspace{0.1ex} n-1}_0(\mathcal{E}_X)$ such that $\psi|_{\mathcal{E}_X}\circ \square_1 = d_0\circ \nu_1$. Then $\nu_1\circ \square_0 = d_0 \circ \gamma|_{\mathcal{E}_X}$ for some $\mathcal{C}$-differential operator $\gamma|_{\mathcal{E}_X}$. Any\footnote{\draftnew{This can also be seen by a diagram chase applied to~\eqref{bicomplex2}.}} such $\gamma|_{\mathcal{E}_X}$ gives rise to the corresponding element of $H^{n-2}(\mathcal{C}^{p-1}\Lambda^{p-1}(\mathcal{E}_X)\otimes E^{1,\hspace{0.1ex} \bullet}_0(\mathcal{E}_X))$.
Similarly, $\ker l_{\mathcal{S}\, (p-1)}^{\hspace{0.15ex} *}$ is isomorphic to the cohomology group $H^{n-2}(\mathcal{C}^{p-1}\Lambda^{p-1}(\mathcal{S})\otimes E^{1,\hspace{0.1ex} \bullet}_0(\mathcal{S}))$, whose direct summand is $E^{p,\hspace{0.1ex} n-2}_1(\mathcal{S})$.
Let us note that $\ker d_1^{\hspace{0.1ex} 0,\hspace{0.1ex} n-2} = 0$ and $\operatorname{im} d_1^{\hspace{0.1ex} 0,\hspace{0.1ex} n-2} = \ker d_1^{\hspace{0.1ex} 1,\hspace{0.1ex} n-2}$ for both $\mathcal{E}_X$ and $\mathcal{S}$. Then it suffices to show that the diagram
\begin{align}
\xymatrix{
H^{n-2}(\mathcal{C}^{p-1}\Lambda^{p-1}(\mathcal{E}_X)\otimes E^{1,\hspace{0.1ex} \bullet}_0(\mathcal{E}_X)) \ar[r] & \ker \square_{1\, (p-1)}^{\hspace{0.15ex} *}\\
H^{n-2}(\mathcal{C}^{p-1}\Lambda^{p-1}(\mathcal{S})\otimes E^{1,\hspace{0.1ex} \bullet}_0(\mathcal{S})) \ar[u] \ar[r] & \ker l_{\mathcal{S}\, (p-1)}^{\hspace{0.15ex} *} \ar[u]
}
\label{TheDiagram}
\end{align}
commutes (or anticommutes) for $p\geqslant 1$, 
where all maps are the isomorphisms described above, except the homomorphism $H^{n-2}(\mathcal{C}^{p-1}\Lambda^{p-1}(\mathcal{S})\otimes E^{1,\hspace{0.1ex} \bullet}_0(\mathcal{S}))\to H^{n-2}(\mathcal{C}^{p-1}\Lambda^{p-1}(\mathcal{E}_X)\otimes E^{1,\hspace{0.1ex} \bullet}_0(\mathcal{E}_X))$, which is given by the pullback of the canonical projection $\mathcal{E}_X\to \mathcal{S}$.

\subsection{The final step}

For the sake of simplicity, let us demonstrate the last step in the case $p = 1$. Choose an element of $H^{n-2}(E^{1,\hspace{0.1ex} \bullet}_0(\mathcal{S}))$ and take the corresponding cosymmetry $\eta\in \ker l_{\mathcal{S}}^{\hspace{0.15ex} *}$. There exists a form $\widehat{\omega}\in E^{1,\hspace{0.1ex} n-2}_0(n-1, m)$ such that $\omega = \widehat{\omega}|_{\mathcal{S}}$ represents the element of $H^{n-2}(E^{1,\hspace{0.1ex} \bullet}_0(\mathcal{S}))$ and the operator $\gamma_{\hspace{0.1ex} \widehat{\omega}}\colon \chi \mapsto E_{\chi} \lrcorner\, \widehat{\omega}$ satisfies
\begin{align*}
\eta\circ l_{\mathcal{S}} = d_0^{\hspace{0.1ex} \mathcal{S}}\circ \gamma_{\hspace{0.1ex} \widehat{\omega}}|_{\mathcal{S}}\,,
\end{align*}
where\footnote{This is formula~\eqref{Cosymmvarform} for $\mathcal{S}$.} $d_0^{\hspace{0.1ex} \mathcal{S}}$ denotes the corresponding differential on $\mathcal{S}$. 

Let us show that the isomorphism $H^{n-2}(E^{1,\hspace{0.1ex} \bullet}_0(\mathcal{E}_X))\to \ker \square_{1}^{\hspace{0.15ex} *}$ relates the pullback of the element of $H^{n-2}(E^{1,\hspace{0.1ex} \bullet}_0(\mathcal{S}))$ to the composition $d\tau\wedge \eta$.
The differential $d_0$ on $\mathcal{E}_X$ can be written in the form $d_0 = d_0^{\hspace{0.1ex} \mathcal{S}} + d\tau\wedge \mathcal{L}_{\partial_{\tau}}$, while the restriction of $\widehat{\omega}\in E^{1,\hspace{0.1ex} n-2}_0(n-1, m)\subset E^{1,\hspace{0.1ex} n-2}_0(n, m)$ to $\mathcal{E}_X$ represents the pullback of the element of $H^{n-2}(E^{1,\hspace{0.1ex} \bullet}_0(\mathcal{S}))$.
One can check that the following relations hold for the operator $\nu_1 = (\eta,  \eta\circ \Gamma -d\tau\wedge \gamma_{\hspace{0.1ex} \widehat{\omega}}|_{\mathcal{E}_X})\in \mathcal{C}(P_1, \Lambda_h^{n-1})$.
\begin{align*}
\nu_1\circ \square_0 = d_0\circ \gamma_{\hspace{0.1ex} \widehat{\omega}}|_{\mathcal{E}_X}\,,\qquad d\tau\wedge \eta \circ \square_1 = d_0\circ \nu_1\,.
\end{align*}
This implies that the isomorphism $H^{n-2}(E^{1,\hspace{0.1ex} \bullet}_0(\mathcal{E}_X)) \to \ker \square_1^{\hspace{0.15ex} *}$ maps the pullback to the composition $d\tau\wedge \eta$, and hence, diagram~\eqref{TheDiagram} commutes if $p = 1$.

\vspace{1ex}

\remarka{If $X$ is a higher symmetry of $\mathcal{E}$, a compatibility complex for a linearization of the corresponding system $\mathcal{E}_X$ can be longer (see, e.g.,~\cite[Example 2]{InvRedII}, where the length of any compatibility complex is $\geqslant 4$).}

\vspace{1ex}

\remarka{If $\mathcal{E}$ is $\ell$-normal, the equation $\square\circ l_F|_{\mathcal{E}_X} = 0$ for a total differential operator $\square\colon P(\mathcal{E}_X) \to \mathcal{F}(\mathcal{E}_X)$ has no nonzero solutions, and \eqref{CompatComp} is a compatibility complex, then Theorem~\ref{Theoremgenerfun} shows that the invariant reduction mechanism can be described as $\psi^{p-1}|_{\mathcal{E}}\mapsto \psi^{p-1}|_{\mathcal{E}_X}$ for $p\geqslant 1$, up to sign (with $\nu_1 = (-\nu|_{\mathcal{E}_X}, \gamma_{\widehat{\omega}}|_{\mathcal{E}_X})$).}

\section{The homotopy}\label{App:D}

Let us recall the classical generalization of integration by parts.

Let $V$ be an $n$-dimensional vector space over $\mathbb{R}$. Denote by $S^r(V)$ the $r$-th symmetric power, $S^0(V) = \mathbb{R}$, while $S^{-1}(V) = 0$, $S^{-2}(V) = 0$, $\ldots$

Choose a basis $e_1, \ldots, e_n\in V$ and its dual $e^1, \ldots, e^n\in V^*$. Similarly, we let $\Lambda^{k}(V^*)$ denote the $k$-th exterior power, $\Lambda^0(V^*) = \mathbb{R}$. Consider the linear operator $\delta\colon S^{r+k}(V) \otimes \Lambda^k(V^*)\to S^{r+k+1}(V) \otimes \Lambda^{k+1}(V^*)$ determined by
\begin{align*}
\delta \colon w \otimes \xi \mapsto e_i w \otimes (e^i\wedge \xi)\,,\qquad w\otimes \xi \in S^{r+k}(V) \otimes \Lambda^k(V^*)\,.
\end{align*}
Here $e_i w$ is the symmetric product. Since $\delta\circ \delta = 0$, there exist the following complexes for $r \geqslant -n$
\begin{align*}
\xymatrix{
0 \ar[r] & S^r(V)\otimes \Lambda^0(V^*) \ar[r]^-{\delta} & \ldots \ar[r]^-{\delta} & S^{r+n-1}(V)\otimes \Lambda^{n-1}(V^*) \ar[r]^-{\delta} & S^{r+n}(V)\otimes \Lambda^n(V^*) \ar[r] & 0
}
\end{align*}
Let us recall that elements of $S^r(V)$ can be interpreted as homogeneous polynomial functions on $V^*$ (of degree $r$ if $r\geqslant 0$). Then we can consider their partial derivatives with respect to the basis vectors in the usual sense. For example, if $w = (e_1 + e_2)e_1 + e_2e_3$, then $\partial w/\partial e_1 = 2e_1 + e_2$. Consider the following homotopy $h\colon S^{r+k}(V) \otimes \Lambda^k(V^*)\to S^{r+k-1}(V) \otimes \Lambda^{k-1}(V^*)$ for $k \geqslant 0$,
\begin{align}
h\colon w \otimes \xi \mapsto \dfrac{\partial w}{\partial e_j} \otimes (e_j\hspace{0.1ex} \lrcorner\, \xi)\,.
\label{HomotopyOperator}
\end{align}
If $k = 0$, we put $e_j\hspace{0.1ex} \lrcorner\, \xi = 0$. Then 
\begin{align*}
h\circ \delta + \delta\circ h\colon w \otimes \xi \mapsto
(n+r) w \otimes \xi\,,\qquad w \otimes \xi \in S^{r+k}(V) \otimes \Lambda^k(V^*)\,,\ k\geqslant 0\,.
\end{align*}

\end{document}